\documentclass[twocolumn]{aastex62}

\usepackage{amsmath}
\usepackage{stix}
\usepackage{xspace}

\newcommand{\mjup}{\ensuremath{M_\mathrm{Jup}}\xspace}
\newcommand{\hst}{\textit{HST}\xspace}

\graphicspath{{./}{figures/}}

\received{2018 Dec 6}
\revised{2019 Jan 23}
\accepted{2019 Jan 29}

\shorttitle{Rotational Modulations of Directly-Imaged Exoplanets}
\shortauthors{Zhou et al.}
\newcommand{\teff}{\ensuremath{T_\mathrm{eff}}\xspace}
\newcommand{\logg}{\ensuremath{\log g}\xspace}
\newcommand{\targeti}{\object{AB Pic B}\xspace}
\newcommand{\targetii}{\object[2MASS 0122-2439 b]{2M0122B}\xspace}
\newcommand{\targetiii}{\object[TWA 27B]{2M1207b}\xspace}
\newcommand{\targetI}{AB~Pic~B\xspace}
\newcommand{\targetII}{2M0122B\xspace}
\newcommand{\targetIII}{2M1207b\xspace}

\begin{document}
\turnoffedit1
\title{Cloud Atlas: High-Contrast Time-Resolved Observations of Planetary-Mass Companions}

\correspondingauthor{Yifan Zhou}
\email{yzhou@as.arizona.edu}
\author[0000-0003-2969-6040]{Yifan Zhou}
\altaffiliation{NASA Earth and Space Science Fellow}
\affiliation{Department of Astronomy/Steward Observatory, The University of Arizona, 933 N. Cherry Avenue, Tucson, AZ, 85721, USA}

\author{D\'aniel Apai}
\affiliation{Department of Astronomy/Steward Observatory, The University of Arizona, 933 N. Cherry Avenue, Tucson, AZ, 85721, USA}
\affiliation{Department of Planetary Science/Lunar and Planetary Laboratory, The University of Arizona, 1640 E. University Boulevard, Tucson, AZ, 85718, USA}
\affiliation{Earths in Other Solar Systems Team, NASA Nexus for Exoplanet System Science.}

\author{Ben W. P. Lew}
\affiliation{Department of Astronomy/Steward Observatory, The University of Arizona, 933 N. Cherry Avenue, Tucson, AZ, 85721, USA}
\affiliation{Department of Planetary Science/Lunar and Planetary Laboratory, The University of Arizona, 1640 E. University Boulevard, Tucson, AZ, 85718, USA}

\author{Glenn Schneider}
\affiliation{Department of Astronomy/Steward Observatory, The University of Arizona, 933 N. Cherry Avenue, Tucson, AZ, 85721, USA}

\author{Elena Manjavacas}
\affiliation{Department of Astronomy/Steward Observatory, The University of Arizona, 933 N. Cherry Avenue, Tucson, AZ, 85721, USA}

\author{Luigi R. Bedin}
\affiliation{INAF – Osservatorio Astronomico di Padova, Vicolo dell'Osservatorio 5, I-35122 Padova, Italy}

\author{Nicolas B. Cowan}
\affiliation{Department of Earth \& Planetary Sciences and Department of Physics, McGill University, 3550 Rue University, Montr\'eal, Quebec H3A 0E8, Canada}

\author{Mark S. Marley}
\affiliation{NASA Ames Research Center, Mail Stop 245-3, Moffett Field, CA 94035, USA}

\author{Jacqueline Radigan} \affiliation{Utah Valley University, 800 West University Parkway, Orem, UT 84058, USA}

\author{Theodora Karalidi}
\affiliation{Department of Astronomy and Astrophysics, University of California Santa Cruz, 1156 High Street, Santa Cruz, CA 95064, USA}

\author{Patrick J. Lowrance}
\affiliation{IPAC-Spitzer, MC 314-6, California Institute of Technology, Pasadena, CA 91125, USA}

\author{Paulo A. Miles-P\'aez}
\affiliation{Department of Physics \& Astronomy and Centre for Planetary Science and Exploration, The University of Western Ontario, London, Ontario N6A 3K7, Canada}
\affiliation{Department of Astronomy/Steward Observatory, The University of Arizona, 933 N. Cherry Avenue, Tucson, AZ, 85721, USA}

\author{Stanimir Metchev}
\affiliation{Department of Physics \& Astronomy and Centre for Planetary Science and Exploration, The University of Western Ontario, London, Ontario N6A 3K7, Canada}
\affiliation{Department of Physics \& Astronomy, Stony Brook University, 100 Nicolls Rd, Stony Brook, NY 11794-3800, USA}

\author{Adam J. Burgasser} \affiliation{Center for Astrophysics and Space Science, University of California San Diego, La Jolla, CA 92093, USA}

\begin{abstract}
  Directly-imaged planetary-mass companions offer unique opportunities in atmospheric studies of exoplanets. They share characteristics of both brown dwarfs and transiting exoplanets, therefore, are critical for connecting atmospheric characterizations for these objects. Rotational phase mapping is a powerful technique to constrain the condensate cloud properties in ultra-cool atmospheres. Applying this technique to directly-imaged planetary-mass companions will be extremely valuable for constraining cloud models in low mass and surface gravity atmospheres and for determining the rotation rate and angular momentum of substellar companions. Here, we present Hubble Space Telescope Wide Field Camera 3 near-infrared time-resolved photometry for three planetary-mass companions, AB Pic B, 2M0122B, and 2M1207b.  Using two-roll differential imaging and hybrid point spread function modeling, we achieve sub-percent photometric precision for all three observations. We find tentative modulations ($<\!\!2\sigma$) for AB Pic B and 2M0122B but cannot reach conclusive results on 2M1207b due to strong systematics. The \edit1{relatively low significance of the modulation measurements} cannot rule out the  hypothesis that these planetary-mass companions have the same vertical cloud structures as brown dwarfs. Our rotation rate measurements, combined with archival period measurements of planetary-mass companions and brown dwarfs do not support a universal mass-rotation relation. The high precision of our observations and the high occurrence rates of variable low-surface gravity objects encourage high-contrast time-resolved observations with the James Webb Space Telescope.
\end{abstract}
\keywords{Planetary Systems --- planets and satellites: atmospheres --- methods: observational}

\section{Introduction}
Among all planet detection methods, direct-imaging \citep[e.g.,][]{Chauvin2004,Marois2008,Lagrange2010,Macintosh2015a} is most fruitful in discovering the targets that are most suitable for atmospheric characterization. Compared to exoplanets that are detected by transiting or radial-velocity techniques, directly-imaged planets have emission photometry/spectra often observed at many times higher signal-to-noise ratios (SNR) and thus result in more precise and higher resolution spectra \citep[e.g.,][]{Patience2012,Ingraham2014,Samland2017}. The development of both extreme ground-based adaptive optics (AO) systems \citep[e.g.,][]{Macintosh2008,Beuzit2008} as well as space-based high-contrast imaging \citep[e.g.,][]{Song2006,Rajan2015,Zhou2016} and image post-processing alogorithms \citep[e.g.,][]{Lafreniere2007,Soummer2012,Hoeijmakers2018} have advanced the field.  The high SNR spectra of directly-imaged planets enables precise measurements of fundamental atmospheric characteristics, such as effective temperatures, surface gravities, and molecular abundances \citep[e.g., ][]{Konopacky2013,Barman2015}. These measurements suffer less confusions from instrumental systematics and stellar activity contamination \citep{Rackham2018} compared to those from transmission spectroscopic observations. The direct-imaging observations also show that, similar to those of brown dwarfs and transiting exoplanets,  the spectra of directly-imaged exoplanets are strongly affected by condensate clouds \citep[e.g.,][]{Barman2011,Skemer2011,Rajan2017}.

Atmospheric models predict the formation of condensate clouds in ultra-cool atmospheres \citep[e.g.,][]{Ackerman2001,Marley2002,Burrows2006a,Helling2008,Saumon2008,Morley2012,Charnay2018,Tan2018}. With their strong opacity, condensate clouds primarily affect the near-infrared (near-IR) spectra of exoplanets and brown dwarfs in two ways. First, clouds redden near-IR emission spectra. Second, clouds reduce the depth of  molecular absorption features. Because of these two effects, there are significant degeneracies between the assumptions of clouds in atmospheric models and the derived values for basic atmospheric characteristics, such as effective temperatures and molecular abundances. These degeneracies intensify the challenges to retrieve the cloud properties and test cloud models.

Such degeneracies may be broken by extending the observations into the time-domain with rotational phase mapping. Brown dwarfs and directly-imaged exoplanets have rotationally modulated photometric and spectral variabilities due to heterogeneous clouds \citep[e.g.,][]{Artigau2009,Radigan2012,Apai2013,Buenzli2014,Biller2015,Metchev2015,Zhou2016}. Long term (five rotation or more) \textit{Spitzer} Space Telescope monitoring of brown dwarfs \citep{Apai2017} demonstrated that the cloud thickness variations in many, perhaps most, BDs are caused by planetary-scale waves, i.e., large scale atmospheric dynamics. In the rotational phase-mapping technique, spectra of the same object in different rotational phases are observed and compared.  The variations in these spectra, which are introduced by differences in clouds,  constrain the cloud properties. High SNR spectral time-series, especially those from Hubble Space Telescope/Wide Field Camera 3 (HST/WFC3) observations, have revealed vertical cloud structures in several highly-variable brown dwarfs \citep[e.g.,][]{Buenzli2012,Apai2013,Lew2016,Biller2017}. These case studies combined together have revealed the dependence of vertical cloud structures on effective temperature (\teff) for objects ranging from mid-L to mid-T types.

Surface gravity (\logg) and stellar irradiation, besides \teff, are other essential characteristics that brown dwarfs and exoplanets. Directly-imaged planetary-mass objects whose surface gravities are intermediate to those of \edit1{typical high surface-gravity brown dwarfs} and transiting exoplanets are critical in exploring the parameter space of surface gravity. The dependence of cloud properties on surface gravity determines whether cloud models that are constrained with brown dwarf observations are applicable to transiting exoplanets that have surface gravity one to two orders of magnitude lower.  Recent studies of field planetary-mass objects \citep[e.g., ][]{Biller2015,Vos2017,Vos2018b} and planetary-mass companions \citep[e.g.,][]{Zhou2016,Manjavacas2017} begun to explore this parameter space. Encouragingly, low surface gravity objects are found more likely to be variable \citep[e.g,][]{Metchev2015,Vos2018b}, thus are better targets for rotational modulation studies. Furthermore, two objects out of three from the sample of \citet{Apai2017} that demonstrated planetary waves are classified as low-gravity or planetary-mass objects. The growing sample of direclty-imaged planetary-mass companions calls for a high-contrast time-resolved observation survey to investigate the relationship between clouds and surface gravity. 

\textit{Cloud Atlas}, an HST large treasury program \citep[Program GO-14241, PI: D. Apai, summary see][]{Manjavacas2019} is the first HST time-resolved observation survey that includes a large sample of planetary-mass companions. Fifteen out of twenty of the targets in the program are intermediate or low surface gravity objects and  seven objects are high-contrast planetary-mass companions. In this paper, we describe the observations and results for two high-contrast targets from \textit{Cloud Altas} program (2M1207b, 2M0122B) and one additional target AB Pic B from HST program GO-13418 \footnote{The usages of lower or upper cases for letter ``b'' follow \citet{Bowler2016}. In this review article, 2M1207b is categorized as ``Planetary-mass Companions Orbiting Brown Dwarfs'' and 2M0122B and AB Pic B are categorized as ``Candidate Planets and Companions Near the Deuterium-burning Limit''.}.

\subsection{The Three Targets}
\begin{deluxetable*}{ccccccccc}
  \tablenum{1}
  \tablecaption{Planetary-mass Information Summary \label{tab:targets}}
  
  \tablehead{
    \colhead{Object} &
    \colhead{Spec. Type} &
    \colhead{Spec. Type} &
    \colhead{Distance\tablenotemark{a}} &
    \colhead{Sep. } &
    \colhead{Sep.} &
    \colhead{J phot} &
    \colhead{J contrast} &
    \colhead{Ref.} \\
    \colhead{} &
    \colhead{Host} &
    \colhead{Companion} &
    \colhead{pc} &
    \colhead{Angular ($\arcsec$)} &
    \colhead{Physical (au)} &
    \colhead{mag} &
    \colhead{mag} &
    \colhead{} 
  }
  
  \startdata
  AB Pic B & K2   & L0$-$L1 & 50.1 & 5.45 & 273  & 16.2 & 8.6 & (1), (3), (4) \\
  2M0122B  & M3.5 & L4$-$L6 & 33.8 & 1.45 & 49.0 & 16.8 & 6.8 & (5)           \\
  2M1207b  & M8   & L5      & 64.4 & 0.78 & 50.2 & 20.0 & 7.0 & (1), (4)      \\
  \enddata
  
    \tablenotetext{a}{Distances are from \citet{Gaia2016,Gaia2018}. }
    \tablerefs{(1)\citet{Chauvin2004}; (2)\citet{Chauvin2005}; (3) \citet{Bonnefoy2010}; (4) \citet{Patience2010}; (5) \citet{Bowler2013}}
    
  \end{deluxetable*}
  
\targeti \citep{Chauvin2005} is a planetary-mass companion to a K2V star. It has an angular separation of $5\arcsec.45$ from its host star corresponding to a projected physical separation of $\sim 273$ au at a distance of 50.1 pc \citep{Gaia2016,Gaia2018}. \edit1{As estimated by the  BANYAN~$\Sigma$ tool \citep{Gagne2018}, AB Pic is likely a member of the $\sim30$ Myr old Carina association with a probability of 99.5\%. } For this age, an isochrone-based estimate suggests a mass of 10-14\,\mjup, and luminosity-based mass estimate results in 1-24\,\mjup  \citep{Bonnefoy2010} \edit1{for AB~Pic~B}. The near-infrared (near-IR) color of \targetI is significantly redder for its spectral-type (L0$-$L1) \citep[e.g.,][]{Bonnefoy2010,Patience2010} comparing to field brown dwarfs, which places it alongside with the directly-imaged exoplanets HR8799 bcde and \targetiii as archetypal low-gravity, near-IR reddened objects. One proposal to explain such near-IR reddening is to introduce thick dusty condensate clouds in these atmospheres \citep[e.g.,][]{Skemer2011}. 

\targetii \citep{Bowler2013} is a planetary-mass companion to a M3.5V star. The system has an angular separation of $1\arcsec.45$ that corresponds to a projected physical separation of $\sim49.0$ au at a distance of 33.8 pc \citep{Gaia2016, Gaia2018}.  The mass of \targetII is estimated to be $10-30 M\mathrm{_{Jup}}$. \edit1{The BANYAN~$\Sigma$ \citep{Gagne2018} estimates that 2M0122 has a 98.6\% probability to be a member of the $\sim120$ Myr old AB Dor moving group}.  Upon the discovery, \citet{Bowler2013} found a  discrepancy of \targetII{}'s effective temperature between derivations from evolutionary models and spectral type classification. This discrepancy suggested that \targetII, too,  has a dusty atmosphere.

Our third target \targetiii \citep{Chauvin2004,Song2006} is a 4-8 \mjup companion to a M8 brown dwarf. The system has an angular separation of $\sim0\arcsec.78$ that corresponds to a projected physical separation of $\sim50.2$~au at a distance of 64.4 pc \citep{Gaia2016, Gaia2018}. \edit1{It is likely a member of the $\sim10$ Myr old TW Hydrae association with a probability of 99.4\% according to the BANYAN~$\Sigma$ tool \citep{Gagne2018}}. Since its discovery, its low mass and red near-IR color indices have drawn great attention. Similar to \targeti and \targetiii, its red NIR color suggests a dusty and possibly cloudy atmosphere. \citet{Zhou2016} discovered that \targetIII had rotational modulations in its \hst/WFC3 NIR light curves. This discovery placed \targetIII as a primary candidate to study the condensate clouds in a directly-imaged, low-surface gravity planetary-mass object.

The summarize of the target information is listed in Table~\ref{tab:targets}

\section{Observations}
We observed \targetI from Oct 16, 2013 15:53:47 UTC to Oct 17, 2013 00:31:13 UTC using Hubble Space Telescope/Wide Field Camera 3 (\hst/WFC3) time-resolved near-IR photometry. The observations were part of \hst Program GO-13481 (PI:Apai). \hst monitored \targetI in two wide-pass NIR filters, F125W (wide J, $\lambda_\mathrm{pivot}=1.2486\micron$, FWHM$=0.2845\micron$) and F160W (H short, $\lambda_\mathrm{pivot}=1.5346\micron$, FWHM$=0.2683\micron$), for six contiguous \hst orbits, over a temporal baseline of 8.5 hr. During the observations, we alternated two filters after every four frames. Here we define an exposure ``group'' as a set of four sequential frames using the same filters. The exposure times were 30\,s and 15\,s for F125W and F160W frames, respectively. We planned the observations to enable two-roll angular differential imaging \citep[2RDI, e.g.,][]{Song2006} to remove the bright host star's PSFs. \hst's roll angle differed by 28 degrees between orbits 1, 3, 5 and orbits 2, 4, 6. Standard $2\times2$-point dither pattern (dither distance: 1\arcsec.375 in $x$ and $y$ directions) was applied in the first three orbits. The remaining three orbits were not dithered and the pointing stayed on the first dither position. \edit1{The dither pattern difference  allowed us to compare the time series photometry performance between the two dithering strategies, in particular, to evaluate the influence of the imperfect flat field correction (i.e., imperfectly characterized pixel-to-pixel sensitivity variations).}

Observations of \targetIII and \targetII are part of \hst large treasury program \emph{Cloud Atlas}. We observed \targetIII on 2016-04-09 from 15:06:48 to 22:04:42 UTC and on 2016-04-10 from to 11:46:39 to 20:19:58 UTC and \targetII on 2017-12-09 from 06:51:26 to 15:30:07 UTC using \hst/WFC3 NIR time-resolved photometry. Telescope and instrument configurations resembled those for \targetI except filter selections. With the primary goal of deriving the condensate cloud deck altitudes, the \emph{Cloud Atlas} observations concentrate on observing the rotational modulations in- and out-of 1.4\,\micron{} water absorption band, because the modulation amplitude difference in these two bands is sensitive to the cloud altitudes \citep{Yang2014,Yang2016}. We thus used medium-pass filters F127M (water band continuum, $\lambda_{\mathrm{pivot}}=1.2740\micron$, FWHM$=0.0688\micron$) and F139M (water absorption band, $\lambda_{\mathrm{pivot}}=1.3838\micron$, FWHM=0.0643$\micron$). As a result of the total transmissions of the medium-pass filters than that of the wide-pass filters, the exposure times for the observations for \targetIII and \targetII were longer than those for \targetI. For the even fainter \targetIII, the exposure time was 179\,s for both filters, and for \targetII, the exposure times were 44.3\,s and 66.3\,s for F127M and F139M respectively. Because of the known rotational modulations for \targetIII, we observed it with two six orbits segments that were separated by 12.8\,hr (8 \hst orbits), which accumulated 60 F127M frames and 96 F139M frames. For \targetII, the observations were obtained in a single six contiguous orbits, which resulted in 96 F127M frames and 108 F139M frames. Telescope roll setups were the same as the observations for \targetI, but dithering was not applied in the observations for \targetIII or \targetII for pointing stability.

\section{Data Reduction}
For high-contrast photometric time-series, the choice of data reduction approach depends on the separation and contrast of the target. For \targetI that has a large separation and moderate contrast (see Figure~\ref{fig:ABPIC-image}), we found the more straight-forward roll subtraction with aperture photometry method is adequate to achieve nearly photon-noise limit photometry. For the higher contrast and closer separated \targetIII and \targetII, we adopt the more complex hybird-PSF photometry method to properly separate the flux of the companion from the brighter host.

The peak pixels in the point spread functions (PSFs) of \targetI exceed a conservatively estimated saturation threshold limit of 70000 e$^-$. The up-the-ramp fitting in the non-linearity regime introduces systematics and degraded the light curve precision measured directly from the \texttt{flt} frames produced by the calwfc3 calibration software\footnote{See appendix E of the WFC3 instrument handbook \citep{Dressel2018}.}. We found that using the last non-saturated reads from the calwfc3 \texttt{ima} frames effectively reduced the standard deviations of the light curves by $\sim 30\%$. This improvement agrees with the comparison of light curves measured in \texttt{flt} and \texttt{ima} frames by \citet{Mandell2013}. We therefore adopted the light curve measured in the last non-saturated reads from \texttt{ima} frames. To ensure that the photometry is not degraded by cosmic rays, we manually examined the up-the-ramp linear fit and confirm no $5\sigma$ outliers.
\subsection{\targetI - Two-Roll Angular Differential Imaging and Aperture Photometry}
\label{sec:ABPIC-reduction}

\begin{figure}[htbp]
  \centering
  \plotone{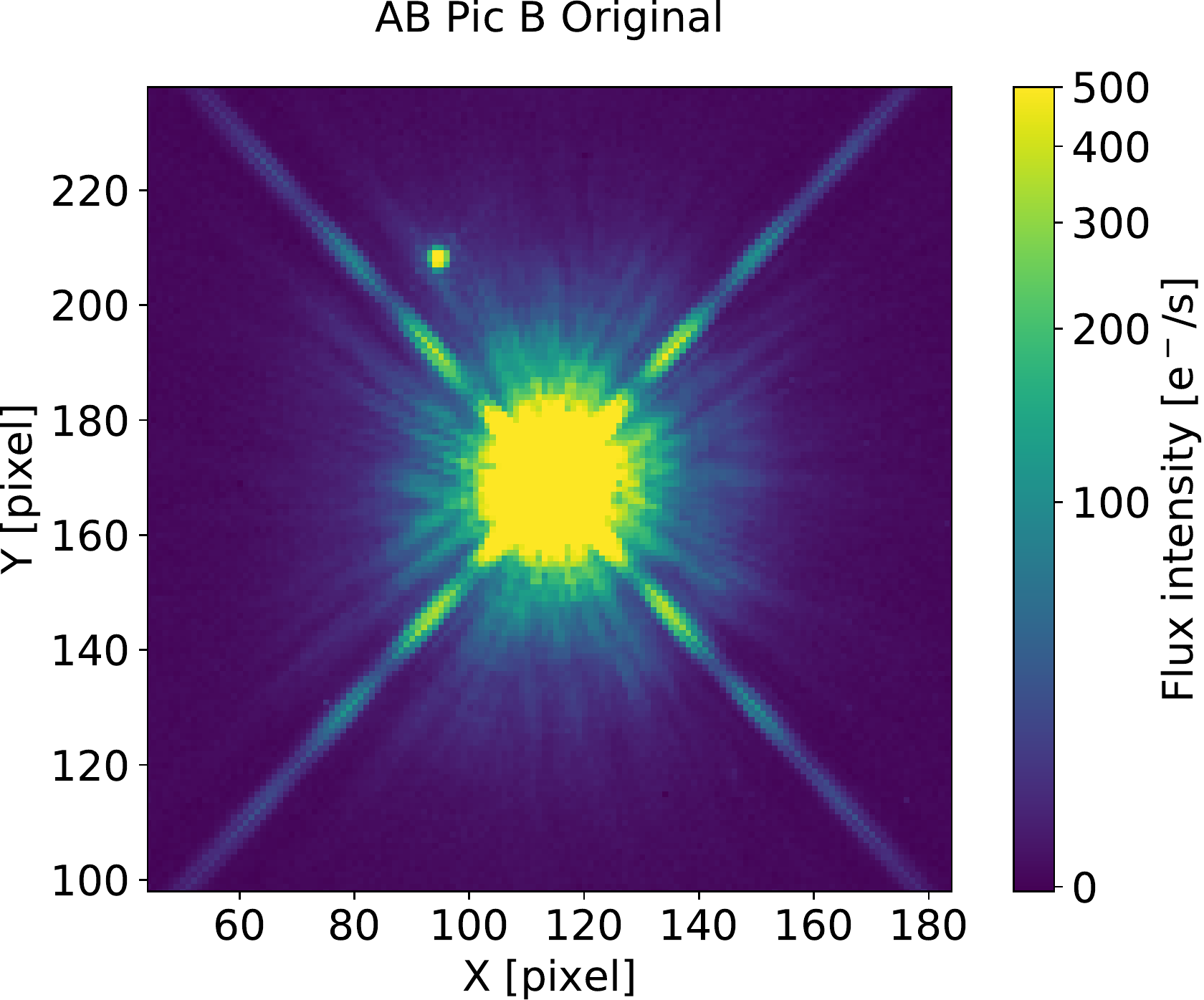}
  \plotone{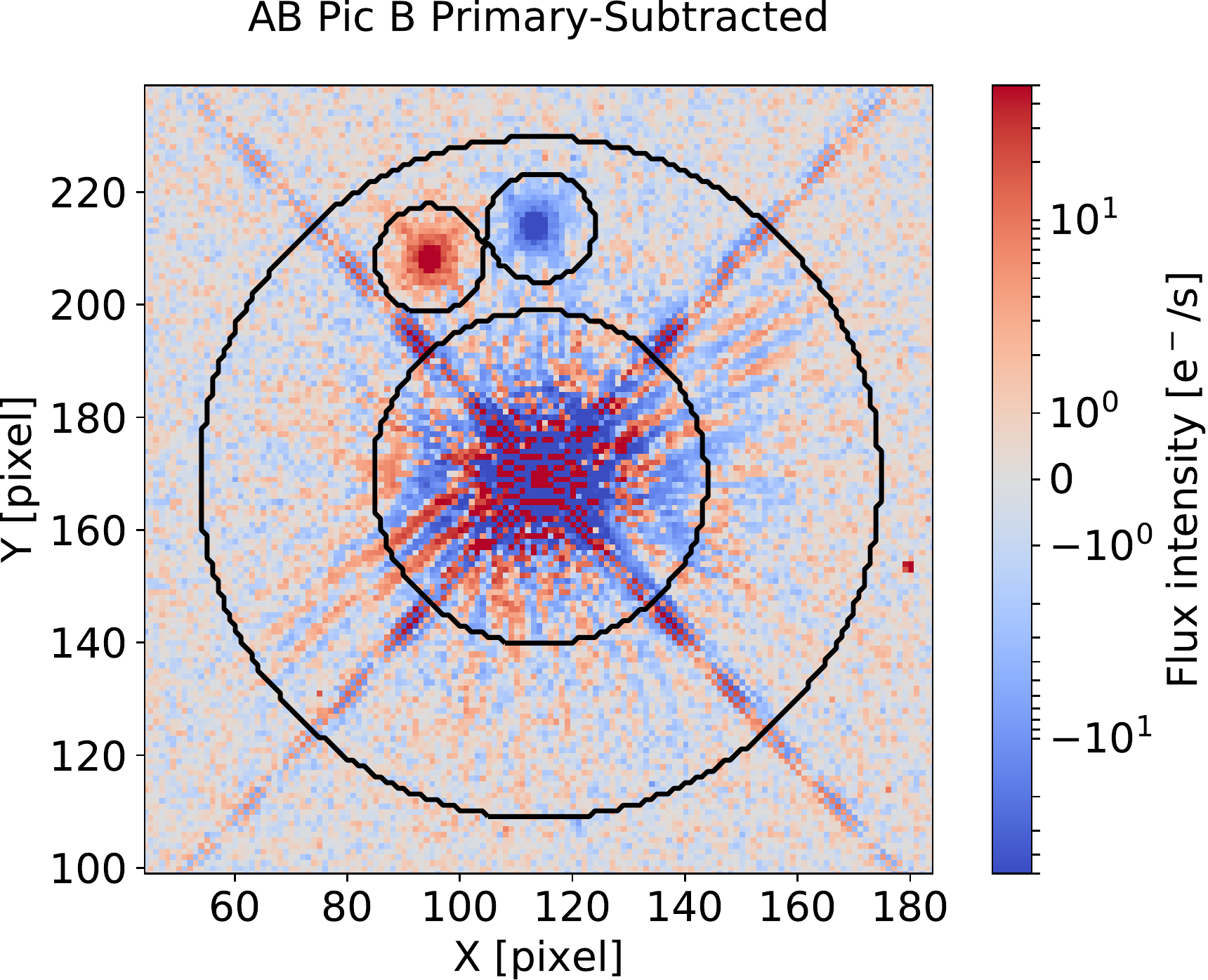}
  \caption{Examples of original image (upper panel) and 2RDI primary subtraction (lower panel) result for \targetI. The negative (blue) PSF in the lower panel are the result of AB Pic B's different angular positions in images with different telescope rolls.  The black contours in the right panel define the region  used to calculate and minimize the residuals for 2RDI.}
  \label{fig:ABPIC-image}
\end{figure}

\targetI, with an angular separation of $5\arcsec.45$ ($\sim\!42$ pixels) to AB Pic A, is visible in the original frames before primary PSF subtraction (Figure \ref{fig:ABPIC-image}) despite the primary star being 8.5 magnitude brighter. In the vicinity of \targetI's PSF,  the background contamination was low in level and spatial fluctuation. These two characteristics ensure robust background-light reduction with two-roll differential imaging \edit1{\citep[2RDI, e.g.,][]{Song2006}}. We removed the PSF of the primary star by 2RDI in four steps. 1) For each filter, we organized the images into two cubes (I and II) by telescope rolls . Images in Cube I are candidate PSFs for images in Cube II and vice versa. 2) We measured the position offset of every frame to the \edit1{reference frame (which was the first one in each image cube)} using two-dimensional cross-correlation \citep[implemented using][]{Ginsburg2014}.  3) We shifted the candidate PSFs by the position offsets using cubic interpolation and aligned the PSF and target images. For a given target image, every candidate PSF was multiplied by a scale factor that minimizes the RMS residuals and then subtracted. The RMS residuals were calculated  within the annular region bounded by the two concentric circles as illustrated in Figure~\ref{fig:ABPIC-image},  excluding two 10-pixel radius regions containing the positive and negative 2RDI imprints of AB Pic B.  4) The PSF that resulted in the least root-mean-squared (RMS) residuals is  selected as the best PSF.   The PSF-subtracted image (Figure~\ref{fig:ABPIC-image}) was the differential between the target image and the best PSF. \edit1{These four reduction steps were applied to every frame in each filter. The final product was PSF-subtracted image cubes for each filter.}

\begin{figure}
  \centering
  \plotone{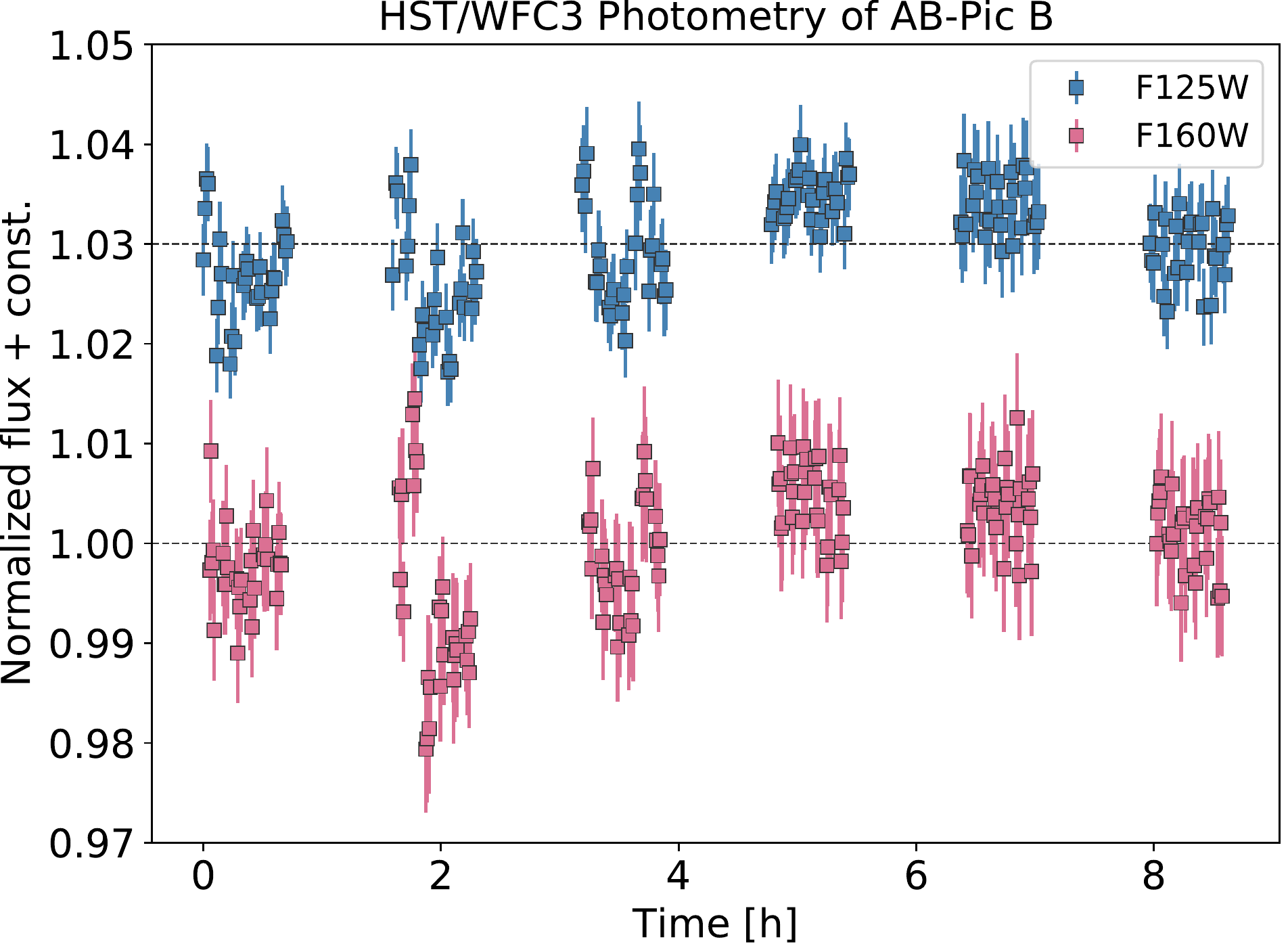}
  \caption{Normalized light curve of \targetI in the F125W (blue) and the F160W (pink) filters before systematics correction. 3\% Vertical offset is applied to the F125W light curve for clarity. Light curves are intermittent because of earth occultations (This effect applies to all HST light curves shown in this paper). Variability is visible particularly in the first three orbits.}
  \label{fig:ABPIC-LCRAW}
\end{figure}

\begin{figure}
  \plottwo{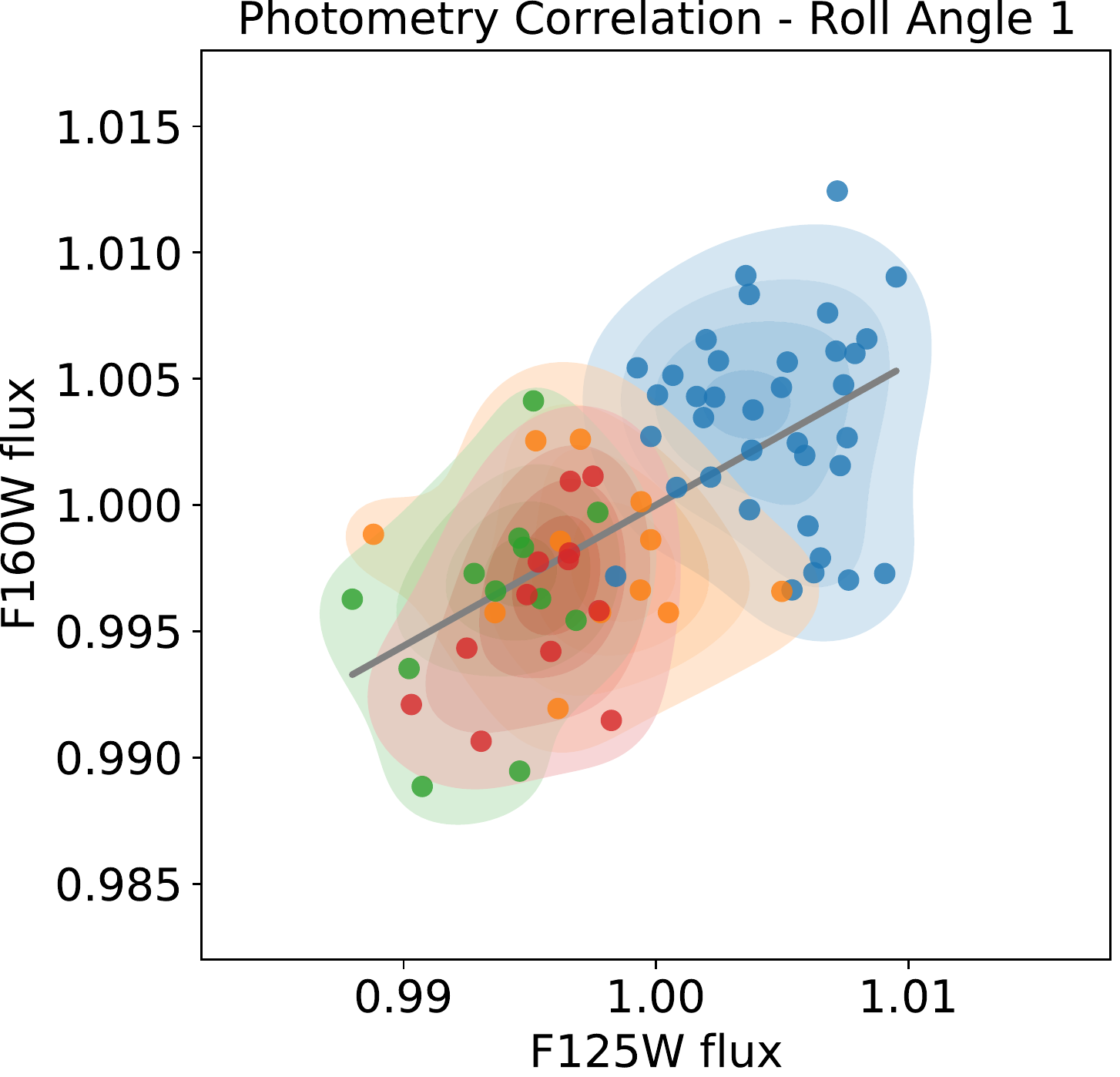}{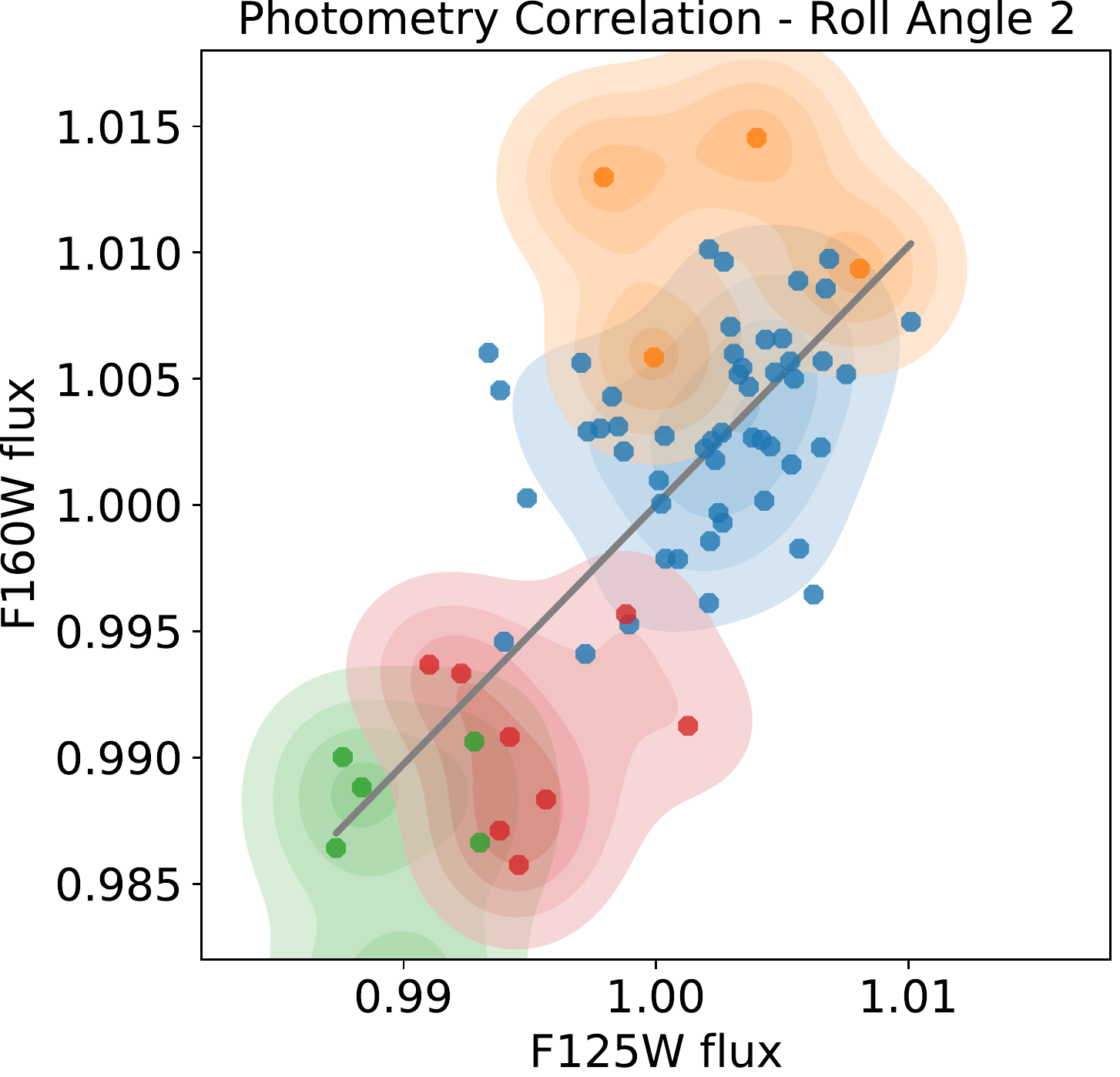}
  \caption{Correlation between the light curves in two photometric bands and the associations between apparent photometric variability and dithering positions. The two panels are for measurements in telescope Roll 1 and Roll 2. Photometric measurements made at dithering positions 1, 2, 3, and 4 are color coded in blue, orange, green, and red, respectively. The contour and shades are kernel density estimates with a Gaussian kernel and a bandwidth of 0.0025. For measurements made in both telescope rolls, the photometric points are clustered by the dithering positions, e.g., in the left panel, flux intensities measured at the first dithering position (blue) are higher in both photoemtric bands than those for other dithering positions.}
  \label{fig:ABPIC-corr}
\end{figure}

We next measured the photometry from the PSF subtracted frames using IDL procedure \texttt{APER} \footnote{The routine \texttt{APER} is available at the \href{https://idlastro.gsfc.nasa.gov/ftp/pro/idlphot/aper.pro}{NASA Goddard ASTRON library}.} with a radius of 5 pixels. The raw light curves demonstrated strong correlations between flux intensity and the dithering positions (Figure \ref{fig:ABPIC-LCRAW}, \ref{fig:ABPIC-corr}). Light curve variations that were associated with dithering positions were as large as 2\%. This correlation reflected the uncertainties in the flat field calibration of the WFC3 detectors, for which the precision is $\sim1\%$. To de-correlate the light curve with dithering positions and reduce systematic offsets, we normalize the photometric points to the median photometry of the images that have the same dithering positions.

The total uncertainty in the \targeti aperture photometry combines photon noise, detector readout noise and dark current, and sky background flux level uncertainties. The sky background uncertainties were calculated as the standard deviations among pixels within the annulus described in \S\ref{sec:ABPIC-reduction} and shown in Figure~\ref{fig:ABPIC-image}, which included the noise introduced by PSF subtraction. We combined noise components based on the assumption that individual components were independent of each other. The average relative uncertainty for F125W frames was 0.28\% and that for F160W frames was 0.38\%.

\subsection{\targetII and \targetIII - Hybrid PSF Photometry}
\label{sec:hybrid_reduction}

\begin{figure*}[t]
  \plottwo{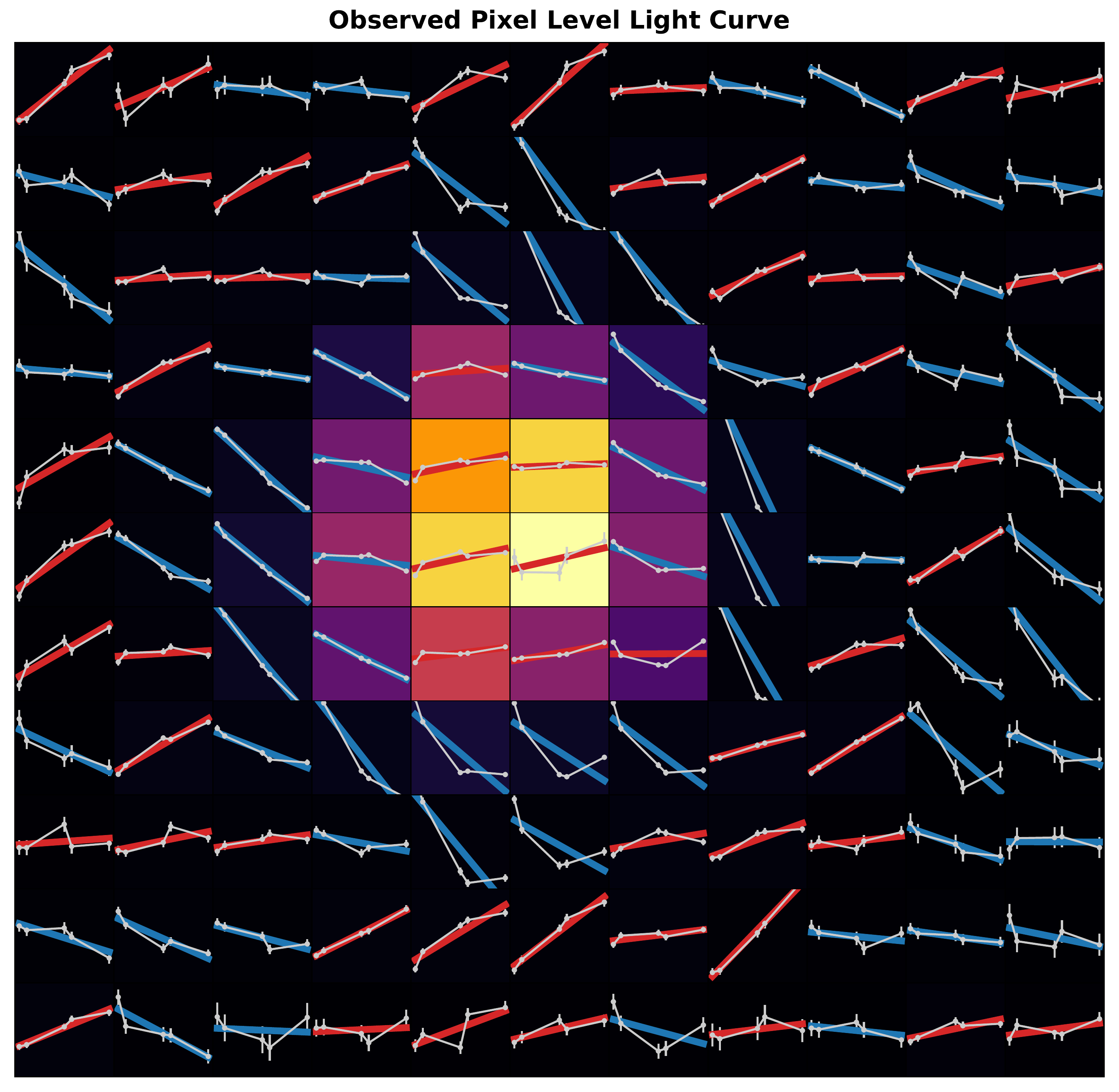}{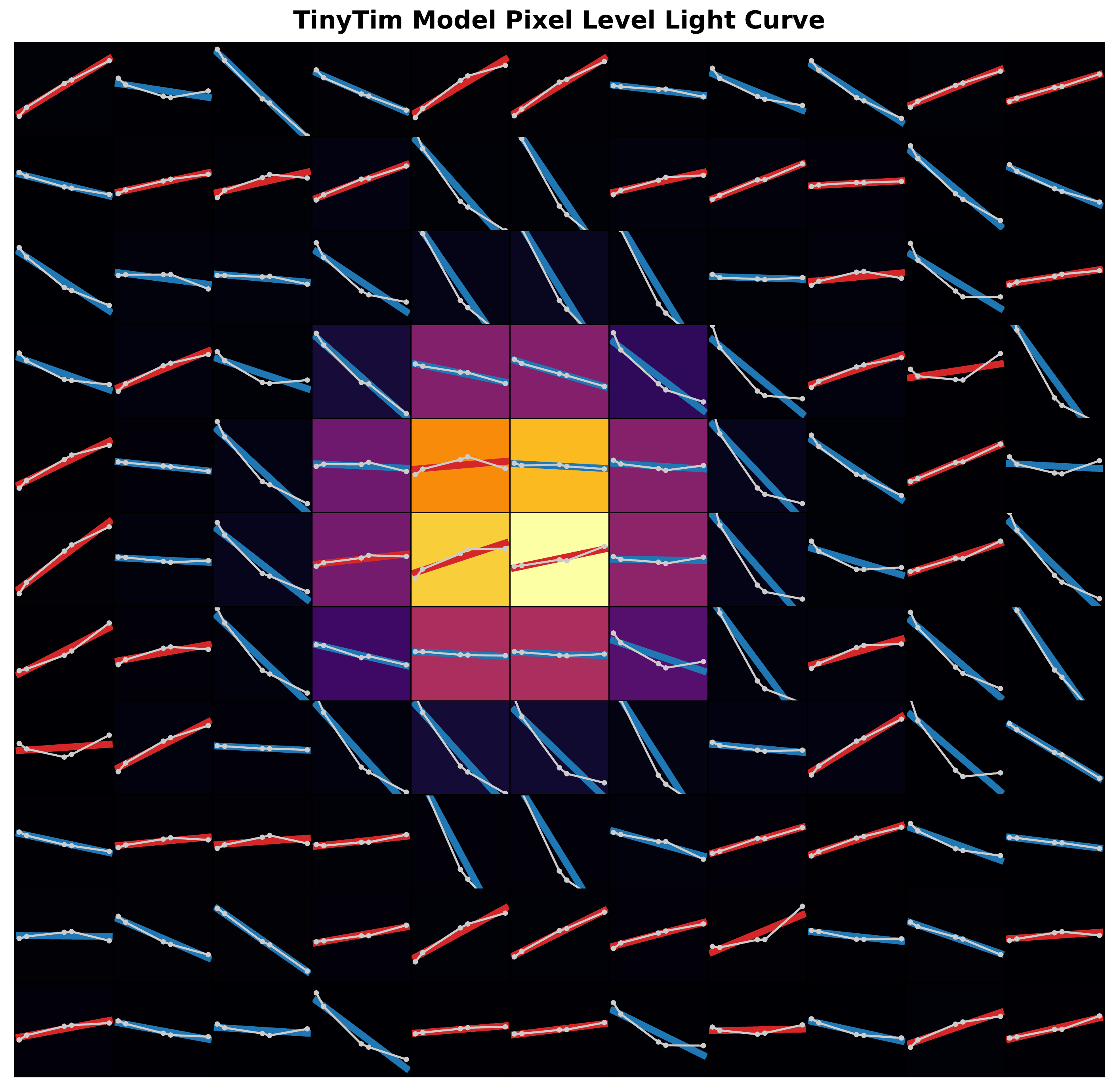}
  \caption{Light curves for individual pixels in a $11\times11$ array around the centroid of 2M1207A in the observed image time-series (\emph{left}) and the best-fit TinyTim model PSF image series (\emph{right}) for one HST orbit. Each square represent one pixel with background colors coding the flux intensities. The gray curves are the observed/TinyTim single pixel light curves. Blue or red lines are the linear fits to the pixel level light curves to indicate the orbital-wise trend. Blue indicates ascending trend and red indicate descending trend. The $y-$axis range for the pixel level light curves is from $-20\%$ to $20\%$. TinyTim PSFs recovers the pixel level trends well. \label{fig:pixeltrend}}
  \end{figure*}

The angular separations of \targetII and \targetIII to their host stars are less than $1\arcsec.5$. Although for \targetII the 2RDI products for the median-combined images were able to marginally reveal the companion, there are substantial residuals in the primary subtracted images. The poor 2RDI performance for companions with smaller angular separations is the result of severely under-sampled WFC3/IR PSFs. Systematics at this level hindered precision photometric measurements. We thus used the hybrid PSF photometry method \citep{Zhou2016} that uses TinyTim \citep{Krist2011} PSFs to perform photometry.

We first assembled TinyTim PSF libraries for the host and companion sources. The input parameters for TinyTim models are filter selections, target centroid coordinates, target spectra, and HST secondary mirror displacements. \edit1{The input target coordinates, which were in integer pixels, determined the position-dependent component of the PSF. This PSF component was introduced by detector geometric distortion.} The exact centroids were determined at one hundredth of a pixel level in the subsequent PSF fitting steps. To ensure no precision loss due to the under-sampled detector, we calculated PSFs that were over-sampled by a factor of nine over the native detector sampling. Target spectra affect TinyTim model PSFs, particularly for the wide-pass filter PSFs. We inserted M3 and M8 spectral template for  2M0122A and 2M1207A, respectively, and L5 for the two companions. The M3 template was from TinyTim's built-in spectral library. The M8 and L5 templates were from the SpeX Prism Library. Displacement of the HST secondary mirror, due to telescope thermal state variations with typical amplitudes of about +/-4 microns, causes quasi-periodic focus changes at the HST instrument focal planes over HST orbital timescale. This displacement induces fluctuations in the light curves for individual pixels as much as 15\% (Figure~\ref{fig:pixeltrend}). To precisely model this effect, we calculate model PSFs with focus parameters ranging between -15\,\micron{} to 15\,\micron{} with 0.5\,\micron{} increment. In total, each source has a pre-calculated PSF library that has 60 frames sampled on the focus grid.

We adopted a two-step fitting procedure to find the best-fit model PSFs. First, we fit two TinyTims PSFs without including PSF correction terms. Here the free parameters are the PSF scaling amplitudes ($A$), $x$ and $y$ coordinates of the centroids of the primary and companion PSFs, and the focus parameter ($f$). For a set of parameters, the model PSF is first linearly interpolated from the focus grid, and then shifted to match the location of the observed PSFs using two-dimensional cubic interpolation, then down-sampled to the native detector pixel scale, and finally scaled with the amplitude parameter to match the flux of the of the observed PSFs. In all cases, the core of the primary object was  (by design) necessarily over-exposed into saturation. For 2M1207A, the most-illuminated pixel saturated after one SPARS25 multi-accum read. We place a mask to exclude these saturated pixels from likelihood calculations (Equation~\ref{eq:chisq_target}). The best-fitting parameters are obtained by maximizing the likelihood function defined in Equation~\ref{eq:chisq_target} using Markov Chain Monte Carlo \citep[implemented with \texttt{emcee}][]{Foreman-Mackey2012}.
\begin{equation}
  \label{eq:chisq_target}
  \begin{split}
    \ln&(\mathcal{L}) \\&= \ln\Biggl(\prod_i\frac{1}{\sqrt{2\pi}\sigma_i}\exp\Bigl(-\frac{\mathrm{res}_i^2(f, A_1, x_1, y_1, A_2, x_2, y_2)}{2 \sigma_i^2}\Bigr)\Biggr)\\
    &= -\frac{1}{2}\sum_i\ln(2\pi) - \frac{1}{2}\chi^2  - \sum_i\ln(\sigma_i)
    \end{split}
  \end{equation}
  in which $\sigma_i$ is the uncertainty of pixel $i$ including photon noise, readout noise, and dark current. 
In all cases, the parameters converged after 1,000 burn-in steps with 64 walkers. Another 500 steps were then applied to sample the parameter space.

\begin{figure*}[t]
  \centering
  \includegraphics[width=0.32\textwidth]{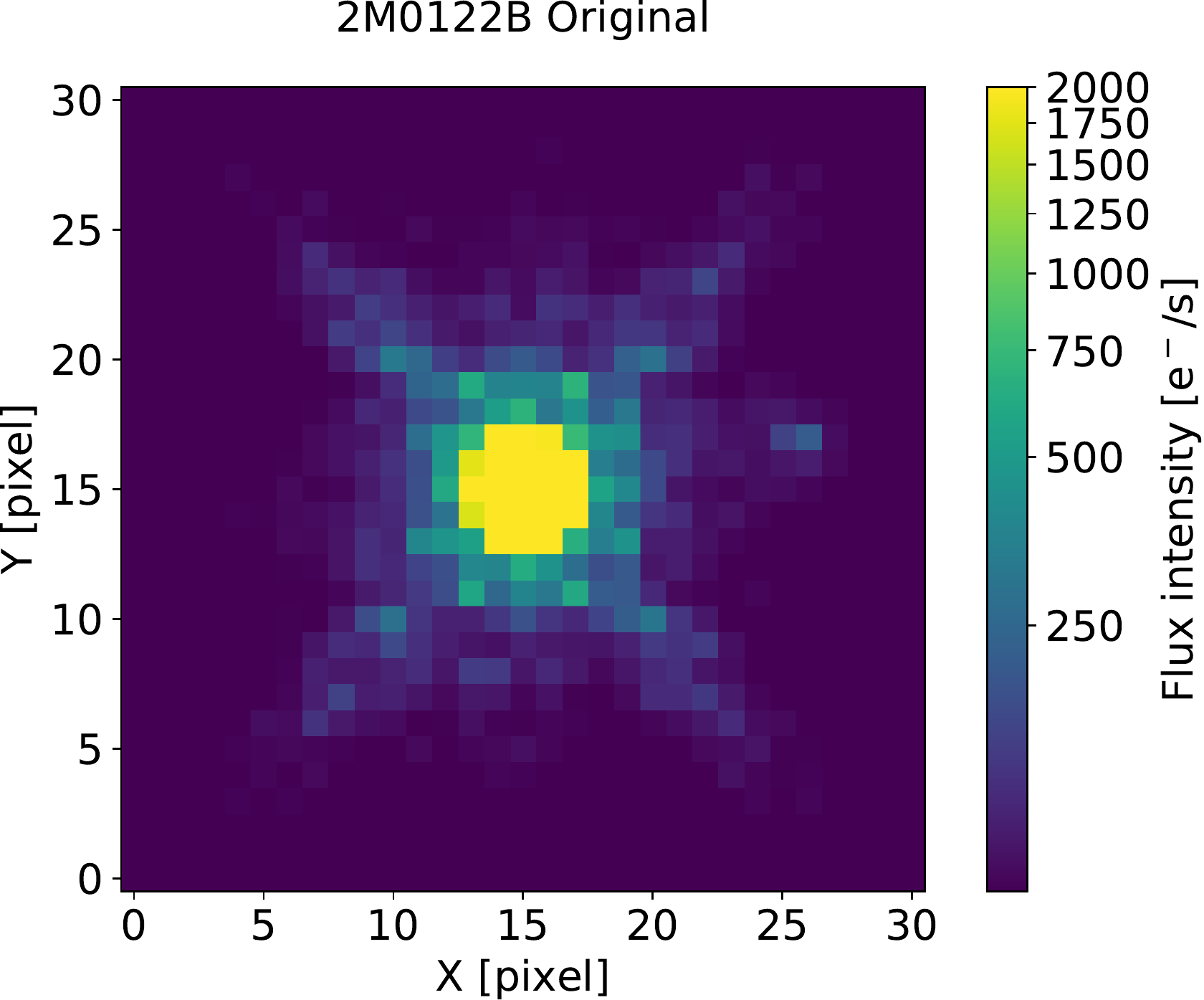}
  \includegraphics[width=0.32\textwidth]{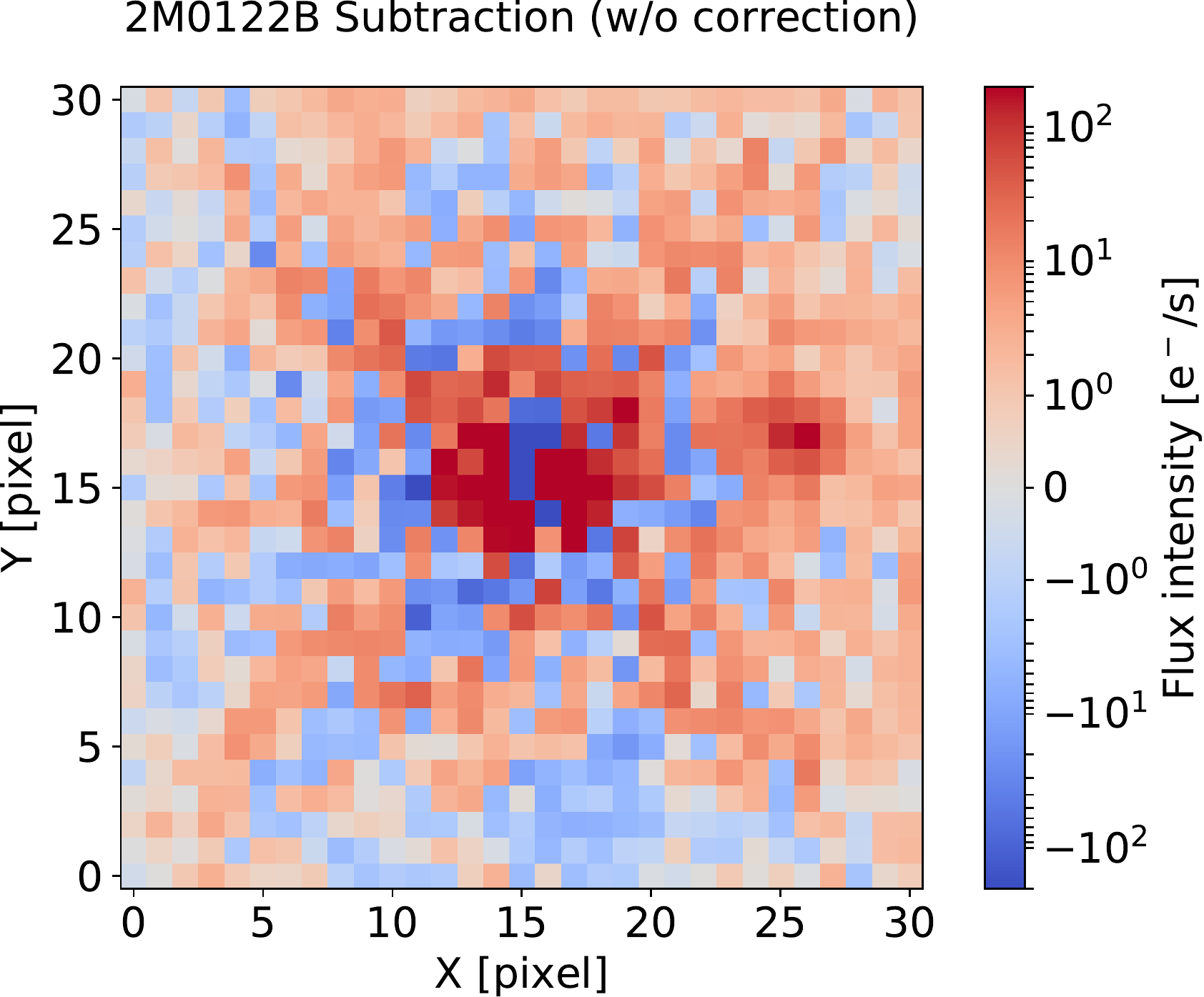}
  \includegraphics[width=0.32\textwidth]{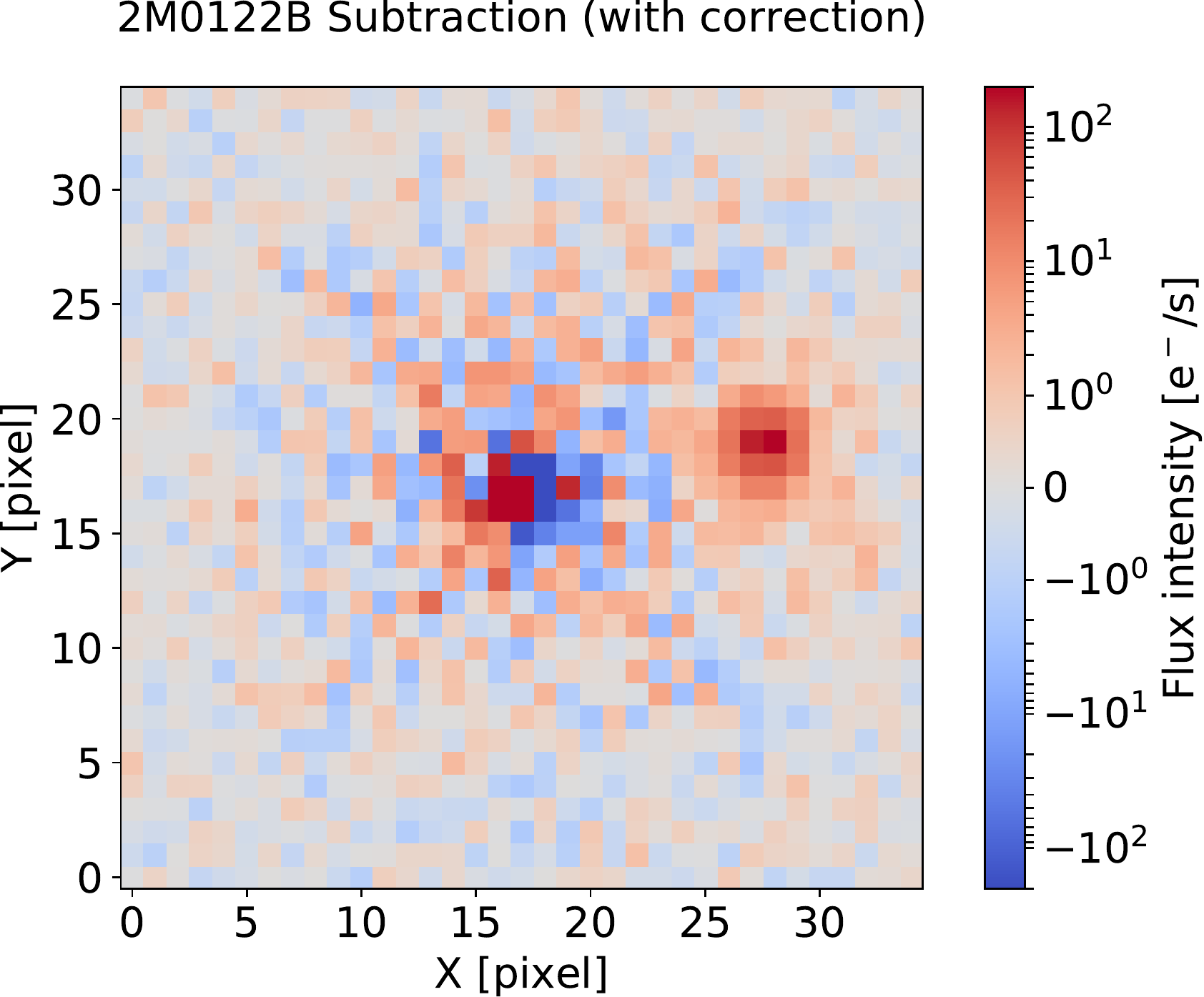}\\
  \includegraphics[width=0.32\textwidth]{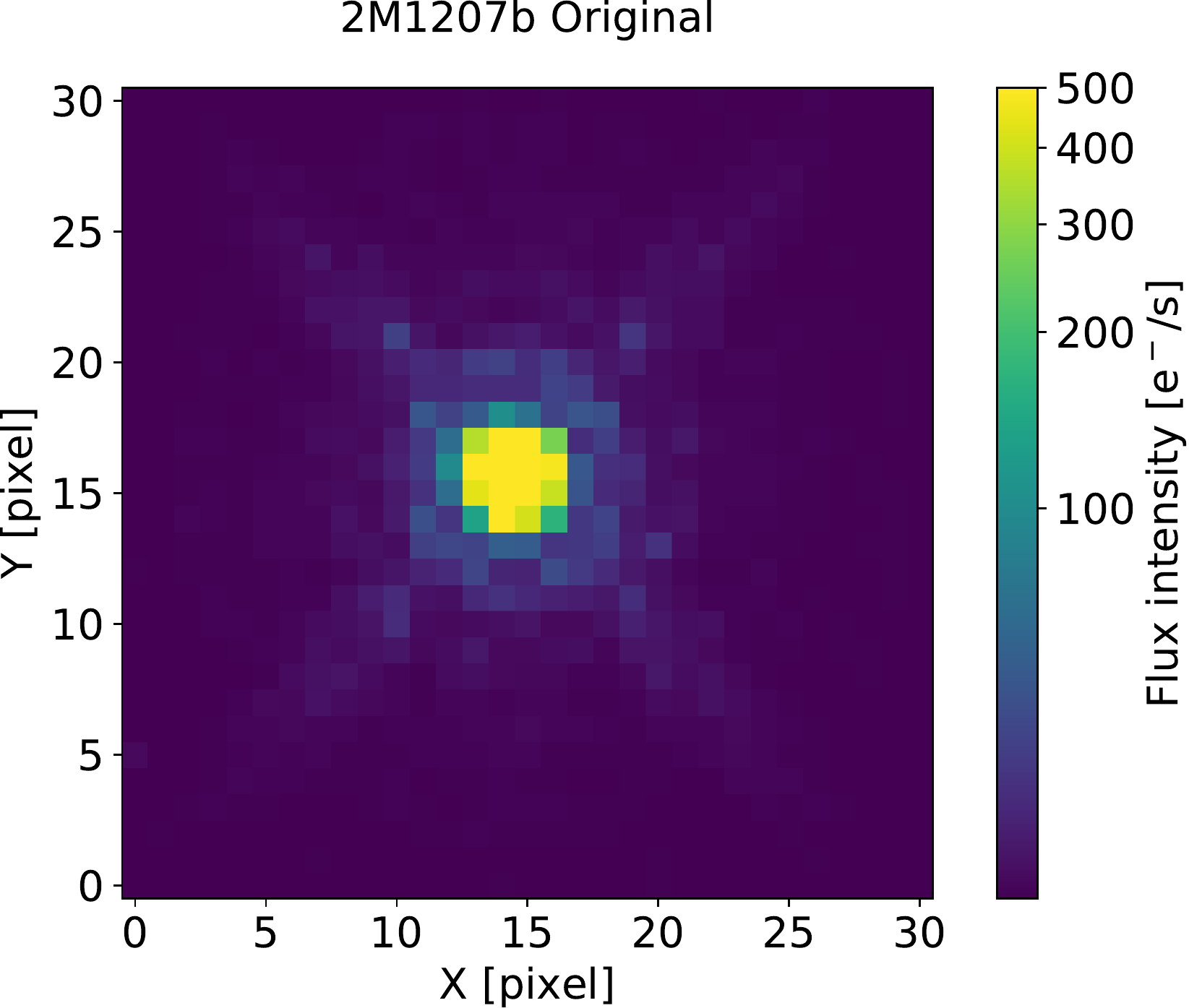}
  \includegraphics[width=0.32\textwidth]{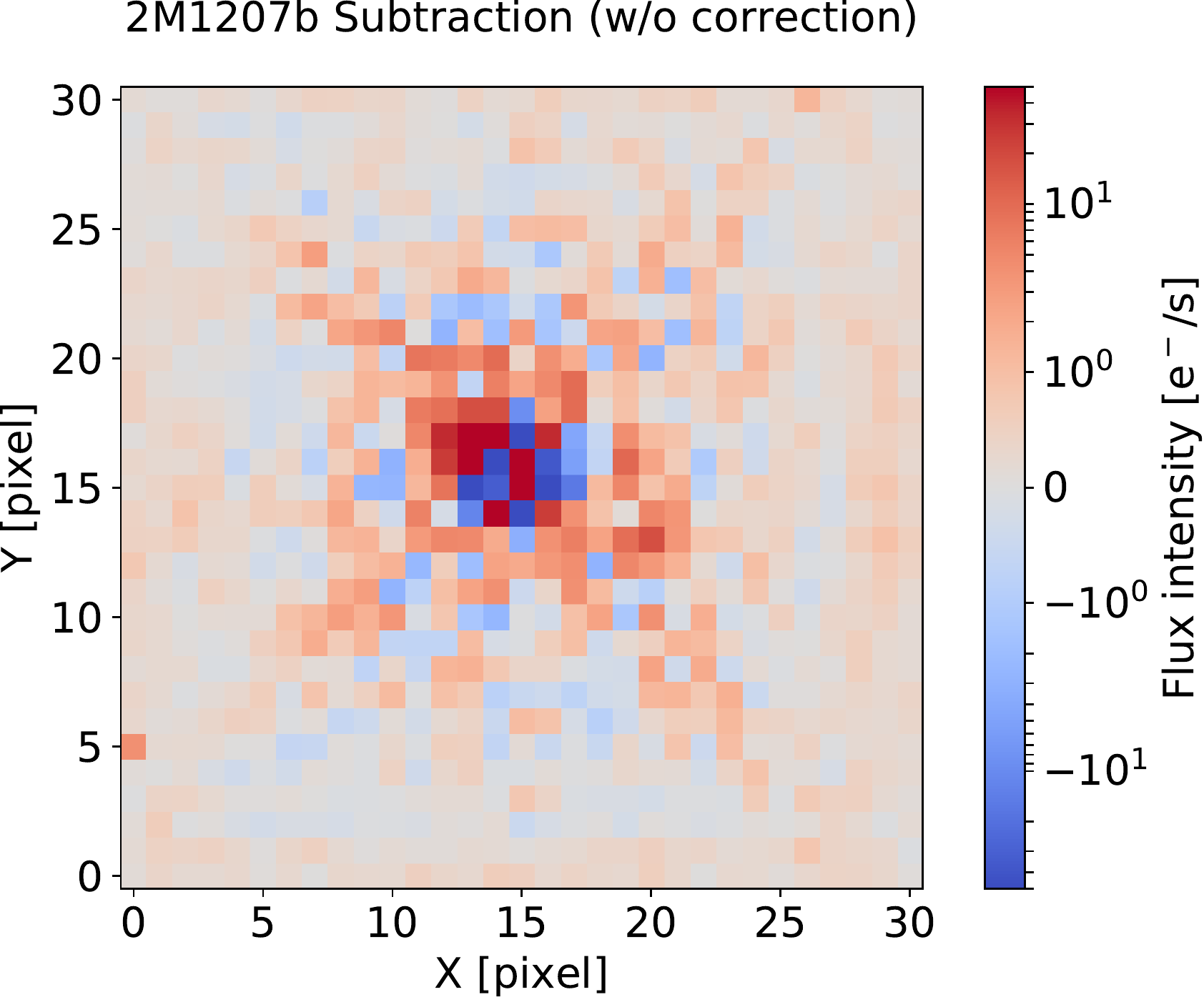}
  \includegraphics[width=0.32\textwidth]{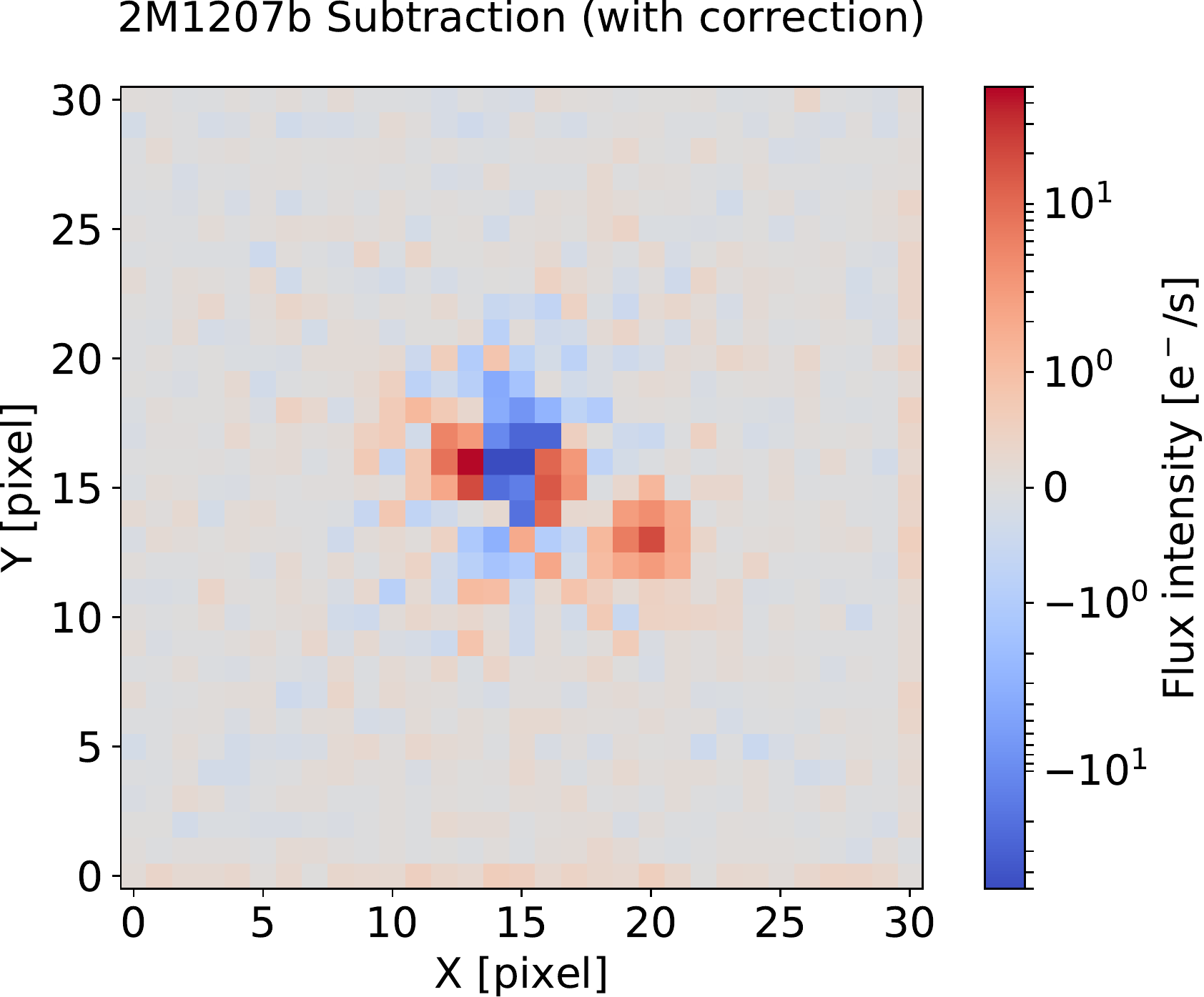}
  \caption{Hybrid PSF subtraction for 2M0122B (upper) and 2M1207b (lower). From left to right, the three columns are for the original images, the PSF subtracted image without the correction term, and the PSF subtracted image with correction term. The correction term significantly improved the subtraction results. For 2M1207b, the companion PSF is only visible after correction term is included in the subtraction.\label{fig:2M0122-subcomp}}
\end{figure*}

We then applied an empirical PSF correction to each image to account for the mismatches between the observed PSFs and the TinyTim model PSFs. This correction is based on the assumption that TinyTim  introduces common mode residuals \edit1{\citep{Zhou2016}} in the PSF fitting, an expected outcome of the procedure given that TinyTim only considers the diffraction-limited component of the images. Applying this common mode image as a correction term reduces the PSF fitting residual amplitudes and increase the PSF photometry precision. We calculated the correction term by median-combining the residual images that share the same filter and telescope roll. We also found that the telescope focus parameter affected the residuals. To account for this effect, we further divided images by the exposure groups, because corresponding groups for different orbits share the same HST orbital phase and similar telescope focus. As a result, there are eight correction terms for each filter, accounting for four exposure groups and two telescope rolls. Then, the essence of PSF fitting was to reduce the residuals in Equation \ref{eq:1}
\begin{equation}
  \label{eq:1}
  \begin{split}
    \mathrm{residual} = &\mathrm{Image} - \mathrm{correction} -\\& \mathrm{PSF_{primary}}(A_1, x_1, y_1, f) - \mathrm{PSF_{companion}}(A_2, x_2, y_2, f)
    \end{split}
\end{equation}
The free parameters were the same as in the non-corrected TinyTim PSF fit. These free parameters were the $x$ and $y$ coordinates for the primary and companion PSFs, amplitudes of the two PSFs, and the telescope focus parameter. The image masks were also identical to the ones used in the non-corrected PSF fit. Pixels that were inside a 3-radius circle around the centroid of the primary PSF were excluded from fitting.

In the hybrid PSF photometry, the uncertainty for individual pixels were propagated through a maximum likelihood calculation. We calculate the 1-$\sigma$ uncertainty of the photometry using the marginal distribution in the MCMC chain. For \targetII, they are 0.76\% and 1.0\% for F127M and F139M frames, respectively. For \targetIII, the average relative uncertainties are 1.6\% and 2.4\% for each individual F127M and F139M frames, respectively. 

The results for hybrid PSF subtraction are in Figure~\ref{fig:2M0122-subcomp}.

\subsection{PSF Fitting Robustness Analysis}
In the PSF fitting photometry, the likelihood function (Equation \ref{eq:chisq_target}) is a non-linear multi-variable function and the optimization processes (MCMC being used here) may settle at the local extrema. To test the robustness of the fit, we calculated the marginal likelihood function for every free parameter in a wide range around the best-fit values found by the optimization processes. For each parameter, we coarsely sample 10 points evenly distributed in the $(-10\sigma, 10\sigma)$ range and finely sample 50 points in the $(-3\sigma, 3\sigma)$ range and evaluate the likelihood function at the sample points. We then examine whether the best-fit values successfully maximize the marginal likelihood functions. We confirmed that the marginal likelihood functions  have maxima at the best-fit values. We used this method to test every fit result and confirm that the automatic fitting routine found the true best-fit values in every case.

\subsection{Ramp Effect Correction}

The WFC3/IR light curves have a common instrumental profile called ramp effect \citep[e.g.,][]{Berta2012}. \citet{Zhou2017} modeled the systematics based on the detector charge trapping effect. The ramp appears as a gradually ascending profile in the light curve with its amplitude  highest at the first orbit of an HST observation visit. The ramp effect is less strong in the subsequent orbits to 0.1-0.3\% as the traps are filled by the charge carriers stimulated by irradiation. The typical amplitudes of the ramp profiles are 1-2\% at the first orbit and 0.1-0.3\% in the rest of the light curves. In this study, the noise for the companion is above the ramp effect amplitude. As a result, the systematics are not visible in the uncorrected light curves. For the brighter primary stars the uncorrected photometry is more than one order of magnitude more precise than that for the companions and consequently the ramp effect is the most prominent feature in the light curves. 

We configured the RECTE model \citep{Zhou2017} for direct-imaging time series. In our direct-imaging time-series, the illuminated areas are in different locations in the odd and even orbits as a result of the telescope roll changes. In successive orbits, at different roll angles, the illumination level for one pixel varies more than two orders of magnitude  in our direct-imaging time series than those in the (G141 grism dispersed) spectroscopic time series. To take this effect into account, we first form two PSF cubes (Cube I for orbits 1, 3, 5 and Cube II for orbits 2, 4, 6), which assumes the illumination levels for one pixel in one orbit is the same as that in the median images. The expected astrophysical variation is below 5\% and has negligible effect on the ramp profile. We then feed the PSF cubes to RECTE to calculate pixel-level ramp profiles for pixels within a 5-pixel radius of the centroid of the PSF. The pixel-level profiles form the correction curve with their sum weighted by the illumination levels. We derived the ramp correction for the companions and their primary in two telescope position angle separately for each target.

\edit1{RECTE corrections improved the light curve precision, particularly for the observations for 2M0122. Ramp systematics that has an amplitude as large as 1.3\% were eliminated in the light curve of 2M0122A. Detailed results are presented in Section \ref{sec:results}.}

\section{Results}
\label{sec:results}

\subsection{\targetI}

\begin{figure*}[t]
  \centering
  \plottwo{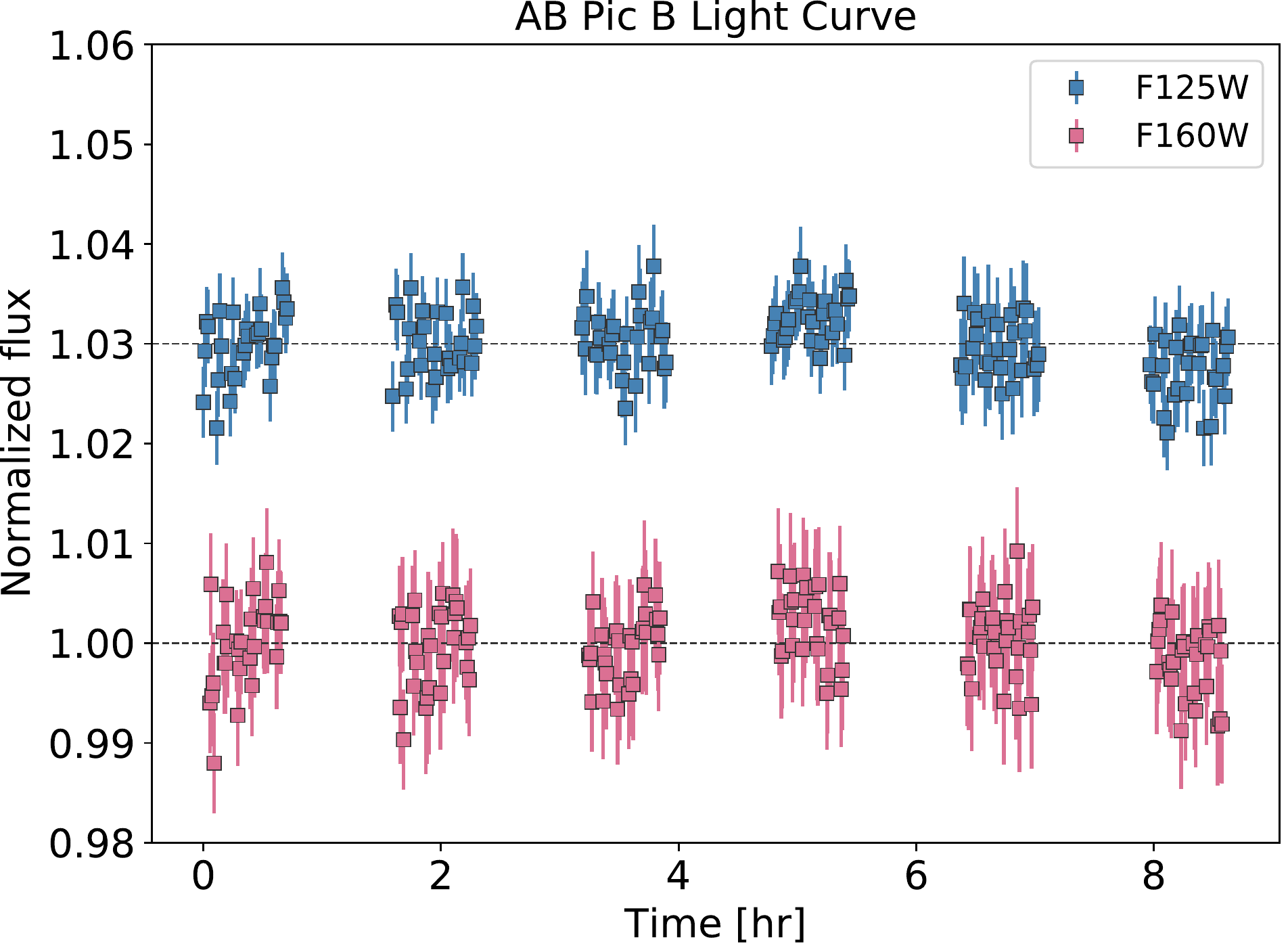}{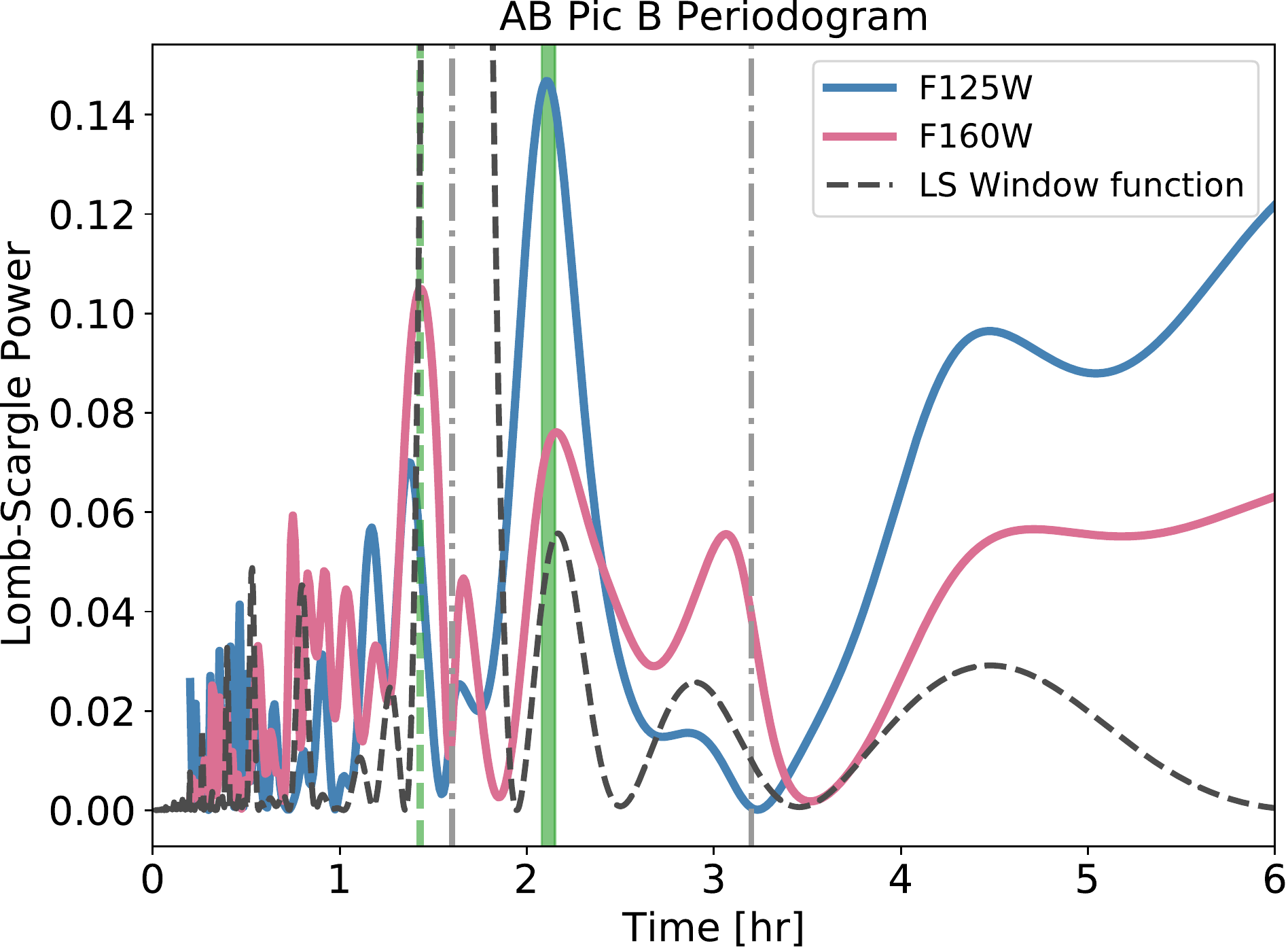}
  \caption{\targeti light curves (\emph{left} panel) in F125W (blue) and F160W (pink) and their LS periodogram (\emph{right} panel). The light curves are normalized separately for different dithering positions. The errorbars represent the photon noise. 3\% vertical offset is applied to F125W light curve for clarity. In the periodogram, the green solid line marks the common 2.12 hr peak in both curves. The green dashed line marks the most significant peak for the F160W periodogram at 1.43 hr. Two gray dashed lines mark the $1\times$ and $2\times$ of the HST orbital periods, where instrumental and data reduction related effects are most likely to introduce signals.}
  \label{fig:ABPIC-LC}
\end{figure*}

\begin{figure}
  \centering
  \plotone{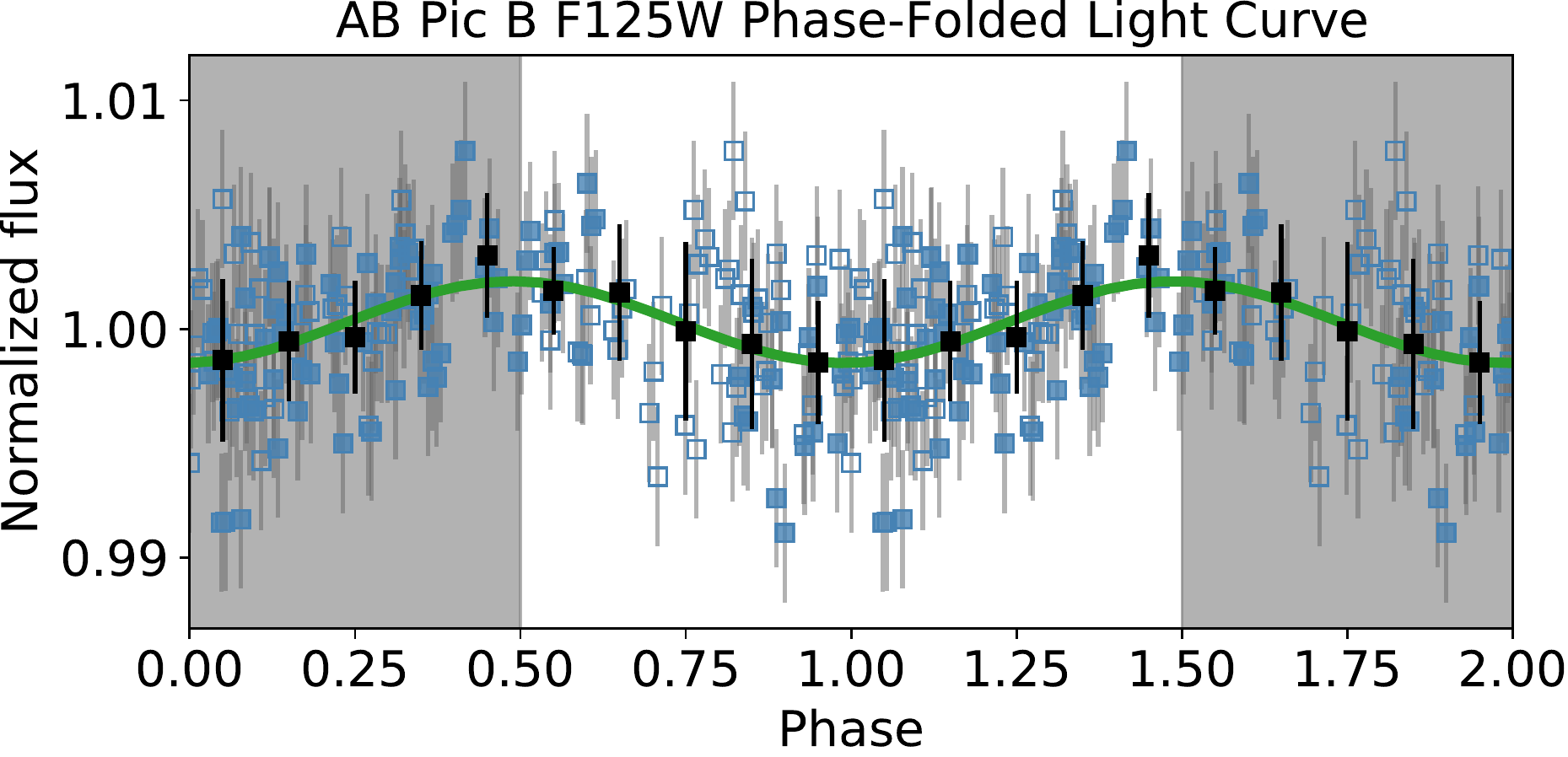}  
  \plotone{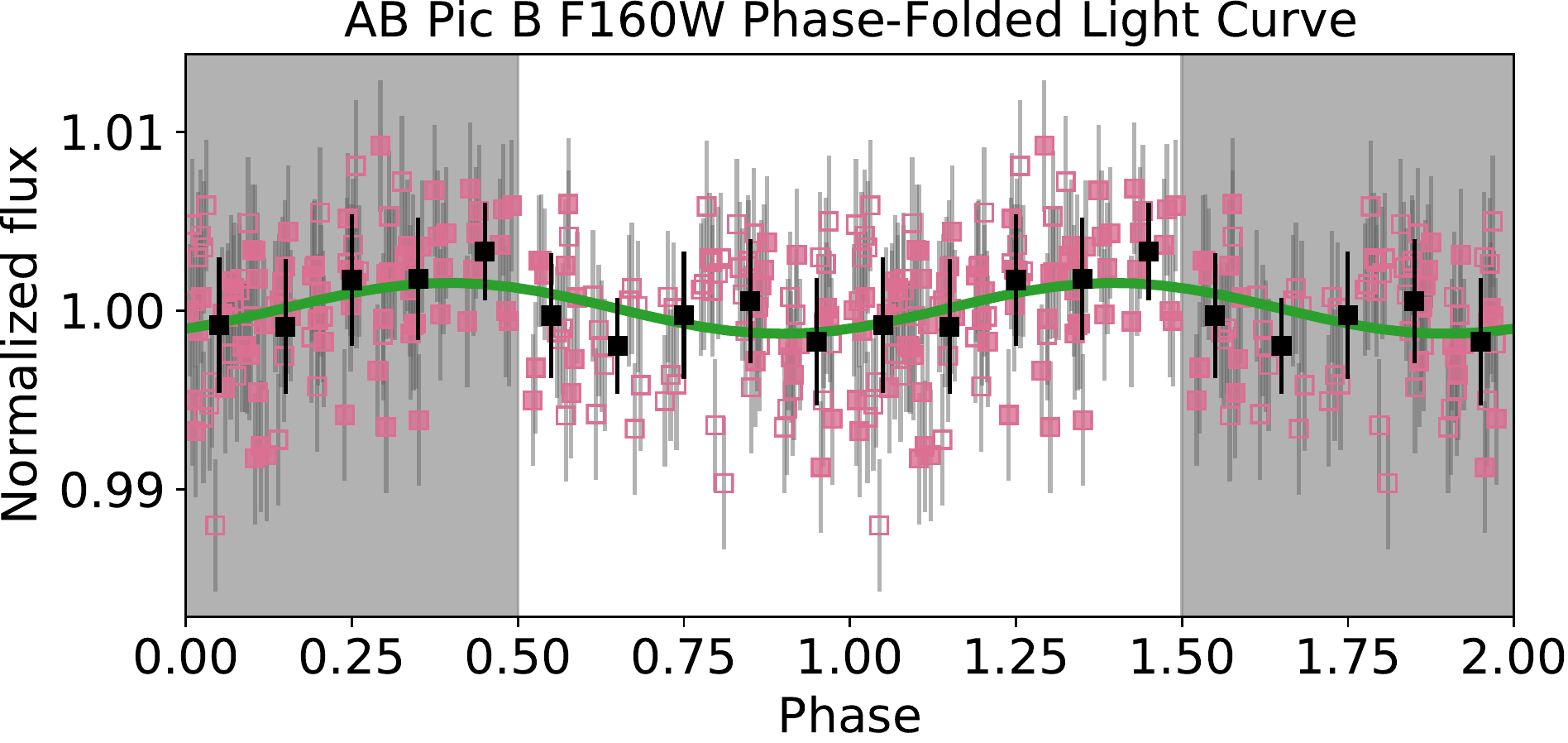}
  \caption{The phase folded light curves for \targetI in F125W (\emph{upper} panel) and F160W (\emph{lower} panel). The light curves include two repeated full phases to illustrate the periodicity. Gray shaded regions mark the redundant data. Open and filled squares are for photometry taken in the first and second halves of the observations. Black squares and errorbars are binned photometry with a resolution of 10 per full phase. Green curves are the best-fit sine waves.}
  \label{fig:ABPIC-fold}
\end{figure}

Differential roll subtraction effectively removes the background flux due to the primary star (Figure \ref{fig:ABPIC-image}). Before 2RDI, the average background estimated in a $r=10$ pixel annulus centered on the  companion's PSF is (a) about 1\% of the companion central pixel brightness, and (b) ten times the brightness of the average sky background. After 2RDI, the average brightness in the same  annulus declined below $1-\sigma$ uncertainty of the sky background, contributing to less than 0.05\% of the total flux inside a 5 pixels circle centered on the companion PSF. The primary star contamination in the roll subtraction products is negligible and further post-processing is unnecessary.

We perform aperture photometry on the time series of primary-subtracted images with a circular aperture of 5 pixels. Figure~\ref{fig:ABPIC-LCRAW} shows the light curves. Both F125W and F160W light curves have apparent variability with a peak-to-trough amplitude of $\sim2\%$. The two light curves are also moderately correlated (Figure~\ref{fig:ABPIC-corr}). We calculated the Pearson correlation coefficient between the light curves in two spectral bands and found the coefficient to be 0.61. We do not attempt to obtain the light curve for the primary star AB Pic A because the saturation at the core of the PSFs.

Closer inspection reveals that the apparent variability is associated with the dithering positions. In Figure~\ref{fig:ABPIC-corr}, while showing the correlation between the light curves in the two filters, we also demonstrate the dithering positions with different colors. If photometric measurements in images taken at a dithering position are systematically off, those measurements will cluster in the correlation plots. Figure~\ref{fig:ABPIC-corr} manifest such clustering. We therefore attribute the $\sim2\%$ level variability to instrumental systematics rather than astrophysical cause. According to \citet{Dressel2018}, the precision of the flat field for the WFC3 IR detector is at 1\% level.  Therefore, flat field errors alone can introduce such level of light curve fluctuations.

We attempt to distinguish \targetI's intrinsic modulation from the systematic-induced variability. We assume that the average photometry in different dithering positions are the same. This assumption allows independent normalization for light curves in different dithering positions. The result of this normalization is shown in Figure~\ref{fig:ABPIC-LC}. The variability seen in Figure~\ref{fig:ABPIC-LCRAW} is no longer appear in Figure~\ref{fig:ABPIC-LC}. Both F125W and F160W light curves demonstrate no visible modulations at the levels of 0.5\% or higher in the 9 hr observation window. The standard deviations for the F125W and F160W light curves are 0.33\% and 0.38\%, which are within 20\% of the intrinsic photometric uncertainties. We note that the typical interval between two sets of observations with the same dithering positions are two HST orbits. This procedure may introduce artifact over that time scale.

We then use Lomb-Scargle \citep[LS,][]{Lomb1976,Scargle1982a} periodogram to investigate the periodic signals in the corrected light curves. We apply LS algorithms to the two light curves separately and plot the periodograms for the two light curves in the right panel of Figure~\ref{fig:ABPIC-LC}. Both periodograms display a peak at 2.12 hr. The 2.12 hr peak is the strongest one in the F125W periodogram and the second strongest in the F160W periodogram. The F160W periodogram has its strongest peak near 1.43 hr, which is close to the 1.60 hr HST orbital period. The 2.12 hr period is neither associated with any known HST instrumental time-scales, nor related to any time-scale that was implicitly introduced in the data reductions.

\begin{figure*}[t]
  \centering
  \plottwo{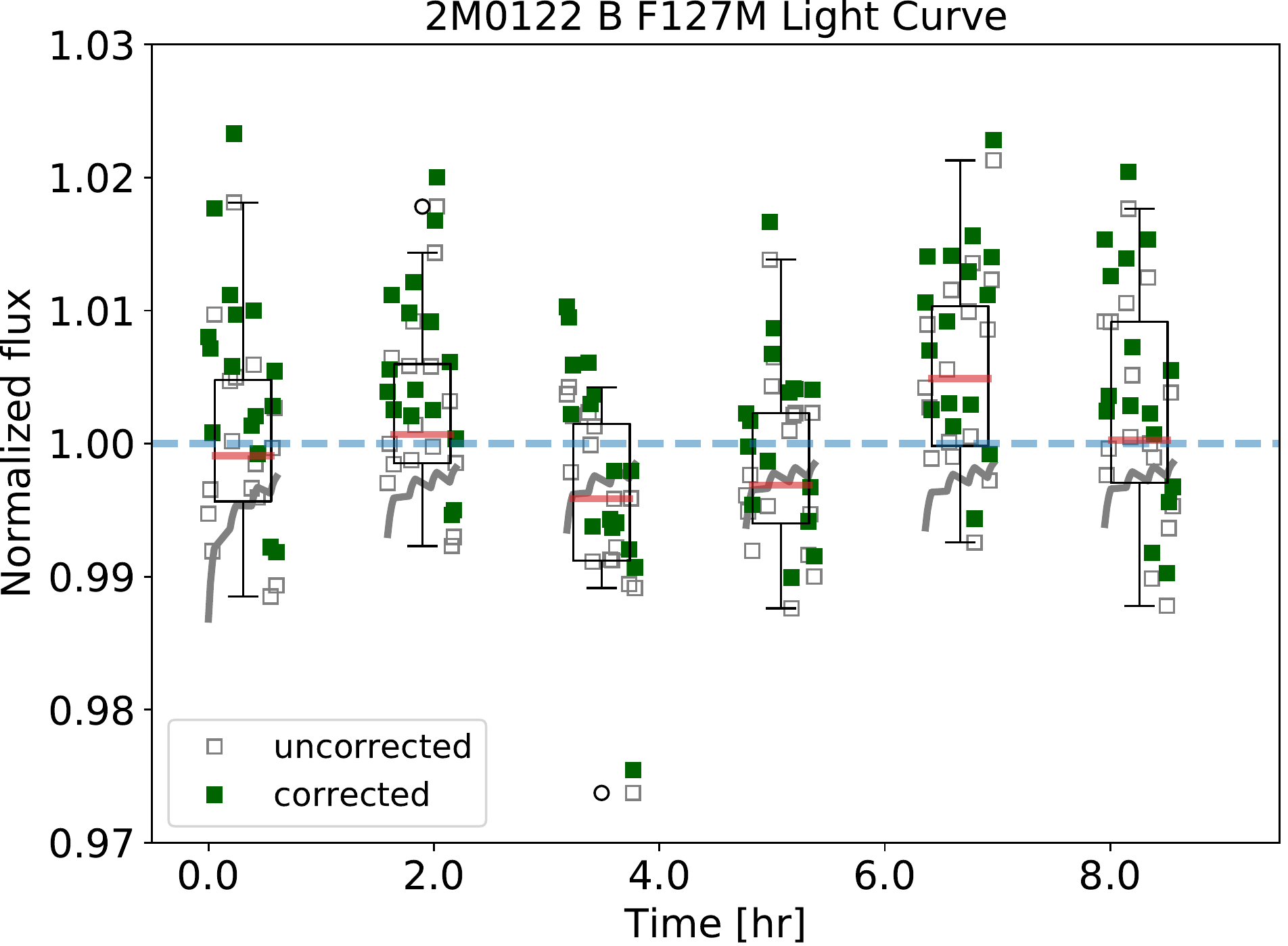}{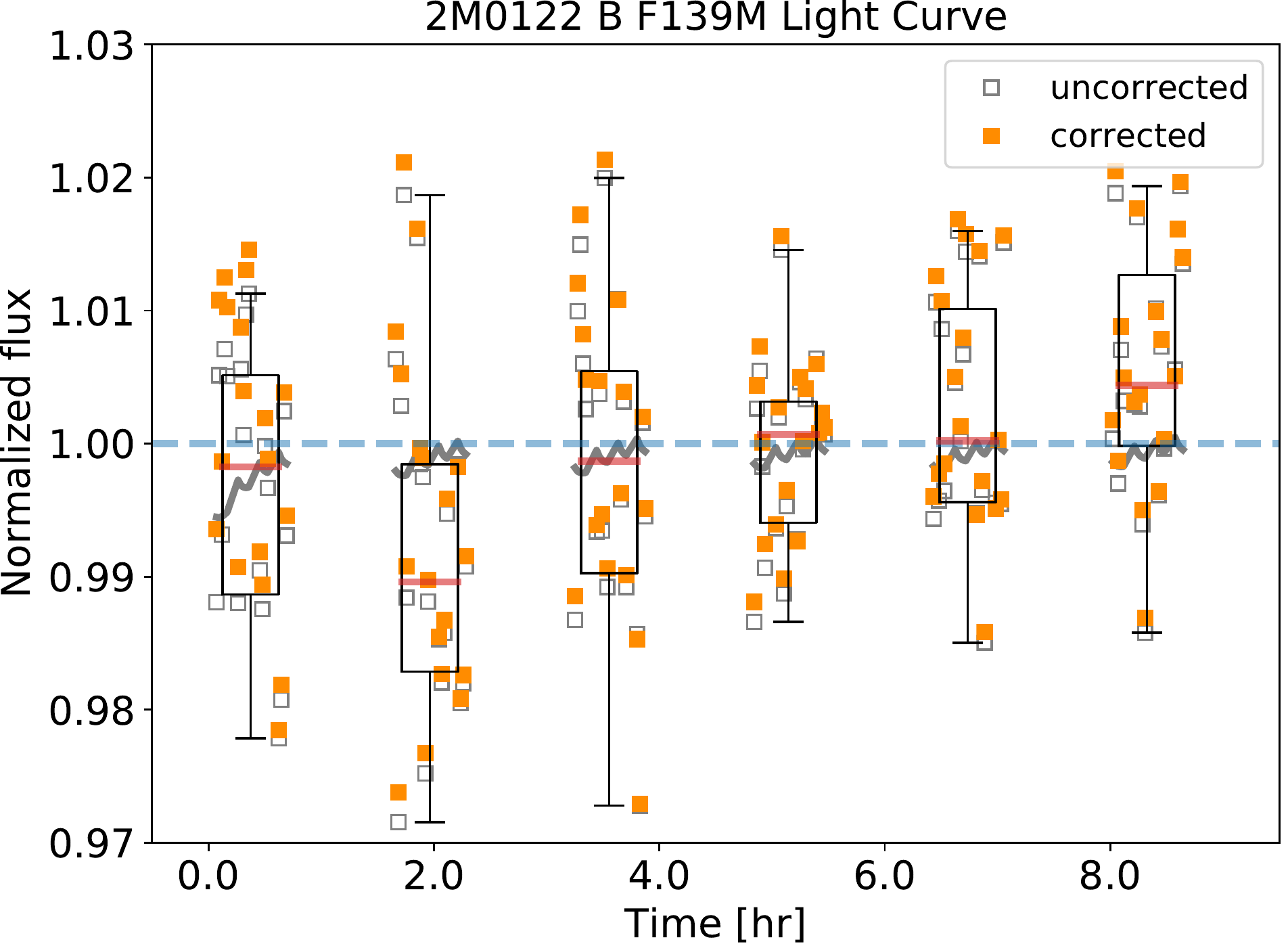}
  \plottwo{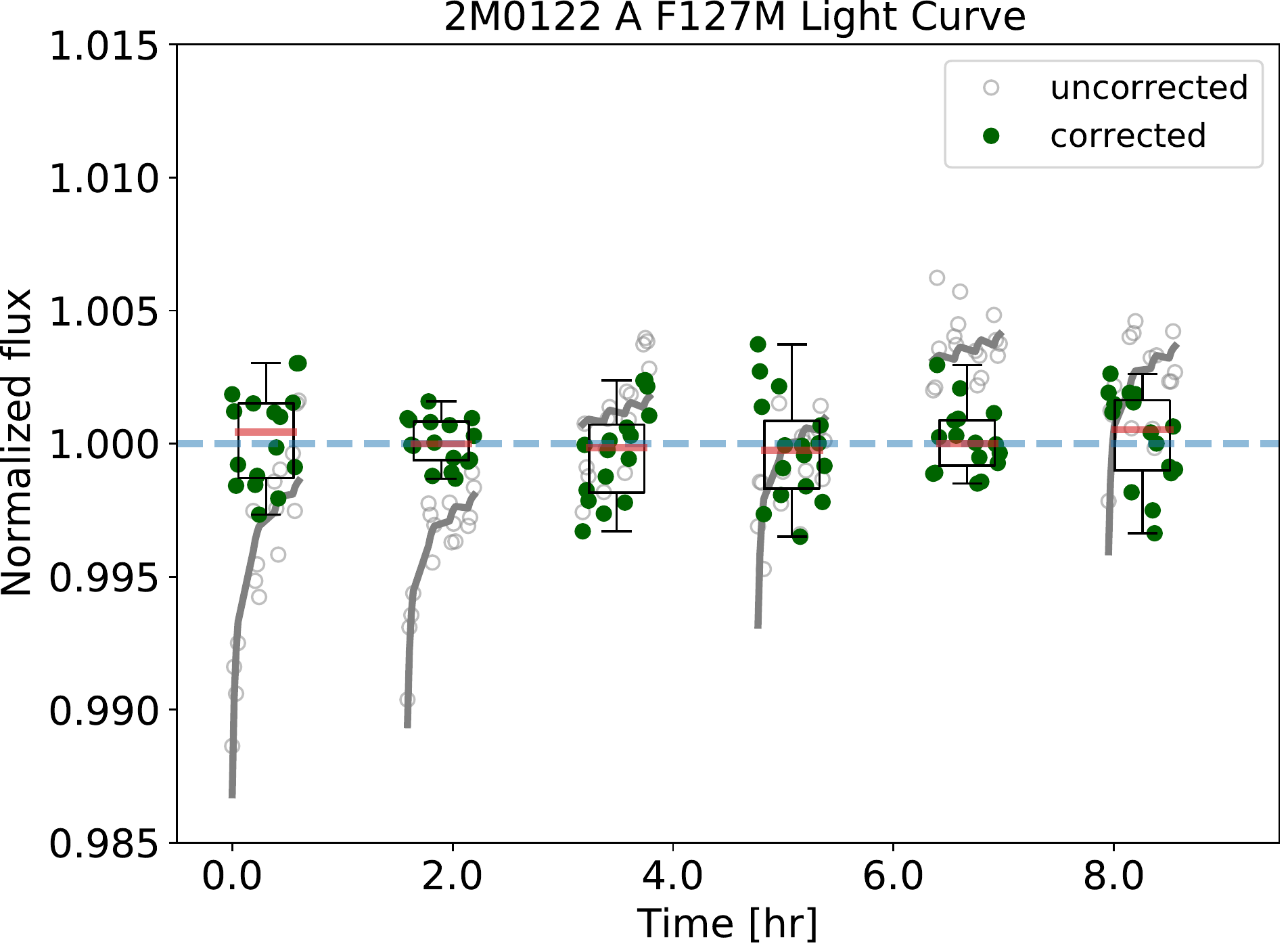}{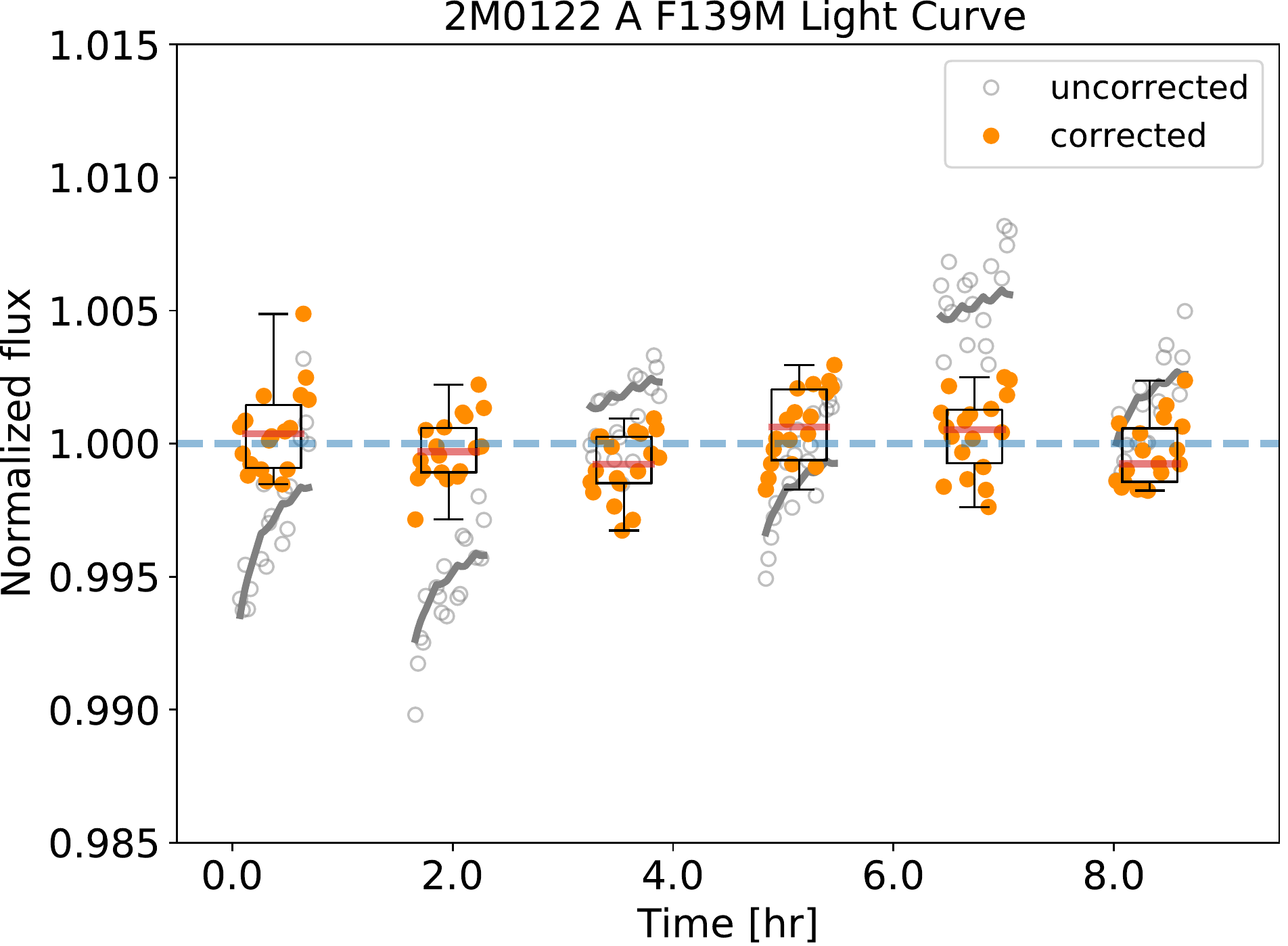}
  \caption{Light curves for 2M0122 B (upper, squares) and A (lower, circles) in the F127M and F139M filters. Upper panel includes the light curves for the companion in the two spectral bands. The light curves in Orbit 1, 3, 5 and 2, 4, 6 are normalized separately. Green and orange circles/squares are for F127M and F139M photometry, respectively. The box plot visualize the orbital combined photometry. The heights of the boxes mark the 25 to 75 percentile and the red lines in the boxes are the median photometry for the corresponding orbits. Errorbars associated with the boxes represent the minimum to maximum photometry range. Light curves for the primary are in the lower panel. The uncorrected photometry is plotted in gray open symbols and the corrected photometry is in filled symbols. Gray curves are the ramp profile for corrections.}
  \label{fig:2M0122-lightcurve}
\end{figure*}

We phase-fold the light curves with the 2.12 hr period and are able to recover a sinusoidal shaped signal in both light curves (Figure~\ref{fig:ABPIC-fold}). We fit single sine waves with fixed period of one rotation to the phase-folded light curves to evaluate the signal properties. The free parameters in the fit are the amplitude, phase offset, and the baseline level. Least-square fit finds the best-fitting parameters are amplitude of $0.180\pm0.034\%$ and phase offset of $-0.21\pm0.03$ fractional period for the F125W light curve, and amplitude of $0.145\pm0.039\%$ and phase offset of $-0.13\pm0.04$ fractional period for the F160W light curve. The signal in F125W light curve is 20\%  stronger than the that in the F160W light curve (at $0.67\sigma$ significance for the difference between two bands). The best fitting sine waves have a slight phase offset between the two light curves. The phase offset is 0.08 fractional period or $1.6\sigma$ level significance.

Being cautious about possible artifacts introduced by the dithering position-based normalization, we calculate the periodogram for the light curves measured only in the second half of the observations, for which dithering was not applied and thus the normalization does not affect the light curves. The resulting periodogram peaks at the same positions as the ones that for the entire light curves. In Figure~\ref{fig:ABPIC-fold}, photometry  in the second half of the observations (filled circles) follow the same best-fit sine waves for the entire light curves, except that they lack phase coverage between phase of 0.6 and 0.75. We therefore conclude that the signals in the periodograms are not from the normalization and use the result from analyzing the entire light curve for better phase coverage.

\subsection{2M0122B}

\begin{figure*}[t]
  \centering
  \plottwo{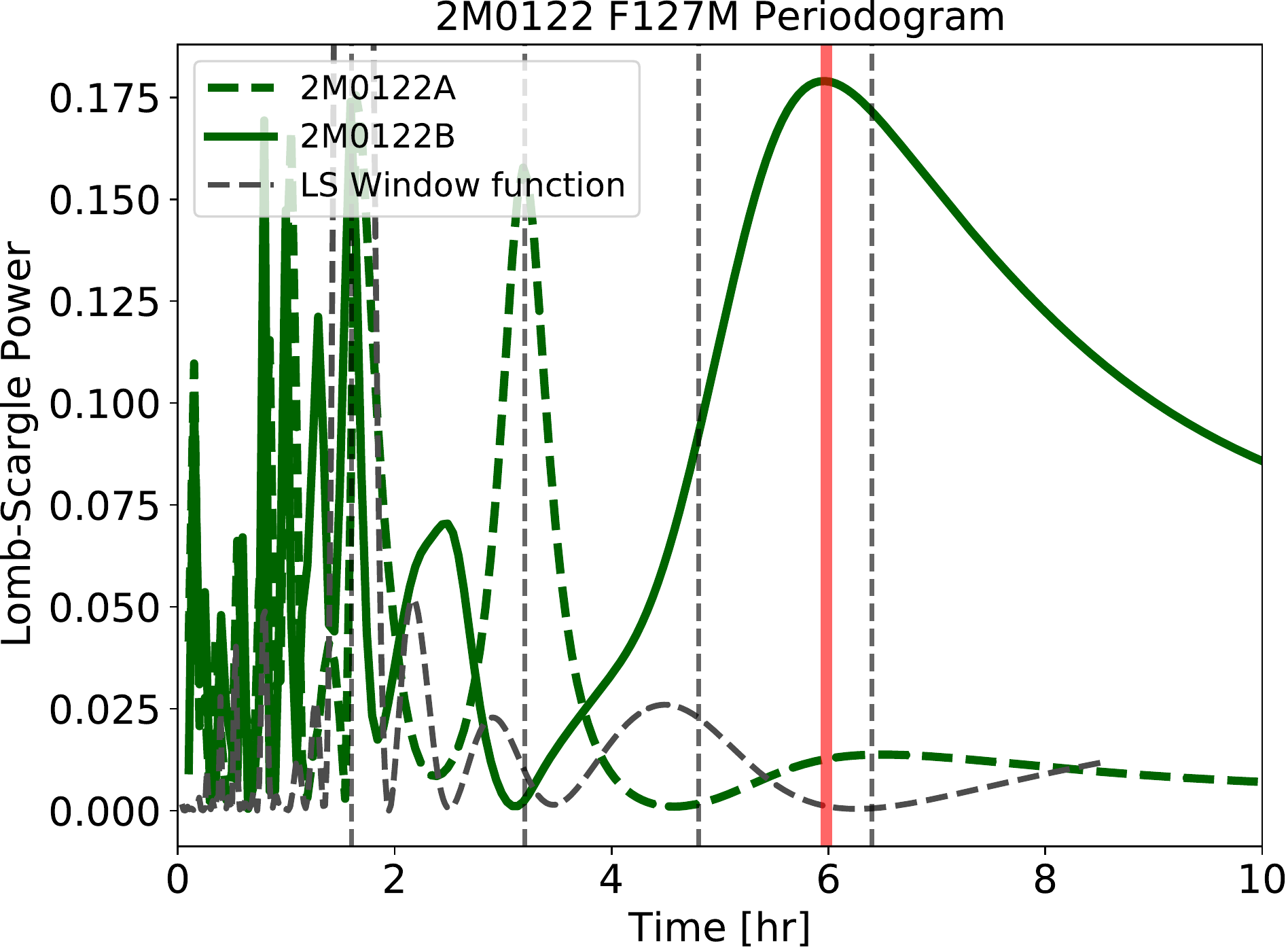}{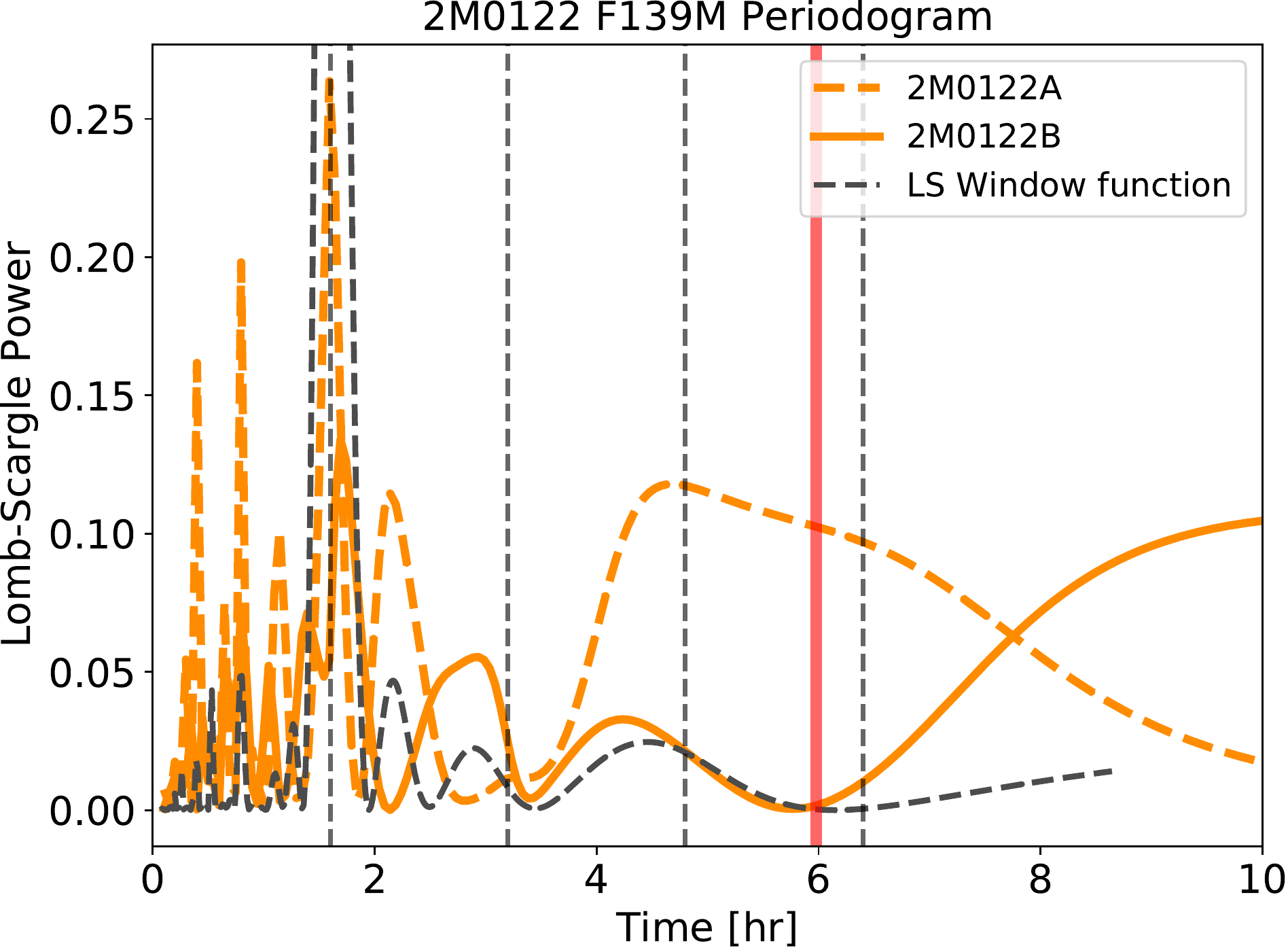}
  \caption{Periodograms for the light curves of 2M0122A and B in F127M (left) and F139M (right) filters. Green curves are the periodogram curves for 2M0122A\&B in F127M, and orange curves are for 2M0122A\&B F139M. Four vertical dashed line marks the $1\mbox{-}4\times$ of the HST orbital period. Red solid line marks the 6.0 hr period --- the location of the only detected peak that does not coincide with any integer multiples of the HST orbital period.}
  \label{fig:2M0122-periodogram}
\end{figure*}

\begin{figure}
  \centering
  \plotone{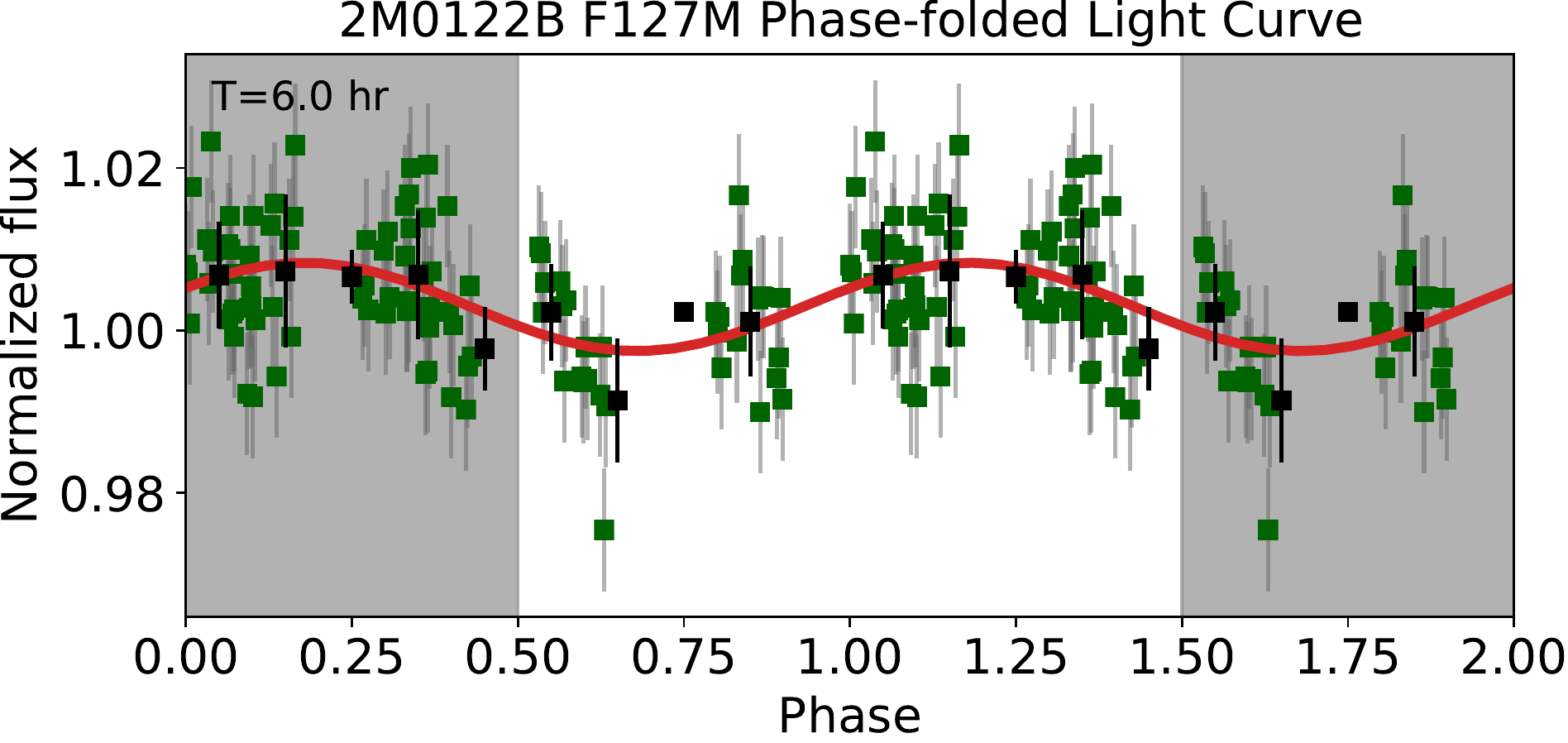}
  \plotone{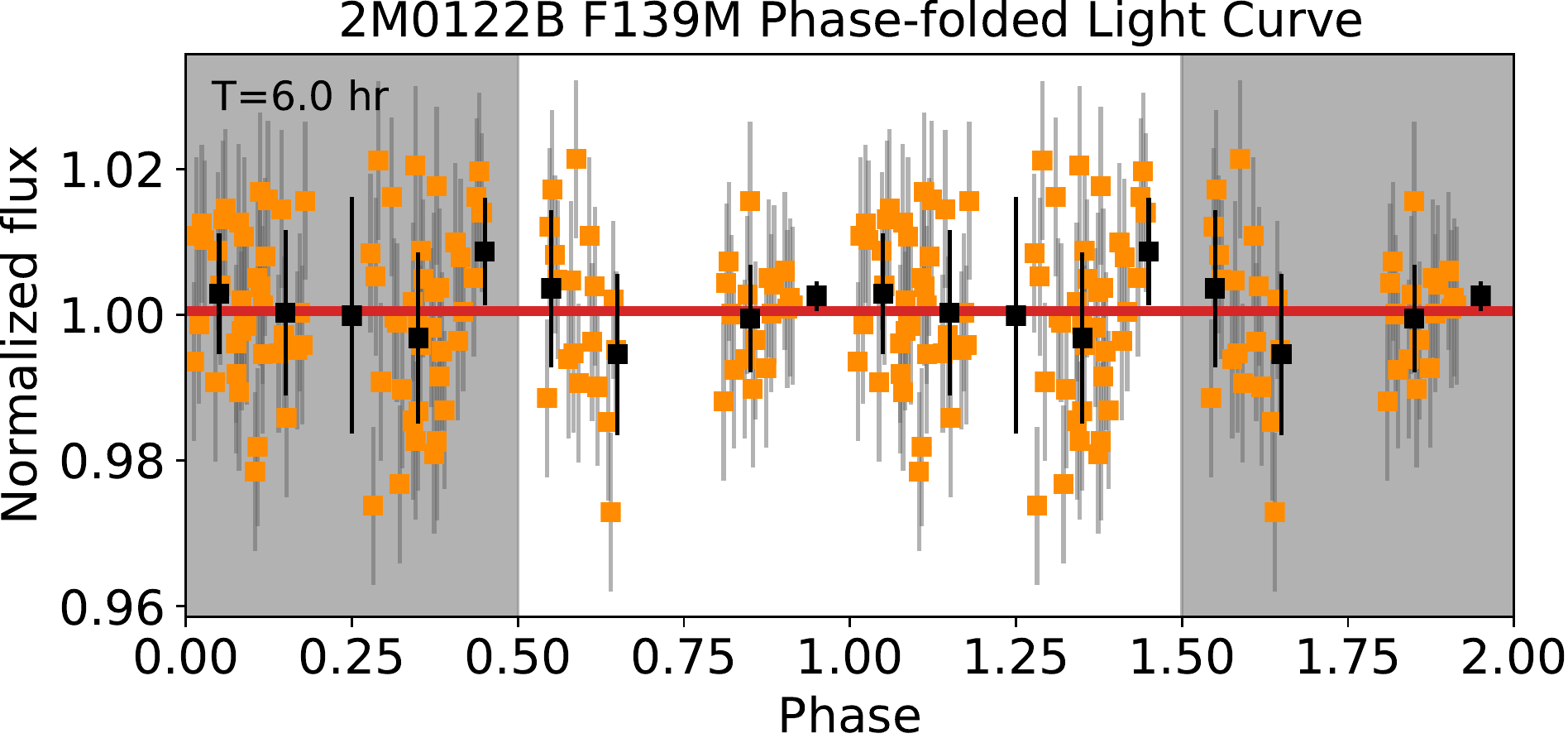}
  \caption{Phase folded light curves of \targetii in F127M (upper) and F139M (lower). The light curves are phase folded by a period of 6.0 hr. These figures are created in the same way as Figure~\ref{fig:ABPIC-fold}. The phase folded F127M light curve has a sinusoidal variation, but lacks phase coverage between 0.6 to 0.8. The F139M light curve agree better with a flat line.}
  \label{fig:2M0122-foldedLC}
\end{figure}

The hybrid PSF reduction removes the flux contamination from the primary star and reveals the image of the companion at high fidelity (Figure~\ref{fig:2M0122-subcomp}).  We simultaneously measured the light curves for both the primary star and the companion using PSF photometry. The results are in Figure~\ref{fig:2M0122-lightcurve}. The raw light curves for the primary star suffer from the HST ramp effect, especially in the first two orbits where the photometry is more than 0.5\% below the visit median. HST/WFC3 light curves usually only have large-amplitude ramps in the first orbits of an observational visit. However, in our case, because telescope rolls alternated between orbits, the star's PSFs fell on different pixels of the detector in the odd and even orbits and charge trapping (that functions in pixel levels) has nearly independent effect on the light curves in the odd and even orbits. To correct for this systematics, we fed the image cubes to the RECTE model \citep{Zhou2017}, calculating the model ramp profiles as well as independent linear trends for the odd and even orbits light curves, and then divided  the observed light curves by the instrumental profiles. The corrected light curves are  in Figure~\ref{fig:2M0122-lightcurve}. They are featureless and have standard deviations of 0.16\% and 0.14\% in the F127M and F139M normalized light curves. The standard deviations are within 15\% of photon noise.

For the companion light curves, the standard deviation is at 0.5-1\% level,  comparable to the typical ramp amplitude. Therefore, the ramp effect in 2M0122B's uncorrected light curves are not as visible as those in 2M0122A's. In the corrected and normalized light curve, the standard deviations are 0.80 and 1.1\% in the F127M filter and the F139M filter, respectively.

Figure~\ref{fig:2M0122-periodogram} shows the result of LS periodogram analysis for 2M0122 A and B in F127M and F139M. There is only one peak that does not coincide with the HST orbital period or its integer multiples in all the four periodograms. It is the 6.0 hour peak in the \targetii F127M curve, which is marked in a red solid line. The periodograms of 2M0122A have a few high-significance peaks that are located exactly at the HST orbital period or its multiple integers. They are likely introduced by the periodic window functions. The F139M periodogram for \targetii does not show any notable peaks with power stronger than 0.15. We then fold \targetii's light curves to the 6.0 hr period and plot the results in Figure~\ref{fig:2M0122-foldedLC}. The F127M phase-folded light curve displays sinusoidal modulations despite the lack of coverage between phase 0.6 to 0.8. Fitting a sinusoid with fixed period of 6.0 hr, we calculated the best-fit modulation amplitude to be $0.52\pm0.11\%$. The F139M light curve agrees better with a flat line with a 3-$\sigma$ upper limit of 0.46\% on the modulation amplitude. In the phase-folded light curves, the primary also have smaller mean squared residuals with flat lines.

\subsection{\targetIII}

\begin{figure*}[t]
  \centering
  \plottwo{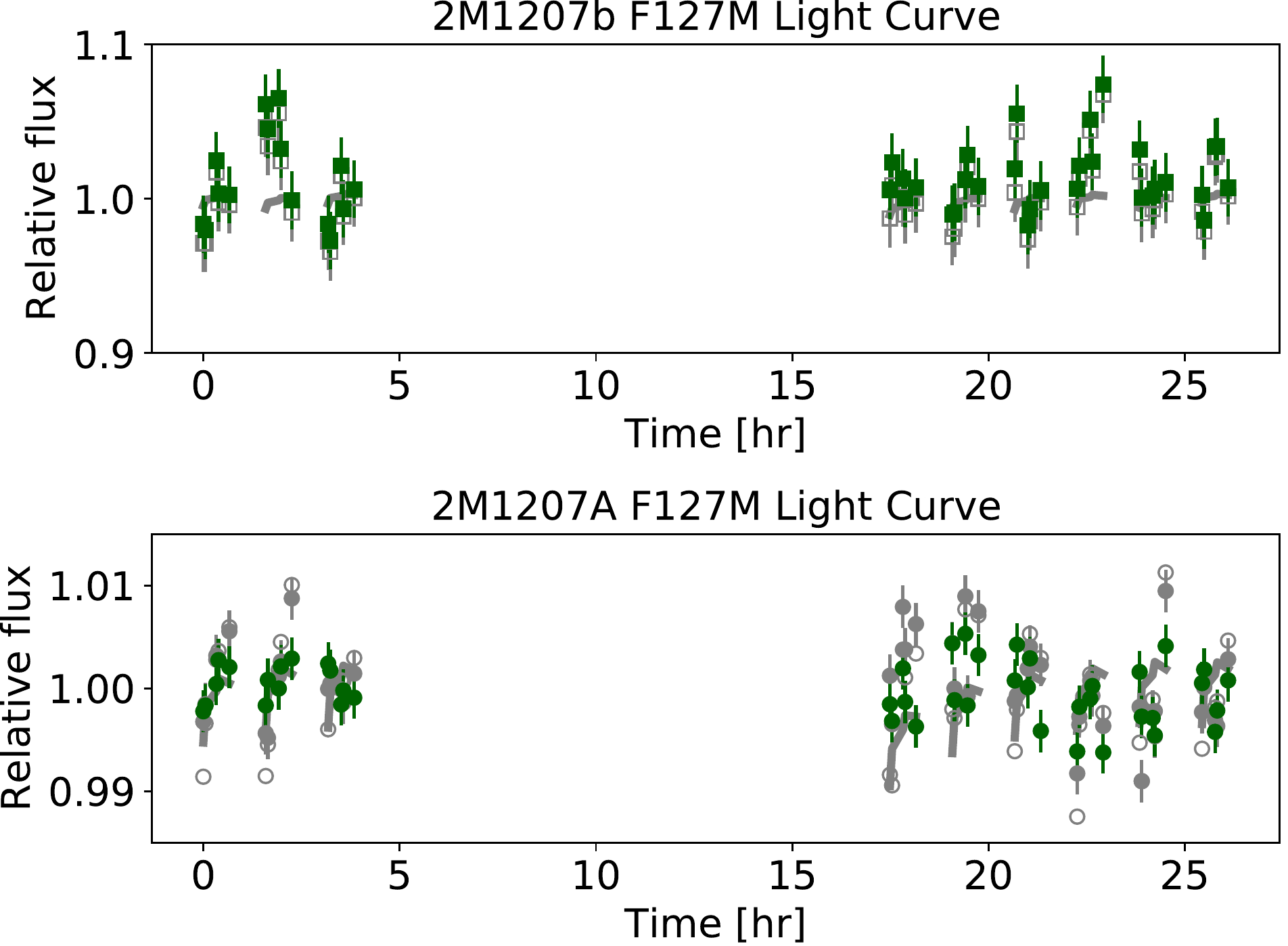}{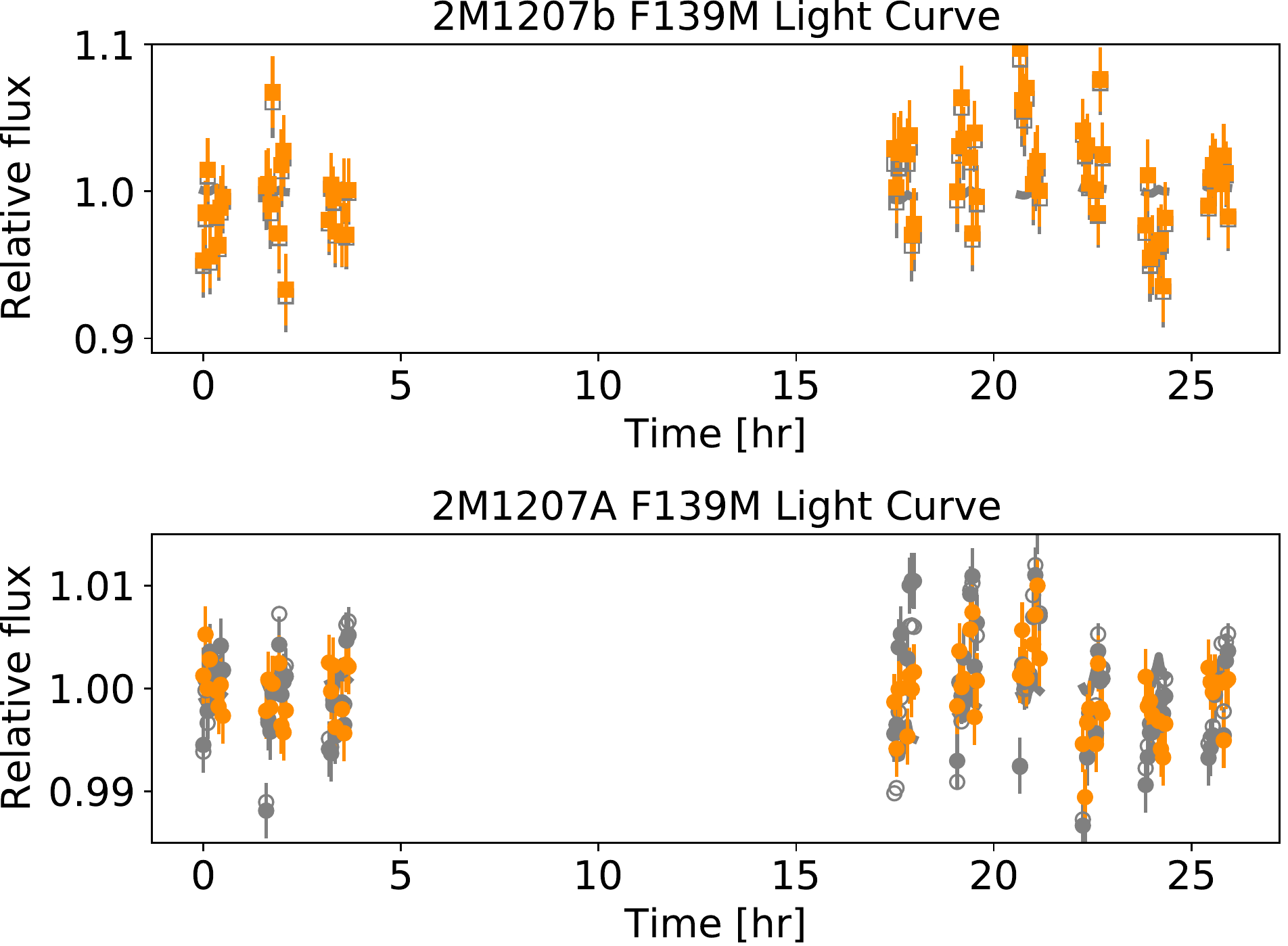}
  \caption{Light curves of 2M1207b (upper panel) and A (lower panel) in F127M and F139M. For 2M1207A, The uncorrected photometries are in gray open symbols. For 2M1207b, Light curves corrected by RECTE are in colored squares. For 2M1207A, light curves corrected by RECTE are in gray filled circles and light curves corrected by orbital-medians are in colored circles. The RECTE ramp profiles, which have amplitude significantly smaller than the scatters of the uncorrected light curves, are plotted in gray curves.}
  \label{fig:2M1207-LC}
\end{figure*}

\begin{figure}
  \centering
  \includegraphics[width=0.5\textwidth]{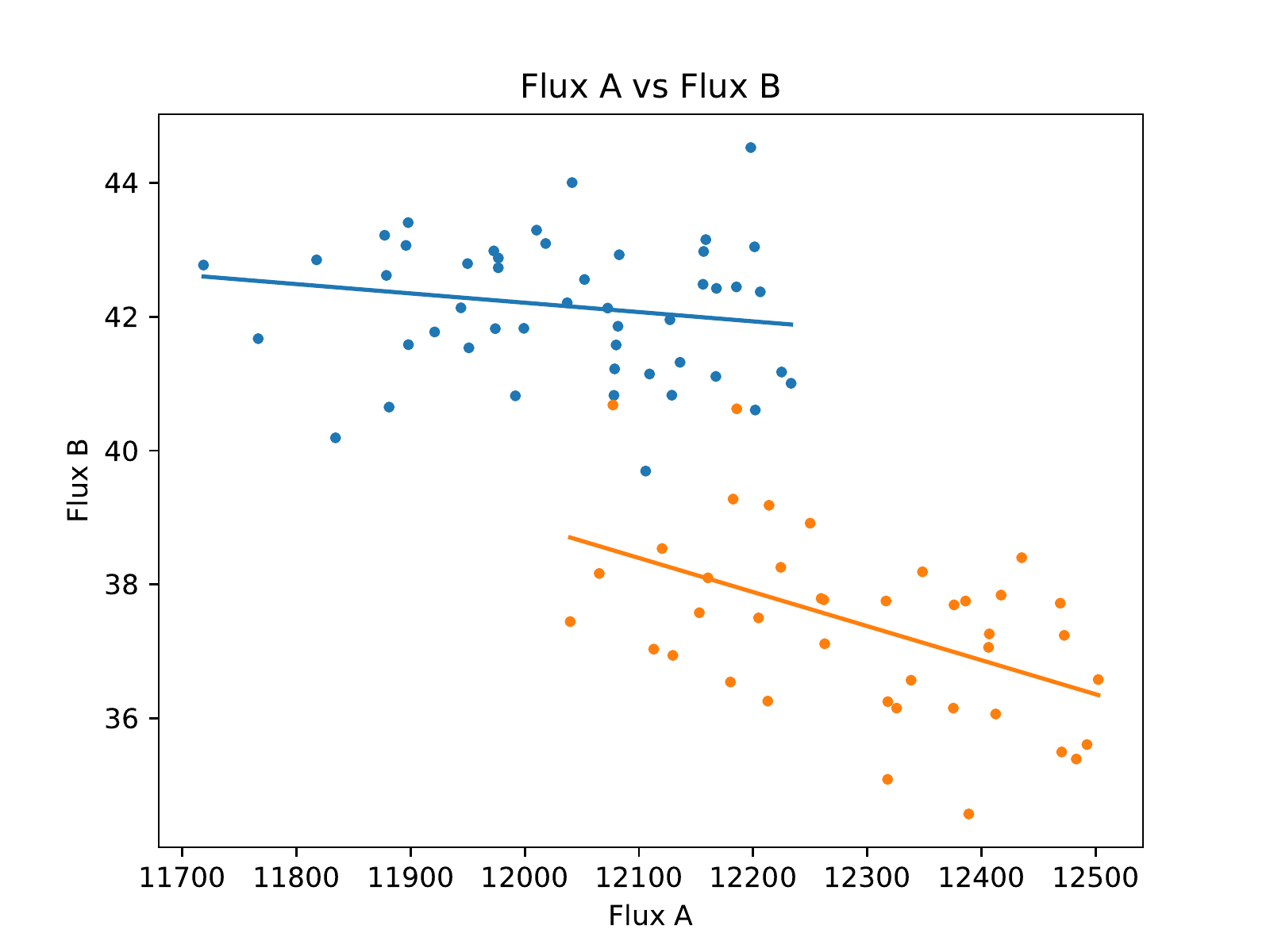}
  \caption{Correlated photometry between 2M1207A and 2M1207b.}
  \label{fig:2M1207-trend}
\end{figure}

\begin{figure}
  \centering
  \plotone{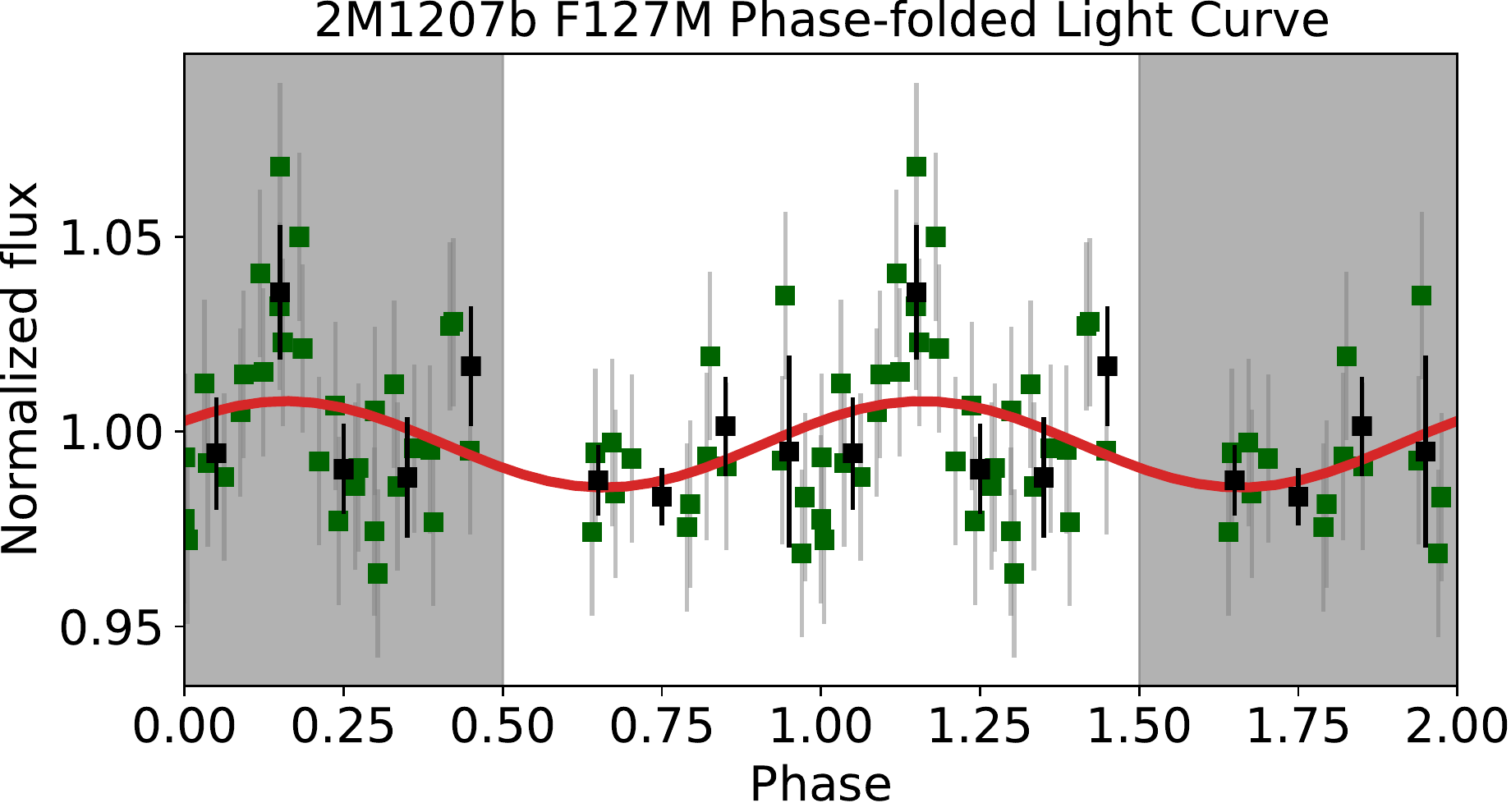}
  \plotone{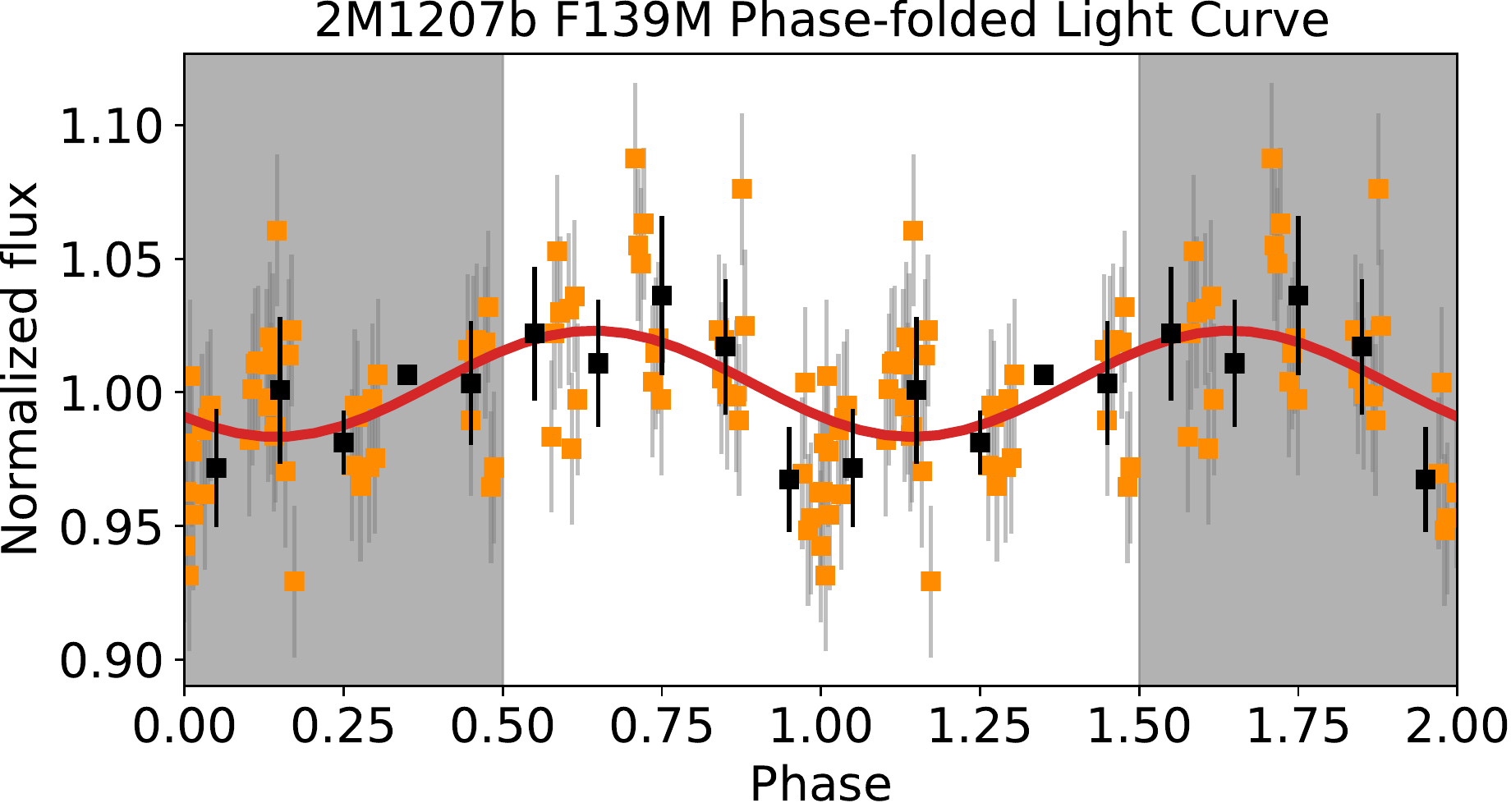}
  \caption{F127M and F139M light curve for \targetiii folded to the best-fit sine wave period.}
  \label{fig:2M1207-folded-LC}
\end{figure}

Systematics significantly affect the light curves in F127M and F139M for both 2M1207A and b. For the uncorrected light curves of 2M1207A, the most prominent systematic feature in every orbit is the 1$-$1.5\% amplitude exponential ramp. These ramps are nearly one order of magnitude larger than those seen in transiting planets observations. We attempted to use RECTE \citep{Zhou2017} to mitigate the ramp effect, but it only marginally reduced the effect. It likely that the a combination saturated PSF core and secondary mirror displacement effect caused these systematics. We made the same measurements and correction on a background star that is $~2.5$ magnitude fainter than 2M1207A in F127M filter and not saturated in our observations. The standard deviation of RECTE corrected light curves for the background star is within 15\% of the photon noise limit. We eventually used the median combination of the light curves for each orbit as the common systematic profile and divided it from the uncorrected light curves for the correction. The corrected result is in Figure~\ref{fig:2M1207-LC}.

For the companion, the most important systematics manifest in the anti-correlation between its intensity and that of the primary as shown in Figure~\ref{fig:2M1207-trend}. Because the companion and the primary contribute comparable fluxes at the region around the centroid of the companion's PSF, an anti-correlation  suggest that the PSF fit does not  break the degeneracy completely. We use the best-fit linear trend to de-correlate the light curve for the companion and provide the corrected light curve in Figure~\ref{fig:2M1207-LC}. In the corrected light curves, the standard deviation of the photometry exceeds the photon noise by 13\% and 43\% in the F127M and F139M light curves, respectively.

The apparent variability in the companion light curves result in rather complex, multi-peaked LS periodograms. The peaks that are closest to the period found in \citet{Zhou2016} are located at 10.6 hr and 12.0 hr for F127M and F139M, respectively. Both periods are within $3-\sigma$ limits of the period reported in \citet{Zhou2016}. Sinusoidal fit to the phase-folded light curves result in amplitudes of $0.57\pm0.44\%$ and $2.45\pm2.67\%$ for the F127M and F139M light curves respectively \ref{fig:2M1207-folded-LC}. The 10.6 hr periodic signal in the F127M light curve is insignificant and the sine fit has almost identical $\chi^2$ to a flat line fit. However, the sine fit to the F139M light curve significantly reduces the residual compare to flat line fit. The residual for the sine fit to the F139M light curve has a standard deviation the same as the photon noise.

\subsection{Assessing the Detection Significance for the Periodic Signals}
\label{sec:result:period}

\begin{figure*}[t]
  \centering
  \plottwo{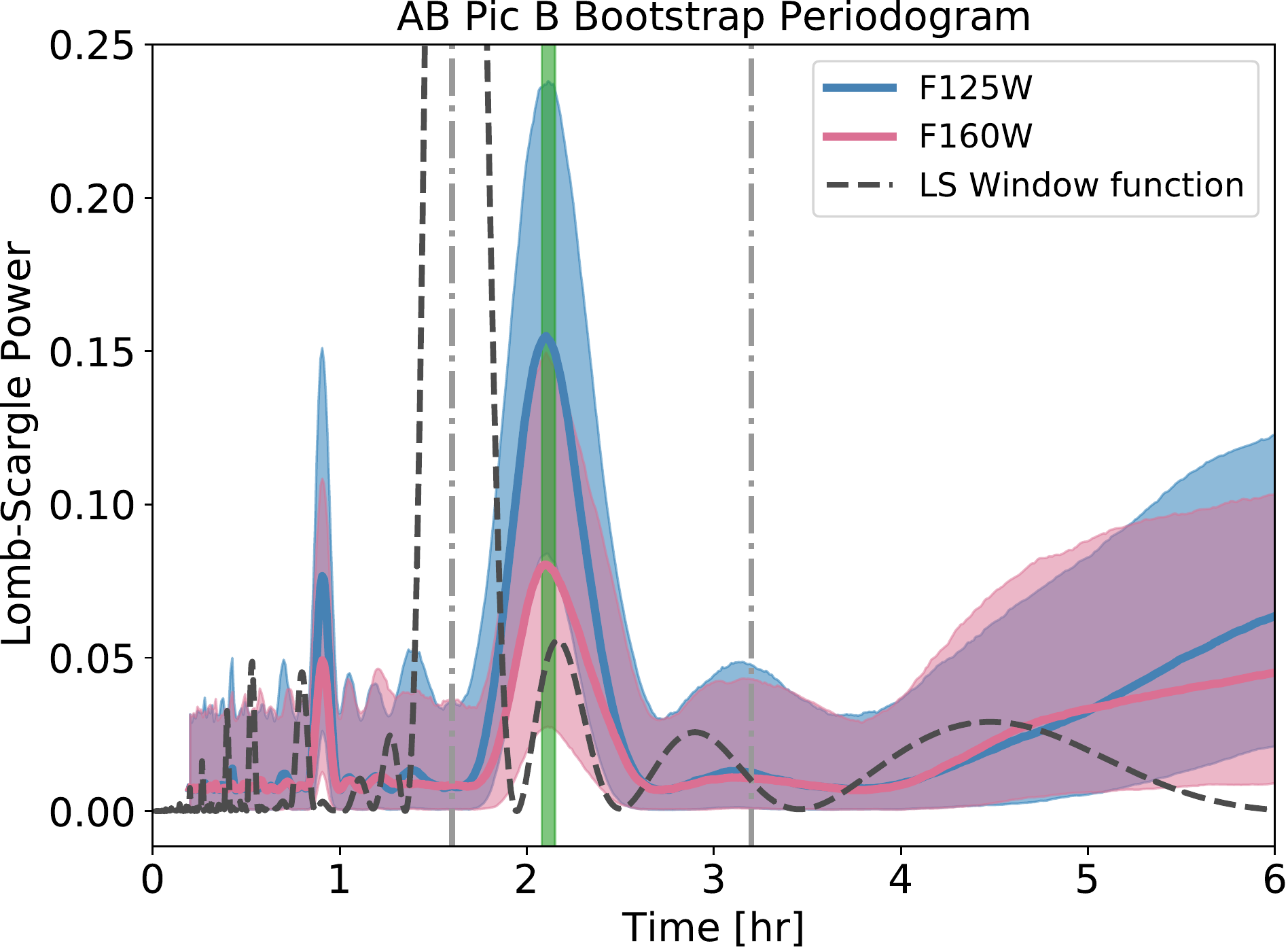}{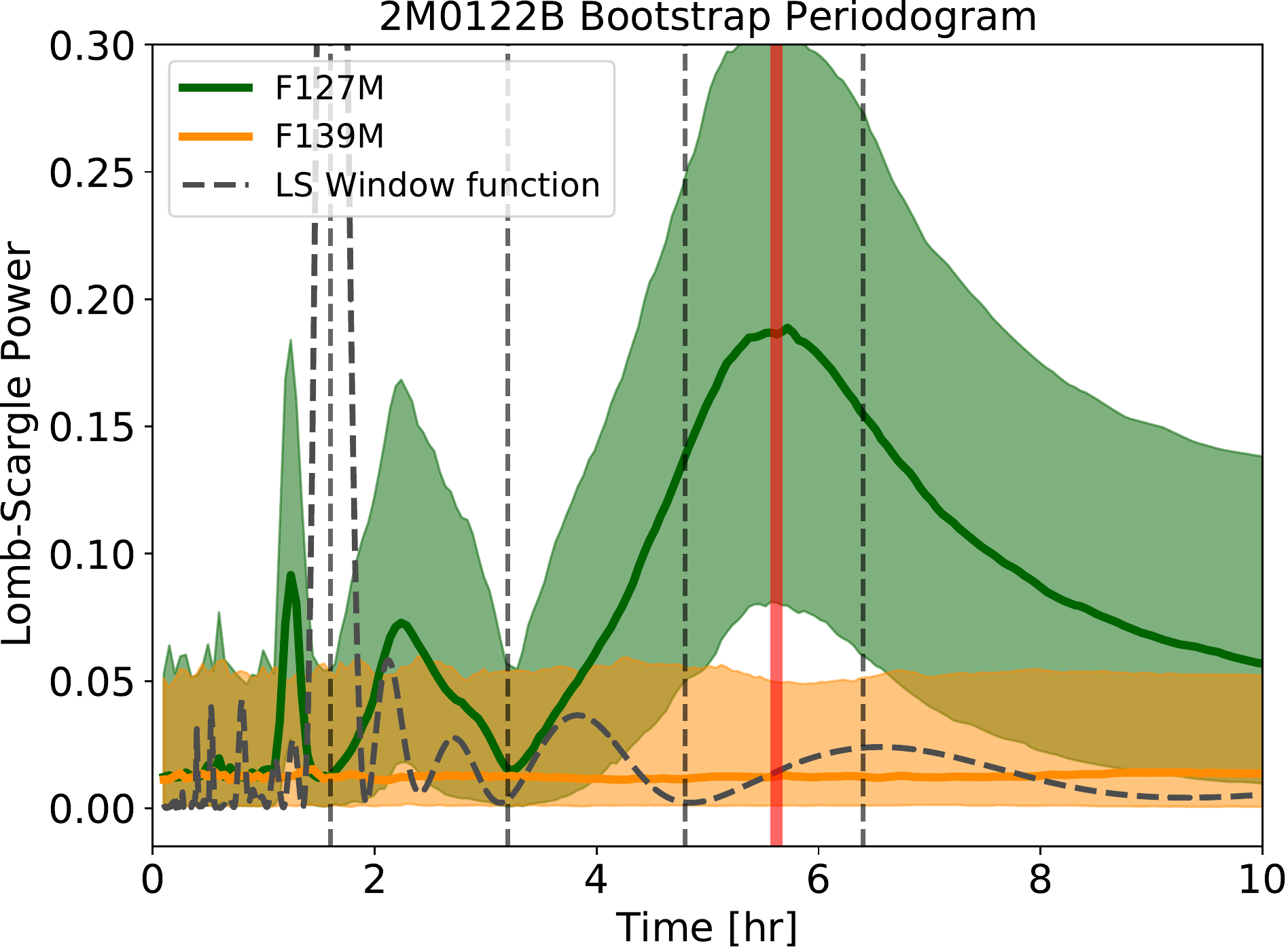}
  \caption{Bootstrap analysis result for LS periodograms for \targeti (left) and \targetii (right). The shaded regions mark the 5 and 95 percentiles ($2-\sigma$) for the distributions of LS power that are calculated from bootstrap. Green/red vertical lines marks the periods being tested and the gray dashed line marks the HST orbital period and its multiples. }
  \label{fig:LSBS}
\end{figure*}

We marginally detected periodic signals in the F125W and F160W light curves of \targeti and the F127M light curve of \targetii using the LS periodogram. We rejected all signals whose periods are close to the HST orbital period or its integer multiples because they are likely introduced by the sample window functions or the instrument/telescope systematics. To further evaluate the robustness of the periodogram detections, we applied a bootstrap analysis that is similar to that used in \citet{Manjavacas2017}. In this method, we generated synthetic light curves by adding randomly permuted residuals of the sinusoidal fit to the best-fit sine wave model and iteratively ran LS and sinusoidal fit with the synthetic light curves.
For each light curve, we iterated  1\,000 times to obtain the distributions of the LS power and assess the detection significance by measuring the ratio of the average peak power and its standard deviation. This test examines the effect of correlated noise on the LS results.

Figure~\ref{fig:LSBS} shows the LS periodograms from bootstrap analysis for \targetI and \targetII. For \targeti, the $P=2.1$ hr signal is above  $2\sigma$ for both light curves: the peak of LS power is $3.1\sigma$ and $2.2\sigma$ above zero in the F125W and F160W periodograms, respectively. For \targetii, the $P=6.0$ hr signal only has detection ($2.7\sigma$) in the F127M periodogram, which is consistent with the simple LS analysis in \S\ref{sec:results}. Therefore, we conclude that the periodic signals that we found in the light curves of \targeti and \targetii are robust and not artifacts or results of correlated noise. However, the detection significance is rather marginal for the signal of \targeti in F160W and \targetii in F127M.

False positives in the periodigram may be introduced by the observational duty-cycle imposed by periodic HST-orbit  target visibility interruptions (the "window function").  To evaluate the possibility that the detected signals are false positives introduced by the orbital visibility window of HST, we calculated the LS periodogram on the window function itself. We modeled the window function effect by creating a light curve in which the target-visible segments have a constant intensity of 1.0 and earth-occulted segments have a constant intensity of 0. We then calculated the LS periodogram on this light curve as the periodogram of the window function and plotted the results in Figure \ref{fig:LSBS}. The periodogram of the window function has a peak close to the 2.1 hr signal we detected in the light curves of  AB Pic B. The overlap of the periodograms peaks of AB Pic B and the window function further reduces the periodic signal detection significance by $1\sigma$. Based on the same analysis, the window function does not interfere the periodic signal detection for 2M0122B. We repeated the same procedures but adding random noise that have standard deviation same as the light curve photon noise to the window function curve and found the noise in the window function light curves did not change their periodograms.

\subsection{Simulated Companion Injection Test for Periodic Signal Recovery from Hybrid-PSF Photometry}
\begin{figure}[htbp]
  \centering
  \plotone{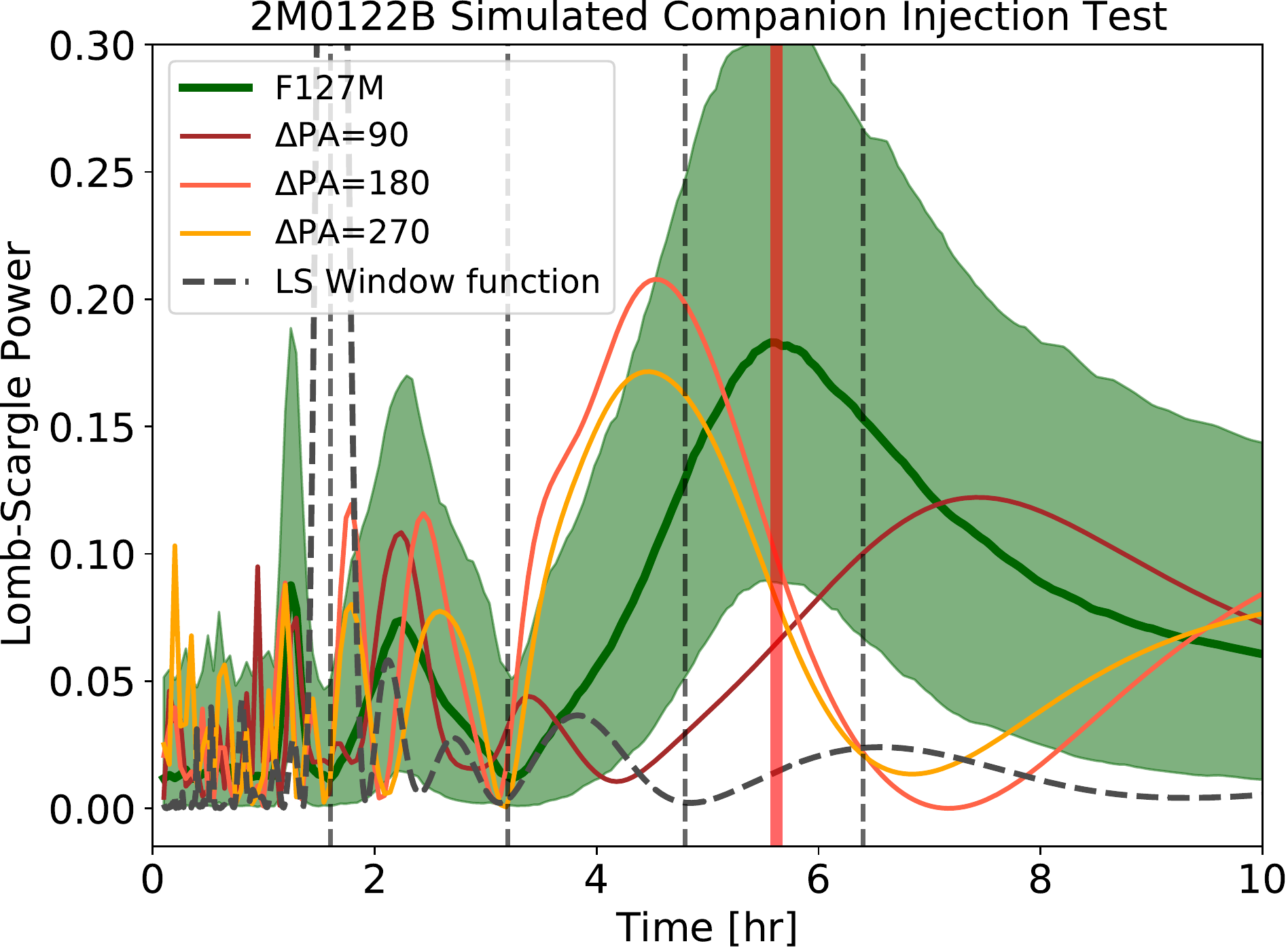}
  \caption{Periodograms for the light curve of  2M0122B in F127M and the three recovered light curves for the injected simulated companions. The simulated companions injections assumed no flux intensity modulations. The periodograms of the injected simulated companions have peaks that have comparable Lomb=Scargle powers as the strongest peak for 2M0122B. This figure highlights the possibility of detecting false positive periodic signals when applying Lomb-Scargle to light curves measured by hybrid photometry.}
  \label{fig:injtest}
\end{figure}
We performed simulated companion injection test to examine if false positive periodic signals could arise in the hybrid-PSF photometry process even for a constant-intensity object. We first injected simulated companions to 2M0122 F127M \texttt{flt} frames. The simulated companions were TinyTim PSFs and had the same intensity as the average intensity for 2M0122B in F127M. We generated three data-sets by injecting the simulated companions at three locations. The injected companions had the same separation to 2M0122A as 2M0122B, but had position angles (with respect to 2M0122A) that were different from that for 2M0122B by $90^{\circ}$, $180^{\circ}$, and $270^{\circ}$ for the three data-sets, respectively. The photon noise of the injected companions were also added to the uncertainty of each pixel for the purpose of uncertainty and likelihood calculations. We then repeated the same reduction procedures detailed in \S
\ref{sec:hybrid_reduction} on the simulated datasets to recover the light curves for the injected companions. We assumed zero intensity modulations for the injected signal and examined whether the periodograms for the recovered light curves have any significant peaks (i.e. false positives).

Figure \ref{fig:injtest} compares the periodograms for 2M0122B and the three simulated companions. We find that the periodograms for the simulated companions have peaks that are comparable to the strongest signal that we detect for 2M0122B. The periodograms for the simulated companions that have $180^{\circ}$ and $270^{\circ}$ position angle difference from 2M0122B have peaks that are similar or even stronger than the 6.0 hr peak in the periodogram for 2M0122B. The highest peaks in these two periodograms for the simulated companions correspond to periods that are close to $3\times$ the HST orbital period. This test demonstrates that low significance periodic signals (false positives) may emerge in the process of PSF subtraction and photometry, particularly when the detected periods are close to HST orbital period or its integer multiple. Our 6.0 hr periodic signal detection for 2M0122B in its F127M light curve, although not being in the vicinity of integer multiples of HST orbital period, may suffer similar systematics. We therefore emphasize the periodic signal detection for 2M0122B is marginal and tentative.

The periodic signals detected in the AB Pic B observations are from aperture photometry for which recovered signal would be the same as the injected signal. Therefore the detections for AB Pic B will not suffer from the systematics revealed in above simulated injection test.

\subsection{A Search for Additional Close Companions}
\begin{figure*}[t]
  \centering
  \includegraphics[width=0.32\linewidth]{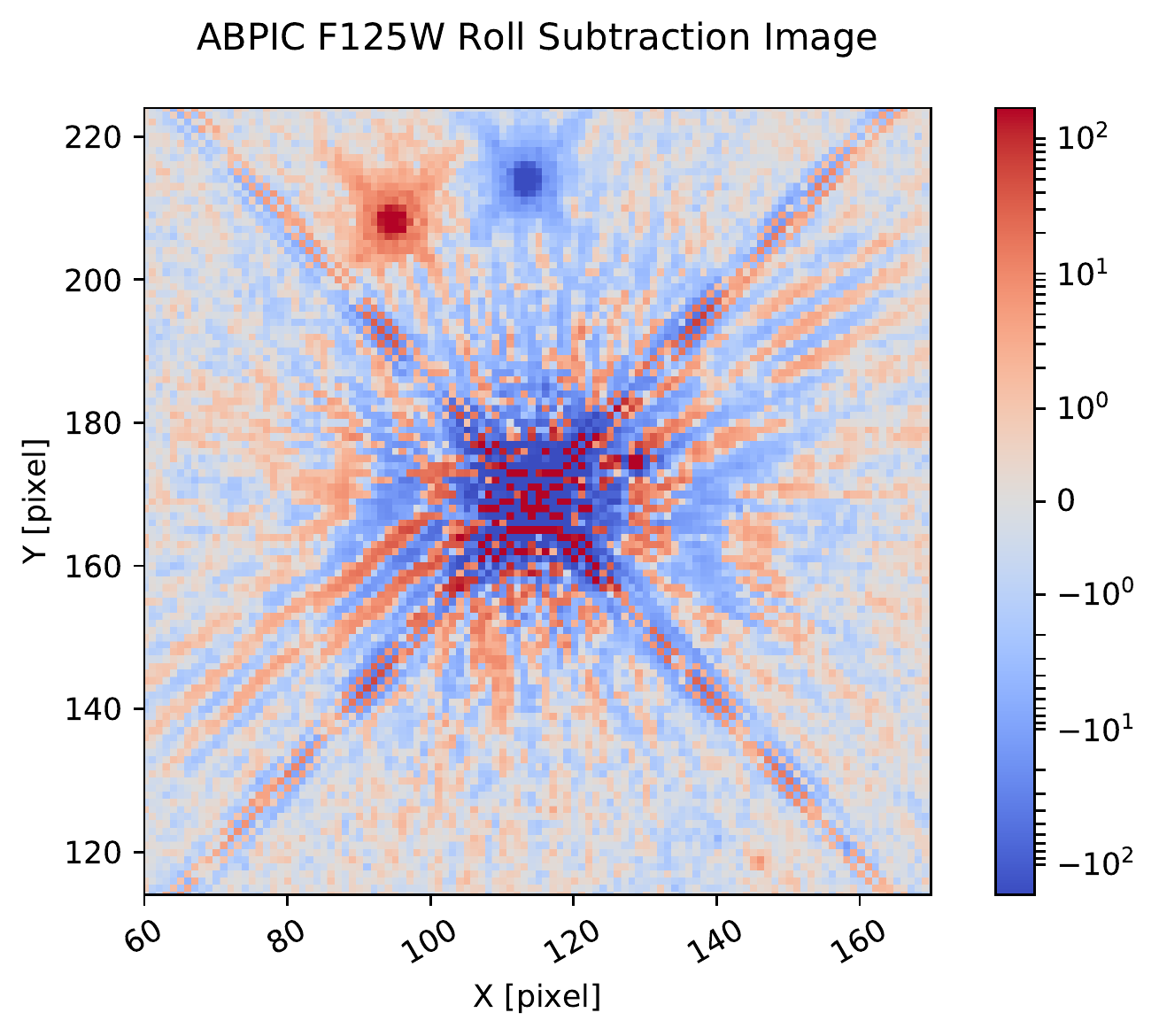}
  \includegraphics[width=0.32\linewidth]{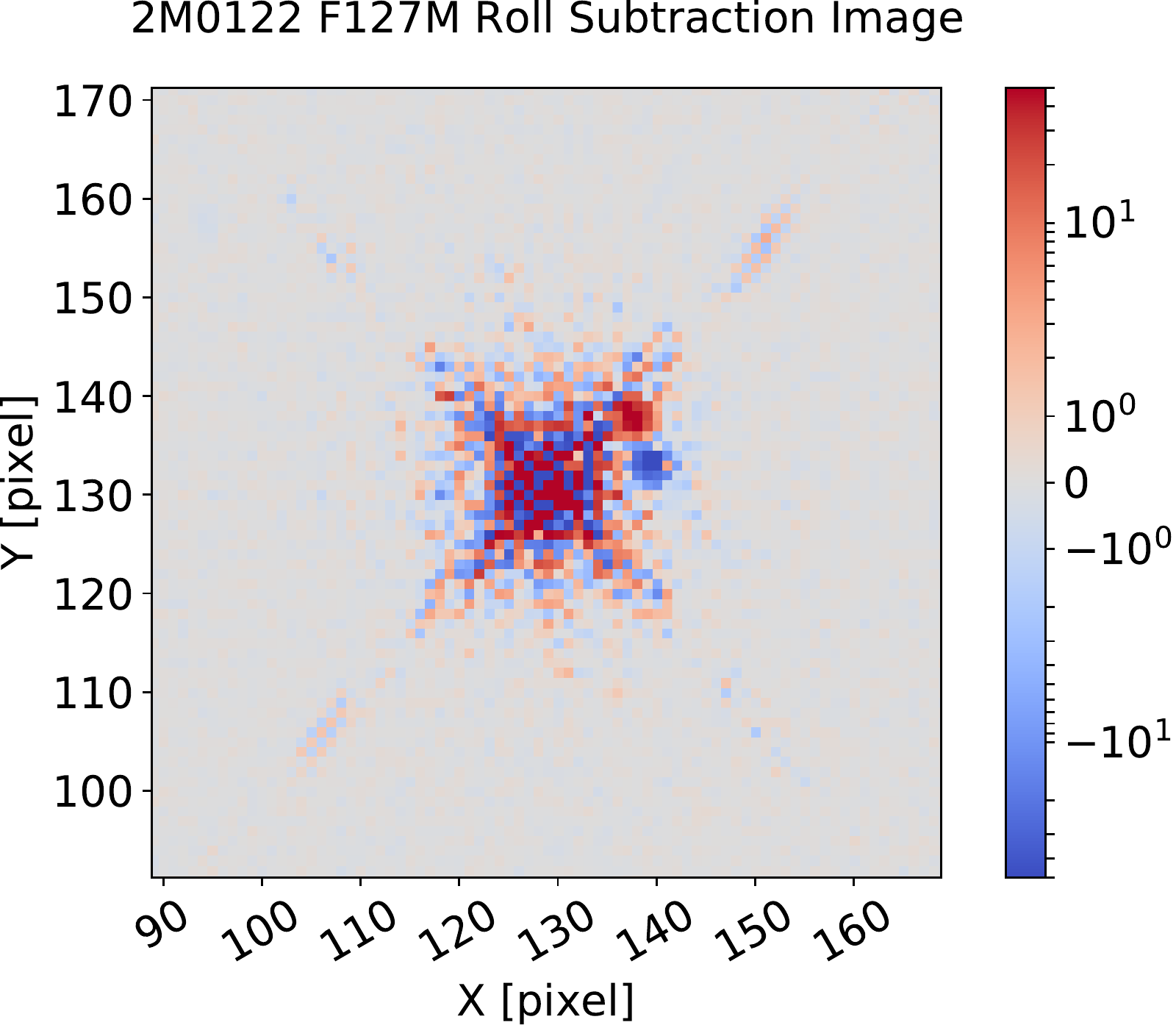}
  \includegraphics[width=0.32\linewidth]{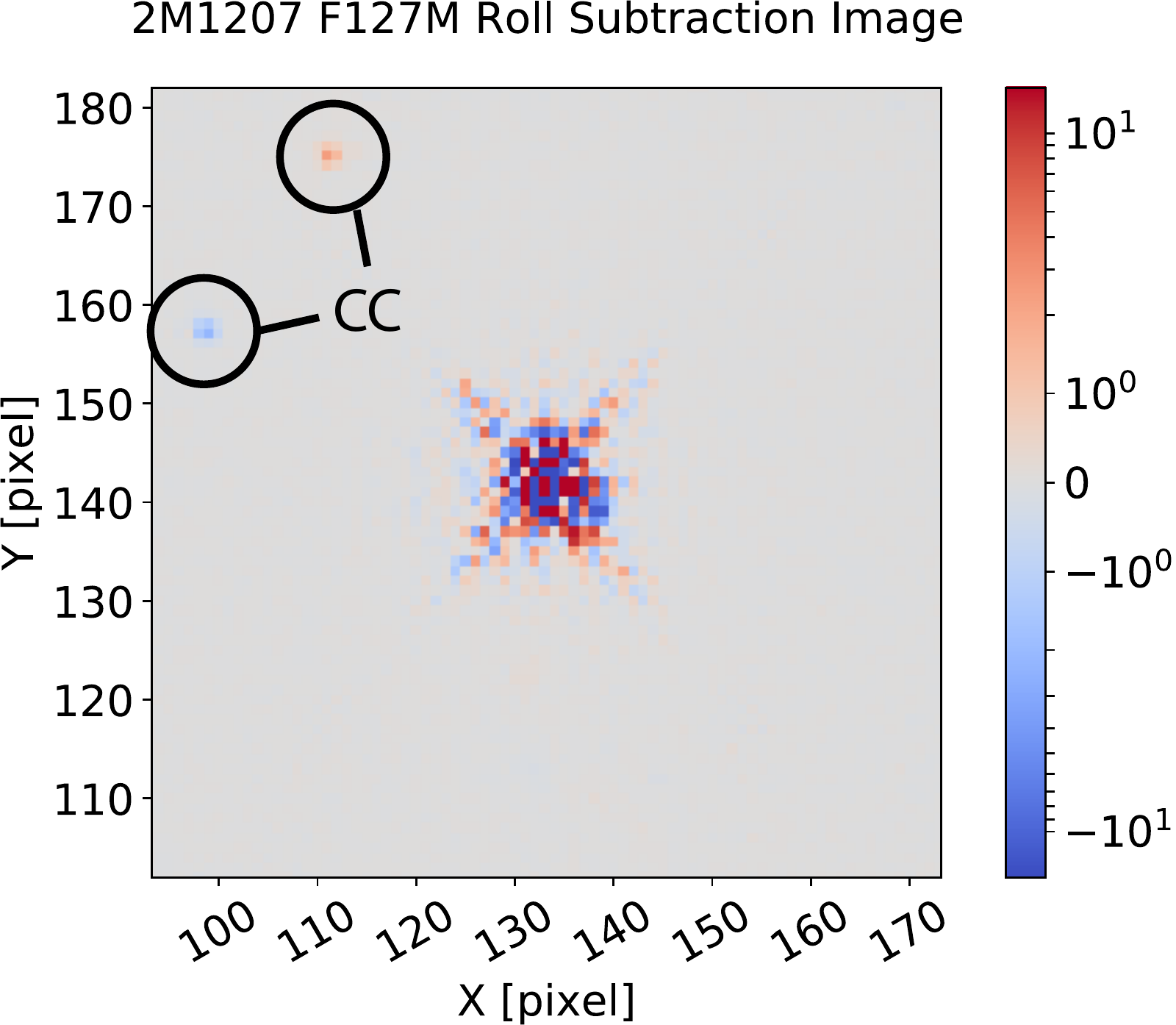}
  \caption{Roll subtracted images for \targeti, \targetii, and \targetiii. The color stretch of the images is set to enhance faint point sources in the background. The only visible close companion candidate in these images is in 2M1207 image, locating 5.03\arcsec away from 2M1207 A in the northeastern position in the image frame. }
  \label{fig:rollsub}
\end{figure*}

\begin{figure*}[t]
  \centering
  \includegraphics[width=0.32\linewidth]{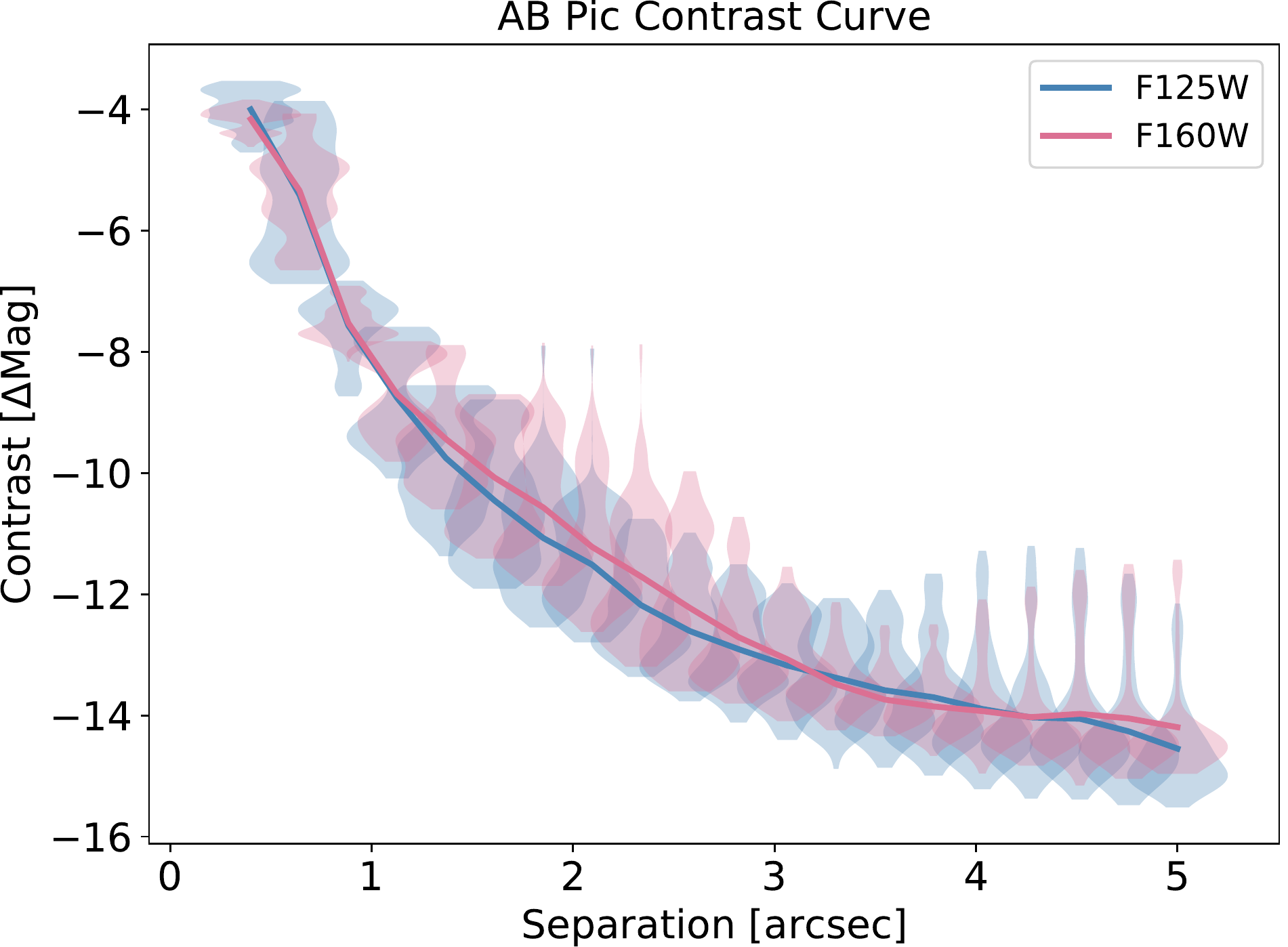}
  \includegraphics[width=0.32\linewidth]{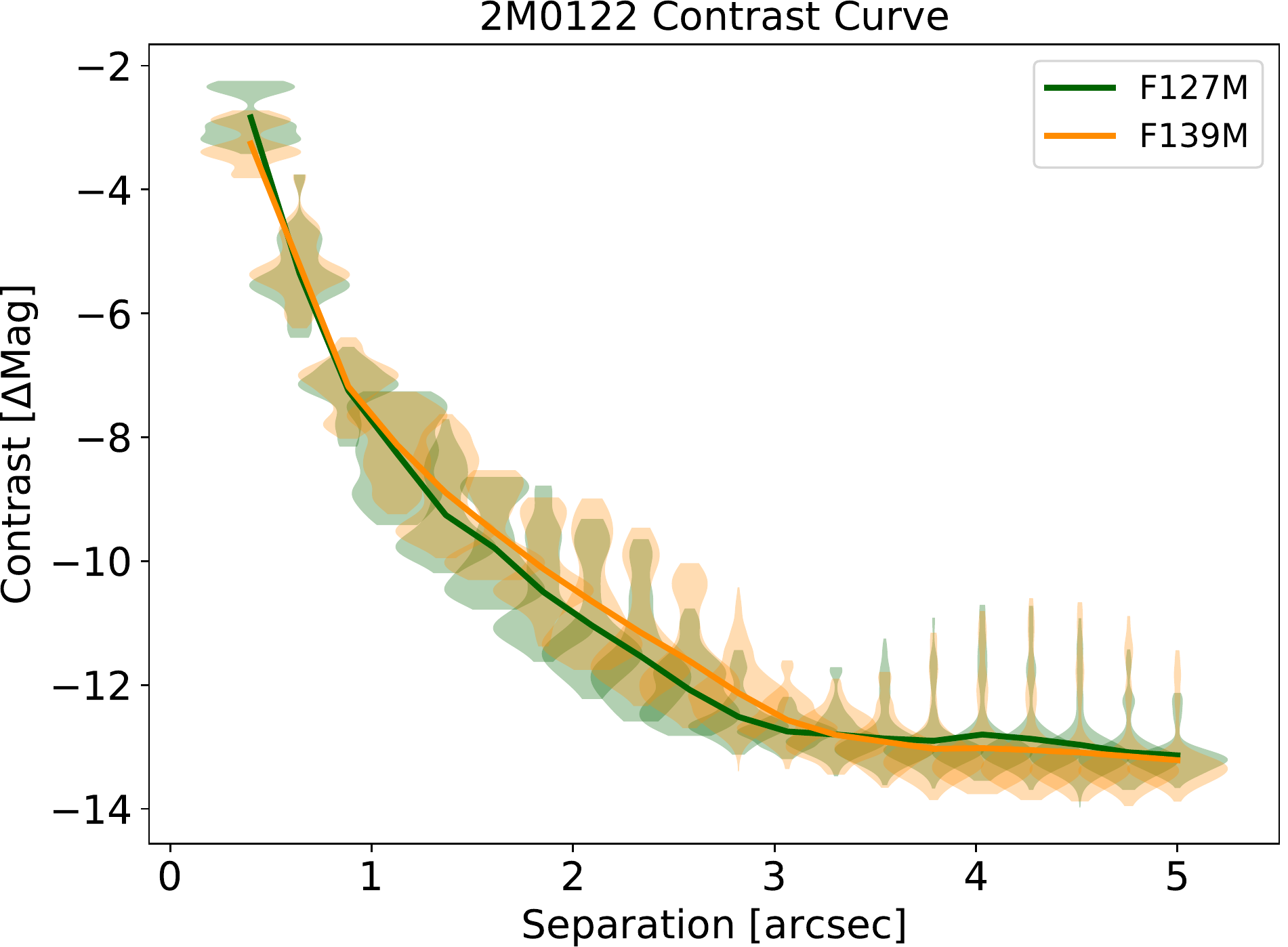}
  \includegraphics[width=0.32\linewidth]{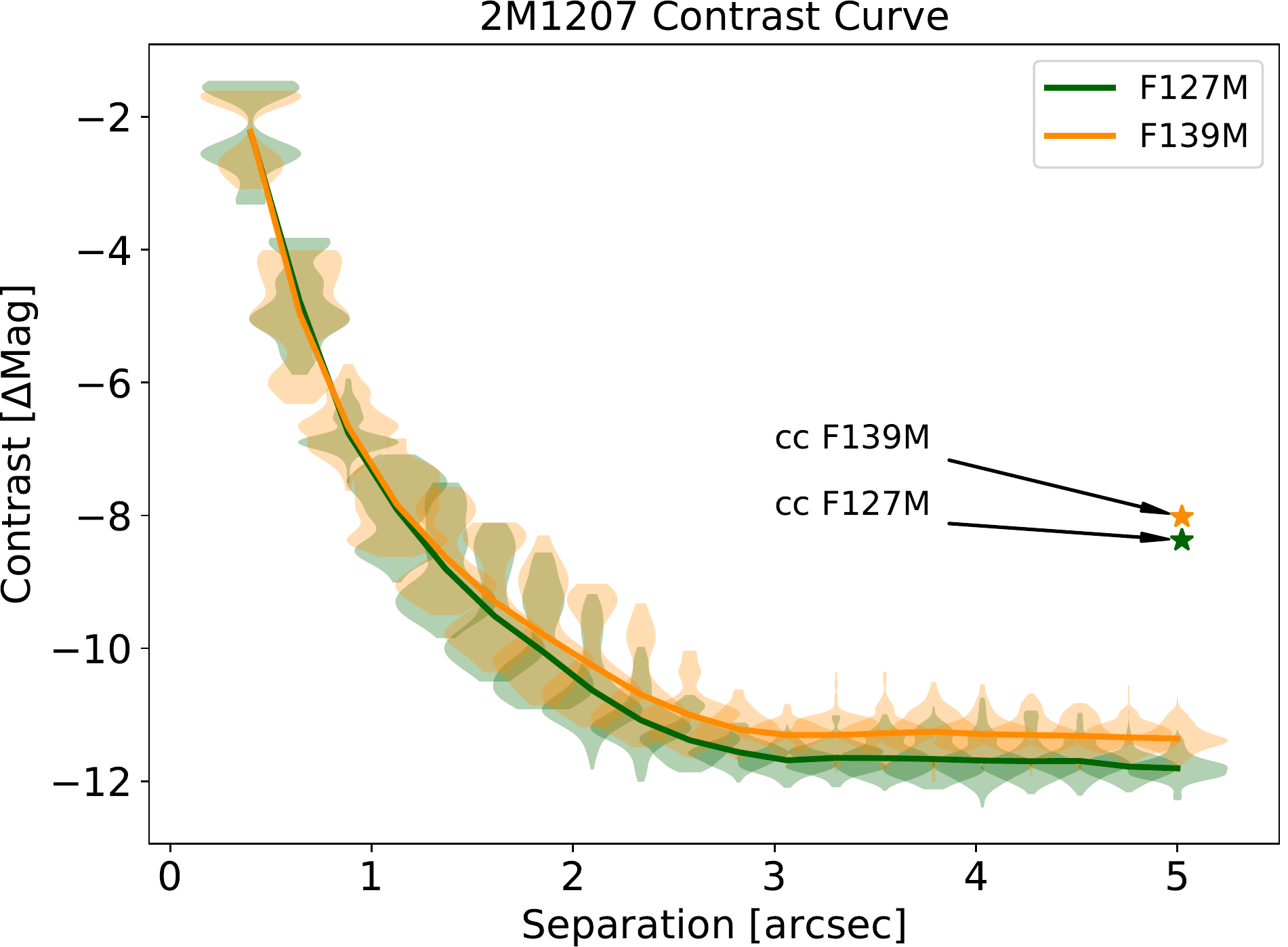}
  \caption{$5\sigma$ contrast curves estimated from roll-subtraction images for the three objects. The lines are for the average contrast measurements, and the violin-plots shows the distributions of the contrast measurements at each sampled separations. The 2M1207 panel also includes the photometry in F127M and F139M for the close companion candidate at 5.03\arcsec. }
  \label{fig:contrast}
\end{figure*}

For each object, we combine the entire sets of 2RDI images to build ultra-deep images to search for planetary-mass companions (Figure~\ref{fig:rollsub}).  We collect color information from multi-filter observations to evaluate the probability of detected close companions being planetary mass objects.

We do not detect any point sources that are closer than $5\arcsec$ to any of the three primary objects. There is a point source $5\arcsec.03$ from 2M1207A with a position angle of 307.9 degrees. It has an F127M flux 17.7\% of that for \targetiii, and a F139M flux that is 16.7\% fainter than its F127M flux. We compared position of this close companion in this observation with those in the archived WFC3/IR images (Program ID: 13418, PI: D. Apai). The astrometry of the close companion agrees with a stationary background star rather than a co-moving companion to 2M1207 \citep[$\mathrm{pmRA}=-64.1\,\mathrm{mas/yr},\, \mathrm{pmDEC}=-23.7\,\mathrm{mas/yr}$,][]{Gaia2016,Gaia2018} with a time baseline of 729 days . The closest point source to 2M0122 is $5\arcsec.58$ away from 2M0122A at a $353.2^{\circ}$  position angle. It is 0.6\% of the brightness of \targetii in F127M and has a F139M flux density 24.6\% fainter than its F127M flux density.

We estimated the point source detection sensitivity through contrast curves. The contrast was calculated at 20 separations evenly sampled from 0.4\arcsec{} to 5\arcsec{} from the host stars/brown dwarf. To estimate the contrast at a specific separation, we injected model PSFs at random position angles with respect to the primary stars and scaled the PSFs with a factor $A$, such that the integrated signal within a three-pixel radius aperture is above $5\sigma$. We defined the integrated SNR as
\begin{equation}
  \label{eq:7}
  \mathrm{SNR} = \sqrt{\Bigl(\frac{\mathrm{PSF}_i}{\mathrm{std}(\mathrm{Pixel}_i)}\Bigr)^2}
\end{equation}
and calculated the contrast as the flux ratio between the injected PSF and the primary star converted to magnitude.
\begin{equation}
  \label{eq:8}
  \mathrm{contrast} = 2.5\times\log\Bigl(\frac{\sum\mathrm{PSF}_i}{\mathrm{Flux_{star}}} \Bigr)
\end{equation}
At each sampled separation, we performed 500 iterations of these calculations to evaluate the effects of position angle on the contrast estimates. Figure \ref{fig:contrast} shows both the average contrast curves and the contrast distribution at each the sampled separations. For every target, the contrast curves for the two filters are nearly identical, which reflects the designs of the observations. For separations larger than 2\arcsec{}, our roll-subtraction images achieve better than 12 magnitude contrast for the observations of \targeti and \targetii, and better than 11 magnitude for \targetiii. Within 1\arcsec, the 2RDI images were limited by subtraction residuals. At such separations, the contrasts were less than 8 magnitude for all three cases. The distributions of the contrast results are non-Gaussian at small separations and gradually becomes Gaussian at large separations. This sequence of contrast measurement distributions demonstrates the limitations of two-roll angular differential imaging in sampling and suppressing the PSF at small separation ($<1\arcsec$). In Figure \ref{fig:contrast}, we also plot the photometry of the close companion candidate identified in the 2M1207 images.

We convert the contrast limits to mass using evolutionary tracks. We derive luminosity using $J$-band brightness and adopt the \citep[cloudy, $f_\mathrm{sep}=2$,][]{Saumon2008} model to map the luminosity and age to mass. Based on these estimates, for \targeti, we can rule out any 10 $M_\mathrm{Jup}$ or more massive companions outside of 40 au or any $5\mjup$ or more massive companion outside of 80 au away from AB Pic A. For \targetii, any companions  more massive than 5 $M_\mathrm{Jup}$ and more than 50 au away from 2M0122A can be detected at $5-\sigma$ significance in our observations. For \targetiii, we can rule out any 2 $M_\mathrm{Jup}$ or more massive companions that are more than 75 au away from 2M1207A.

\section{Discussion}

\subsection{(Quasi-)Periodic Signals in the Light Curves}

The Lomb-Scargle periodogram that we used to search for periodic signal in \S\ref{sec:result:period} is based on least-square fittings to sinusoidal series \citep{Lomb1976,Scargle1982a}. A key assumption of this method is that the intrinsic form of the light curve is a sinusoid \citep{VanderPlas2018}. Certain intensity maps lead to this light curve form (e.g., longitudinal sinusoidal basis map \citealt{Cowan2008}), however, these maps may not represent the underlying physics of brown dwarfs or directly-imaged exoplanets that have heterogeneous clouds. The deviation of light curves from single sine waves is best demonstrated by the high SNR long-term \textit{Spitzer Space Telescope} monitoring of brown dwarfs \citep{Apai2017}. Nevertheless, considering the low significance of the modulation signal detections, single sinusoids are the most practical models for the analysis of this paper.  To emphasize this approximation, we refer to these signals as \emph{quasi-periodic.}

Additionally, rather than being interpreted as the rotational periods, the quasi-periodic signals can also be higher order ($k>1$) planetary-scale waves. \citet{Apai2017} found that the light curves of three L/T transition brown dwarfs cannot be explained by spot-like features, but are better-fit by combinations of $k=1$ and $k=2$ sinusoids, which could be the results of planetary-scale waves. In the planetary-scale wave dominated scenario, the power spectra of the light curves peak at both the full rotational period and the half-period. If the periodic signals correspond to a $k=2$ wave, the rotational periods for \targeti and \targetii should be 4.2 and 11.2 hr, respectively. In this regard, the quasi-periodic signals measure a lower limit for the rotation period or upper limit for the spin velocity.

If the $P=2.1$ hr signal is indeed the rotational period of \targeti, it would be among the fastest rotators for the ultra-cool brown dwarfs. Such fast rotational periods have thus far been found in only a few objects \citep[e.g.,][]{Metchev2015}. Given its mass and radius estimates of $M=13 M_\mathrm{Jup}$ and $R=1.55R_\mathrm{Jup}$ \citep{Chauvin2005, Bonnefoy2010, Patience2012}, the breakup velocity defined as the spin velocity when centrifugal force balances surface gravity corresponds to a rotational period of 0.25\,hr. This break up limit is consistent with the estimates that assumes hydrostatic equilibrium of polytropic gas \citep[Equation 2 of][]{Marley2011,Chandrasekhar1933}. Thus, a period of 2.1 hr is equivalent to a rotational velocity 12\% of the break up limit. In the extreme case where \targeti shrinks to $R=1R_\mathrm{Jup}$ as it cools without losing its angular momentum, the spin rate of \targeti will increase by 55\% and the break up limit will increase by 15\%. In the final stage of this extreme rotational evolution scenario, the rotational velocity of \targeti will be 16\% of the break up limit. Based on the derivation in \citet{Marley2011}, AB Pic B, due to its fast spinning, could have an oblateness over 0.4. But regarding rotational breakup, the quasi-period signal for \targetI{} is well within the physically allowed range.

\subsection{Rotation and angular momentum evolution of planetary-mass objects}

\begin{figure}
  \centering
  \plotone{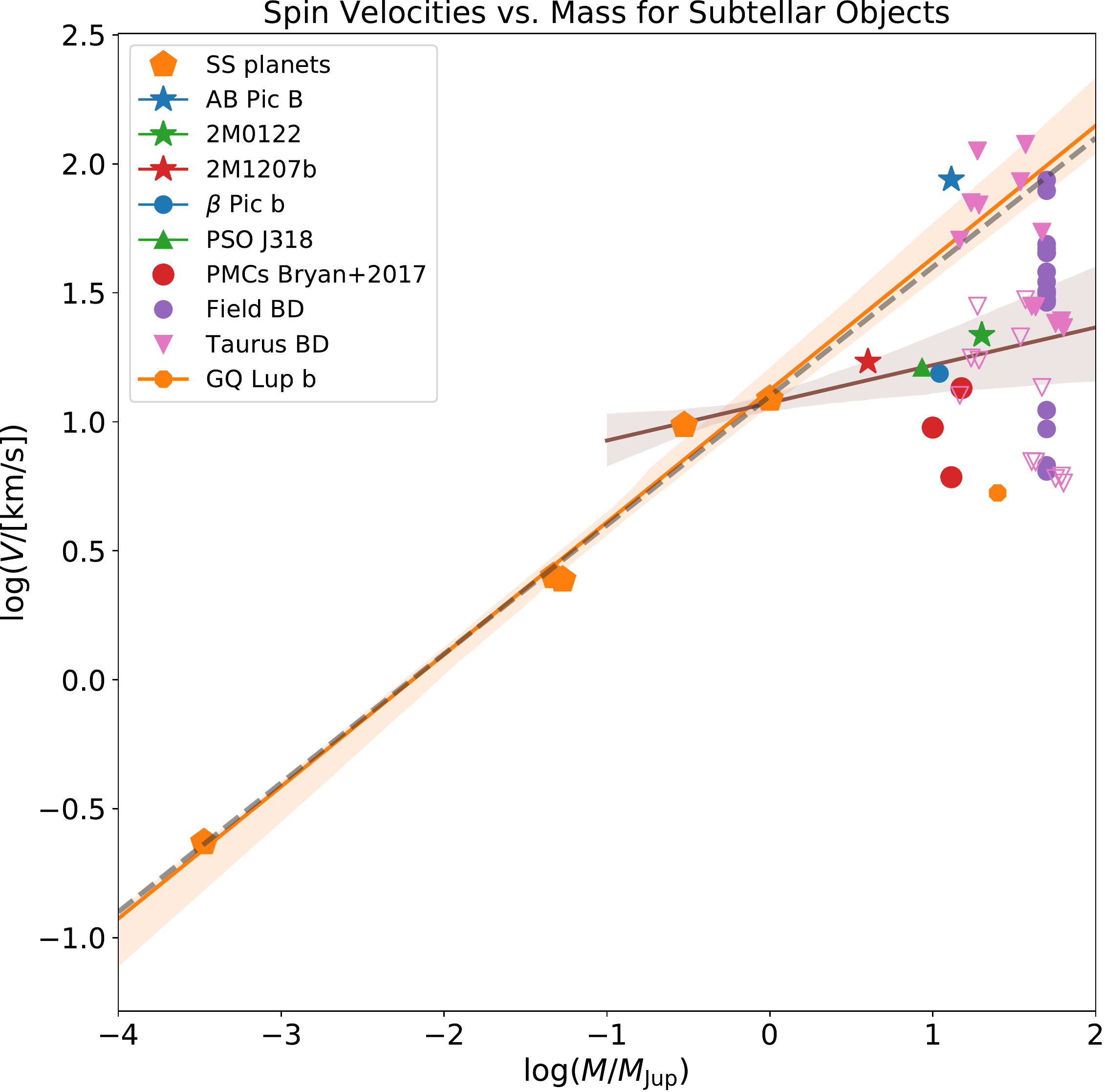}
  \plotone{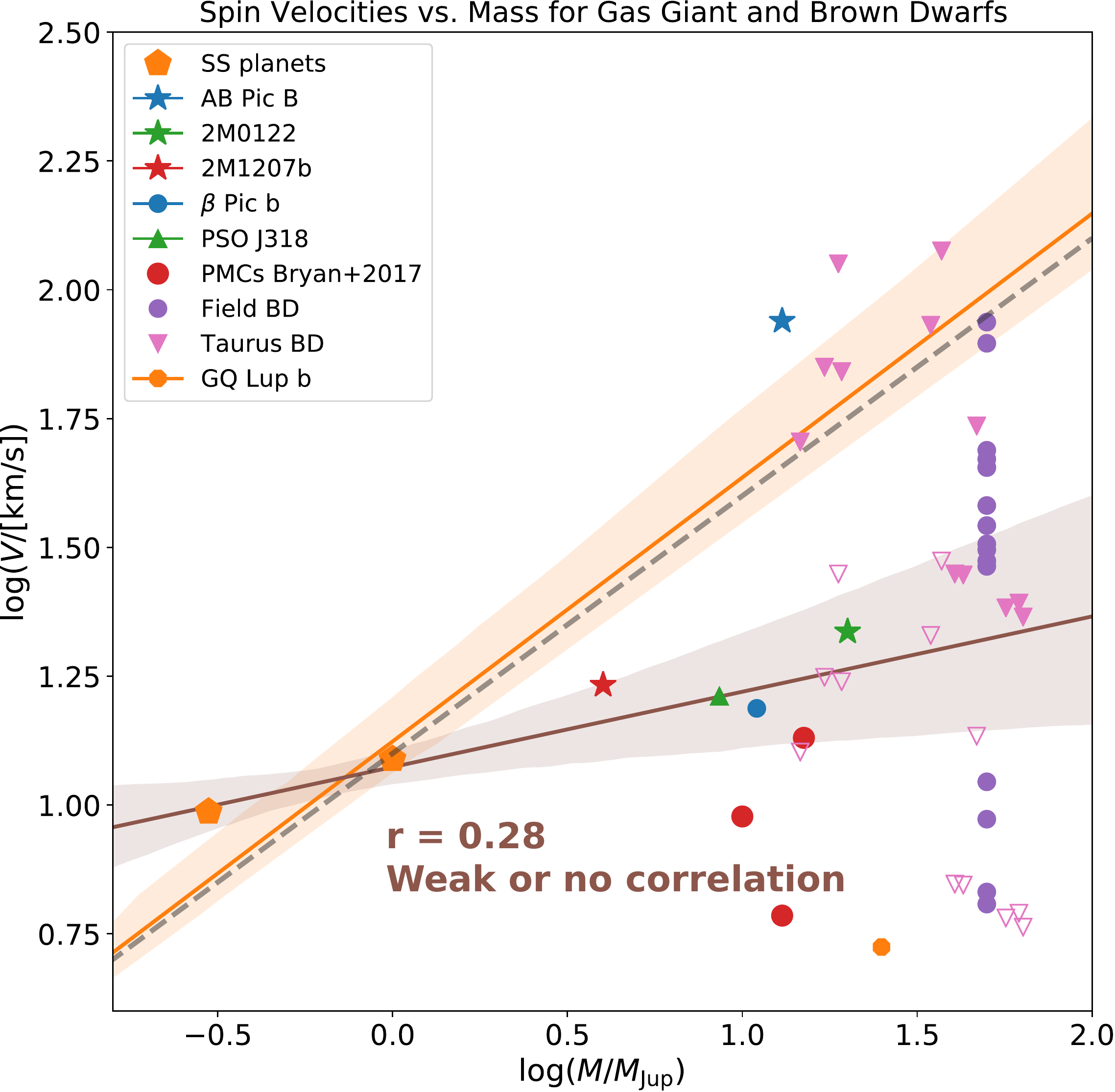}
  \caption{Spin velocity and mass relationship for solar system planets, exoplanets, planetary-mass companions, and brown dwarfs.  The lower panel zooms in for gas giants or more massive substellar objects. \edit1{For the brown dwarfs that are in sample of \citet{Scholz2018}, we plot both the observed spin velocity (pink open triangles) and the derived final spin velocity (solid open triangles) assuming angular momentum conserved contraction.} The orange line and shade are the best-fit linear relation between rotational period and $\log(M)$ for  five solar system planets (Mars, Jupiter, Saturn, Uranus, Neptune) and the $1\sigma$ uncertainty. This line represents the universal spin-mass relation claimed by \citet{Scholz2018}. The brown line and shade are the same as the orange ones but for companions with mass below 20\,\mjup. \edit1{The linear fit that excludes Mars, Uranus and Neptune shows much weaker correlations than the one that includes all the substellar objects.}}
  \label{fig:BDrotation}
\end{figure}

The rotational rates of planetary-mass companions are the results of their angular momentum evolution. planetary-mass companions gain and lose angular momentum primarily through accretion and interaction with the disks. Hints about the formation history of these objects could be revealed by studying their rotational rates. E.g., \citet{Snellen2014,Biller2015,Zhou2016} pointed out a linear trend between mass and rotational rate --- the more massive the planet, the faster it spins. \citet{Scholz2018} found a $v\propto M^{1/2}$ relationship between spin velocity and mass that fits simultaneously for solar system planets, planetary-mass companions, and brown dwarfs and claimed it as a universal relation between rotational rate and mass. This relation requires young planetary-mass companions and brown dwarfs of ages between 1\,Myr to a few tens of Myr to spin up without angular momentum loss to fit in. Contrarily, \citet{Bryan2018} found no correlation between rotation rate and mass for companions and brown dwarfs with mass less than 20\,\mjup.

To investigate the spin-mass relation for planetary-mass companions and brown dwarfs, we collected the entire sample of planetary-mass objects that have rotation measurements available as well as two representative samples for brown dwarfs \citep{Metchev2015, Scholz2018} and plot the spin velocity mass relation in a $\log-\log$ plot (Figure\ref{fig:BDrotation}). We reaffirm the caveat and limitation of our Lomb-Scargle periodogram based measurements and the possibility that the detected quasi-periodic signals are lower limit of the real rotational periods. For following discussion, we only consider the case that these signals are the rotation period. We convert the period measurements to spin velocity for \targeti, \targetii, and \targetiii with radius of 1.5, 1.5 and 1.0 $R_\mathrm{Jup}$, respectively. For the field brown dwarf sample from \citet{Metchev2015}, we use a radius of $1R_\mathrm{Jup}$ for the conversion. For the young brown dwarf sample from \citet{Scholz2018}, we use $4R_\mathrm{Jup}$ as their current radii and $1R_\mathrm{Jup}$ as the final radii and plot spin velocities for both calculations. We note that the $4R_\mathrm{Jup}$ is an estimate from substellar evolution models assuming ``hot start'' formation scenario. Different formation and evolution model may result in smaller radius estimates. But reducing the current radii for the young brown dwarf sample do not change these discussion and conclusion. Following \citet{Scholz2018}, for the final spin velocities we assume conservation of angular momentum during the contraction. For all the spin velocity converted from rotation periods, we assume the spin axis is perpendicular to the line-of-sight.

Figure~\ref{fig:BDrotation} suggests a trend between spin velocity and mass. This trend is tight for solar system planets but becomes obscure for objects of mass over $1M_\mathrm{Jup}$. To demonstrate this transition, we provide two linear fits: one for solar system planets (excluding Mercury, Venus, and the Earth), and the other for giant planetary-mass companions with mass between 0.1 to 20\,\mjup (including Jupiter and Saturn). The first fit is similar to the relation of $\log(v)\propto \frac{1}{2}\log{M}$ found by \citet{Scholz2018}. This relation predicts faster rotation for almost all planetary-mass companions than the observed values except \targeti and requires all the planetary-mass companions have to spin up while conserving angular momentum. However, for field brown dwarfs, many of which have reached the final stage of angular momentum evolution, are also over-predicted by the trend. Therefore, we conclude that the universal trend set by \citet{Scholz2018} sets an upper limit of the rotation rates under angular momentum conserved scenario and the rotation rates of field brown dwarfs suggest they lost angular momentum after the first 1\,Myr of their evolution. \citet{Bouvier2014,Moore2019} have shown the evidence of disk braking and wind braking for brown dwarfs during their evolution.
\edit1{The second linear fit is flatter and has larger uncertainty than than the first one. The Pearson $r$ coefficient for the second linear fit is 0.28, suggesting very weak or no correlation. The lack of strong correlation between spin velocity and mass for gas giants, planetary-mass companions, and brown dwarfs agrees with  the findings of \citet{Bryan2018}.}

\subsection{Wavelength dependence of the modulation signals}
\begin{figure}
  \centering
  \plotone{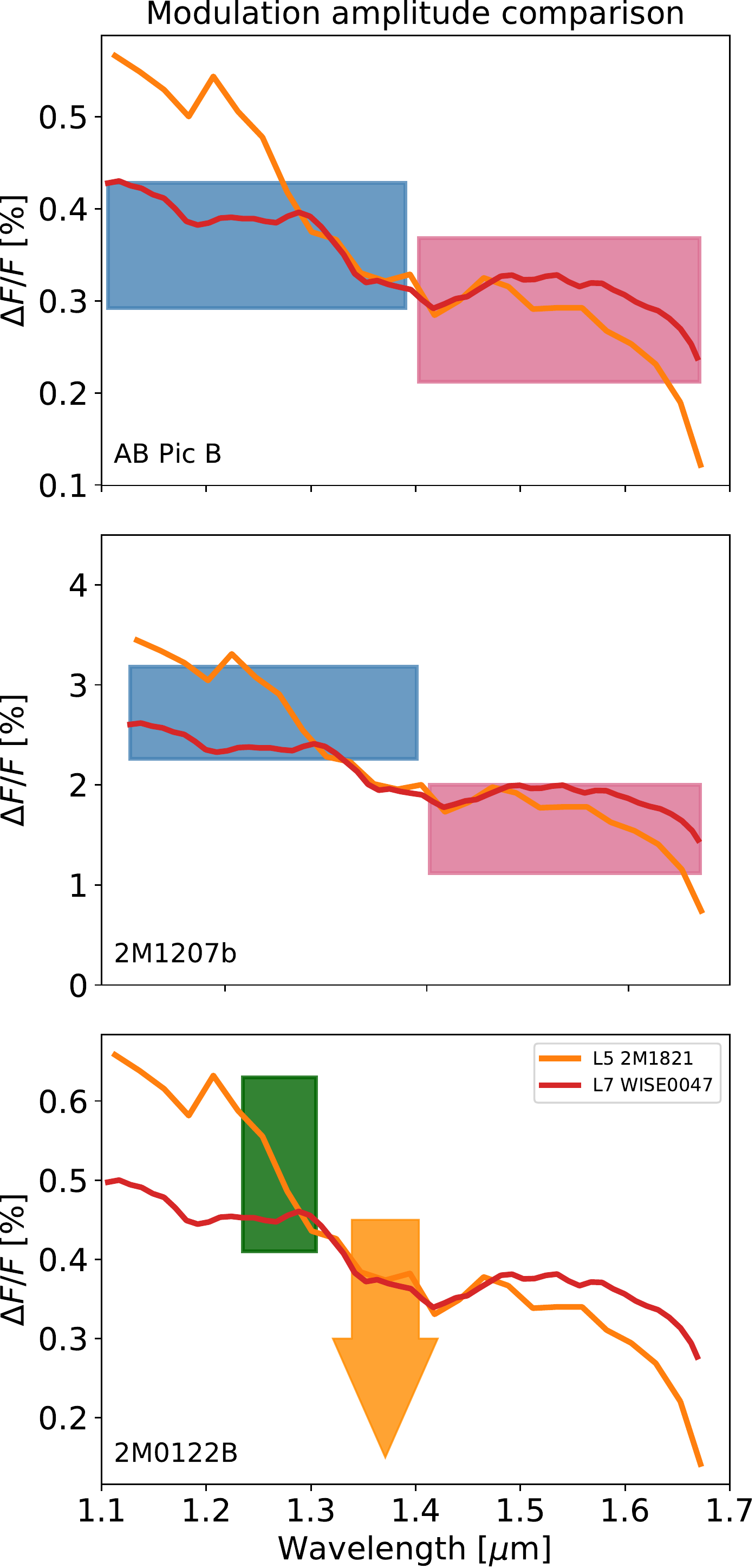}
  \caption{Comparison of spectral modulations among planetary-mass companions/objects. The measurements for 2M1207b are from \citet{Zhou2016}. The heights and widths of the rectangle patches represent modulation amplitude uncertainties and wavelengths, respectively. The arrow in the 2M0122B panel is the $3\sigma$ upper limit for rotational modulation amplitude in F139M (water absorption) band. \edit1{For comparison, We also included two rotationally modulated brown dwarf amplitude-wavelength relation. The brown dwarf modulation amplitude-wavelength relations are scaled to match the modulation amplitudes for the planetary mass companions. At the observed precision, these three planetary-mass companions do not differ from brown dwarfs in the wavelength dependence of their modulations.}}
  \label{fig:ratio_compare}
\end{figure}

\edit1{Spectrally-resolved rotational modulations probe cloud opacity changes as a function of wavelength. Considering the corresponding relation between wavelength and atmospheric pressure level, the wavelength dependence of the modulation essentially probes vertical cloud profiles. Low surface gravity affects cloud structures in at least three aspects  \citep{Marley2012}, all of which may influence the wavelength dependence of the modulation amplitude. (In low surface gravity atmospheres) These three aspects are: 1) Clouds tend to form at lower pressure levels; 2) There is larger amount of condensible material in an atmospheric column above a given pressure level; 3) Condensate dust particles are larger. However, the combined effect on the wavelength dependence of rotational modulation remains unclear. }

\edit1{All three planetary companions studied here belong to a peculiar group of ultra-cool objects that have redder-than-usual near-infrared colors. Thick condensate clouds that are associated with the intermediate to low surface gravity of these objects are often suggested to be the cause of their near-IR colors. They offer a valuable opportunity to examine whether these young, low-surface gravity, planetary-mass companions have the same wavelength dependence of the rotational modulations as their higher surface-gravity counterparts.}

Figure~\ref{fig:ratio_compare} compares the spectrally-resolved modulation amplitudes for the three planetary-mass companions in this study. For comparison, we also include three most precisely measured WFC3/IR modulation amplitude-wavelength curves for brown dwarfs and scale them to the match the average amplitudes for each planetary-mass companion. For \targeti and \targetiii which have modulation amplitudes measured in the wide-pass filters, we can compare overall amplitude-wavelength slope with those for the brown dwarfs. Both objects have larger amplitude modulations in the bluer F125W band than the F160W band. \targetiii shows steeper slope and agrees better with brown dwarf 2M1821. \targeti has weaker slope and agrees better with brown dwarf WISE0047. For \targetii, for which the modulations are measured in mid-pass filters, we explore the modulation amplitude in and out of 1.4\,\micron{} water absorption band. Due to the limits of the observational precision, \edit1{the upper limit on the water  modulation amplitude} is not strong enough to reject the hypothesis that 2M0122B's rotational modulations have the same wavelength dependence as brown dwarfs with the same temperatures at $3\mbox{-}\sigma$ level. \edit1{In summary, with our observational precision, the wavelength dependence of the rotational modulations for our three targets are not distinguishable from those of selected brown dwarfs.}


\subsection{Hybrid PSF Photometry for Space Telescopes}

\begin{figure*}[t]
  \centering
  \includegraphics[width=0.32\textwidth]{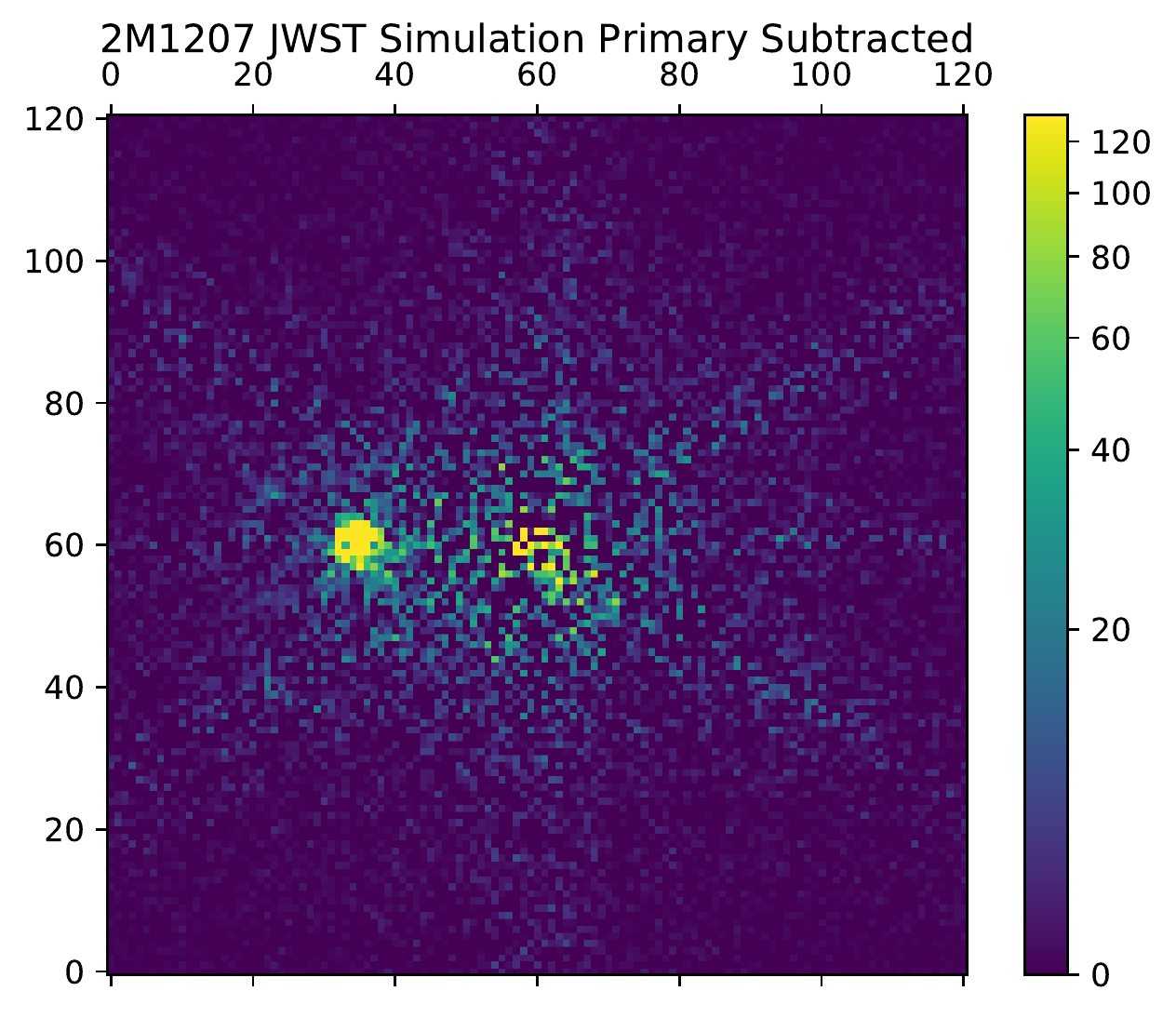}
  \includegraphics[width=0.32\textwidth]{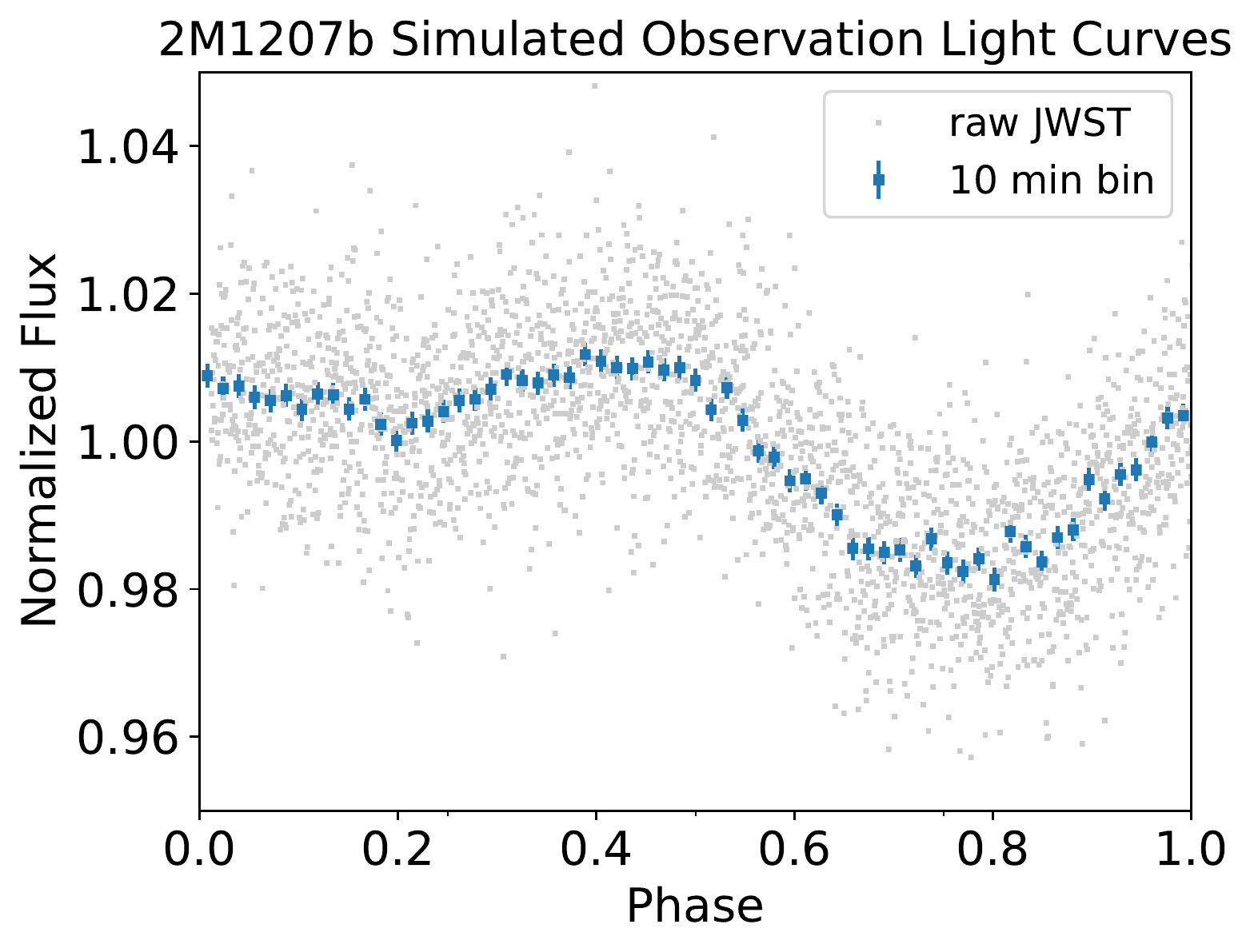}
  \includegraphics[width=0.32\textwidth]{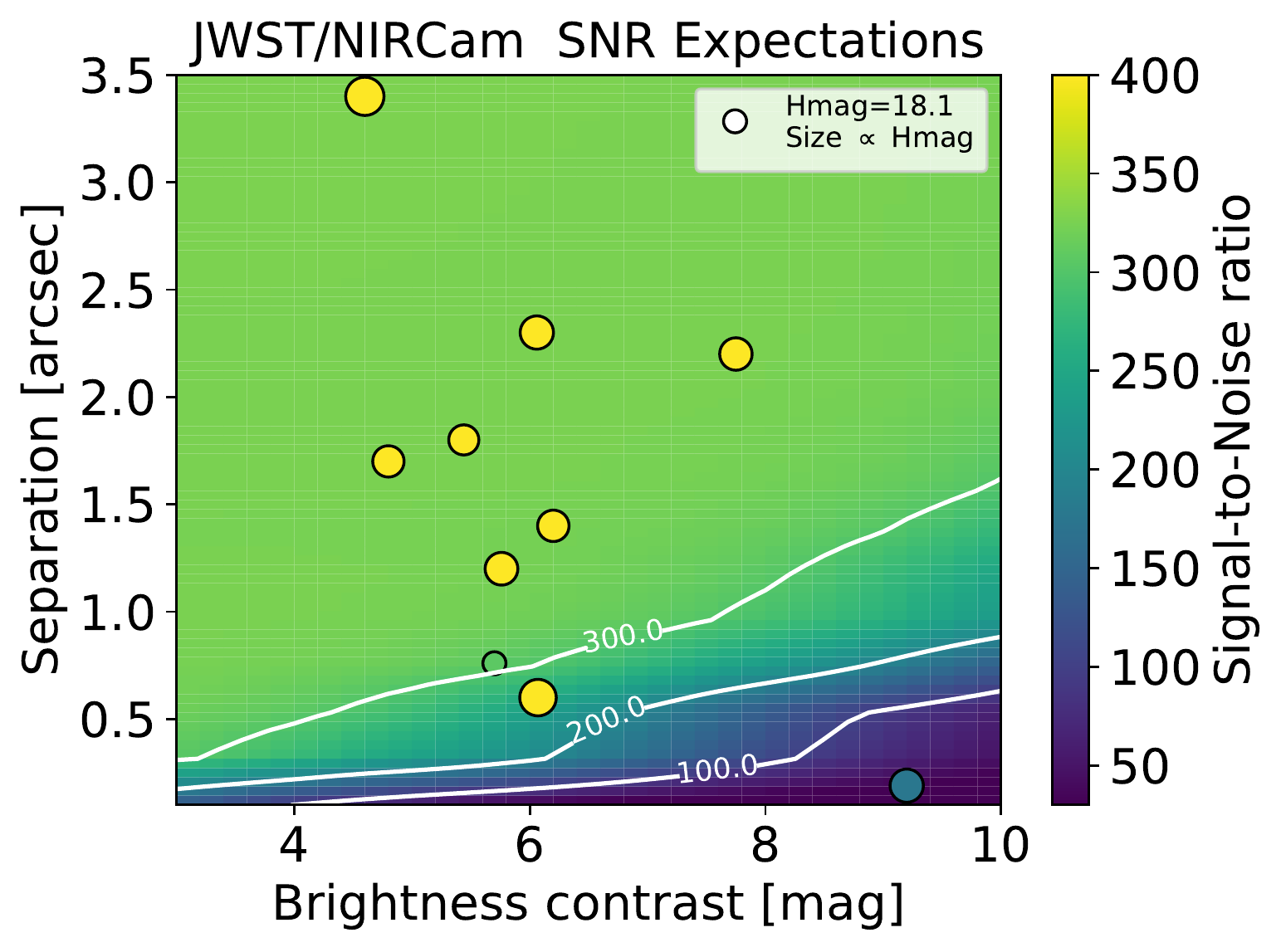}
  \caption{Simulated JWST/NIRCam observation results and expected SNR. \label{fig:NIRCam-Image}. \emph{Left:} simulated primary subtracted image for 2M1207b observed by JWST/NIRCam. \emph{Middle:} The light curve for 2M1207b observed by JWST/NIRCam $1.2\times$ photon noise uncertainties and $\sim3\%$ peak-to-peak modulation amplitude. \emph{Right:} The light curve precisions for planetary-mass companions observed by JWST/NIRCam as a function of their brightness, contrast, and angular separations to their hosts. The sizes and the colors of the circles represent the brightness and estimated SNR for the planetary-mass companions.}
\end{figure*}

We used a hybrid PSF photometry method to achieve better than $1\%$ precision photometry for planetary-mass companions that have contrast ratios of $200\mbox{-}500$ and angular separations of $0.7\arcsec\mbox{-}1.4\arcsec$. For time-resolved high-contrast photometry time-series, the primary noise source additional to the photon noise is from the variable contamination of the host star PSFs. The stability of the PSF, which is unique to HST/WFC3 observations, allows precise characterization of the systematics in the PSF models. The median variance in the subtraction residual images is more than one order of magnitude smaller than the one without PSF corrections. In the hybrid PSF photometry fit, the scale of the PSF correction term is a free parameter and have a standard deviation of only $\sim0.1\%$ in the best-fit results. The PSF correction terms captures $\sim90\%$ variance among all the frames and stay stable across the whole time series. This precise characterization of the PSF eliminates the contamination to the companion photometry by the primary star PSF. 

Telescope focus change is the primary sources of systematics that compromise the stability of the PSF. The focus of  HST is affected by the thermal state of the telescope and modulates in the same period as the HST orbital period of 96 minutes. We refer proximity in HST orbital phase to group the images when we calculate the PSF correction term to reconcile the focus change problem. Comparing to the corrections derived using the entire image set regardless of the telescope focus status, those calculated for groups of images with similar telescope focus have smaller and less variable  PSF fitting residuals.

Telescope pointing drift can also downgrade the hybrid PSF photometry precision. This effect is exacerbated by the under-sampled WFC3 IR detector. HST maintained stable and precise pointing for the observations of AB Pic B and 2M0122B. The pointing difference was less than 0.1 pixel for frames in the same orbits and less than 0.2 pixel for frames in different orbits. However, the observation of 2M1207b suffered from large intra-orbit pointing variance. The median pointing position of the second orbit of the observation was off by more than one pixel compared with the that for the entire observations. The less than ideal pointing performance prevented us to draw stronger conclusions from the 2M1207b observations.

For ground-based observations, when the PSF is strongly affected by atmospheric turbulence and adaptive optics correction, hybrid PSF photometry is not an ideal strategy for precision photometry. For comparison, in a VLT/SPHERE high-contrast imaging time-series of the \object{HR8799} system, the primary component in the principal component analysis of the host star PSF explains 47\%  of the variance among PSFs \citep{Apai2016}. Therefore, modeling PSF with a combination of PSF and correction term will leave more than 50\% of the residual unexplained and result in poor photometry precision. Satellite-spot-based calibration and multi-planet self-calibration proved to be better strategies for these observations\citep{Apai2016}.

We note that the adopted hybrid PSF photometry deals with images from a rather simple optics system. Future development of this method in systems with spectrographs and coronagraphs will extend its scientific applications.

\subsection{Future studies with HST and JWST}

Time-resolved observations of exoplanets and brown dwarfs will be even more powerful in the JWST era \citep{Kostov2013}. JWST will have more than seven times larger collecting area than HST and  advanced instruments that are designed with  time-resolved and high-contrast imaging observations as priorities. The observing strategy and data reduction techniques we discussed previously in this paper will be applicable to JWST/NIRcam direct-imaging time-series observations. In this mode, the incoming light is split into long wavelength and short wavelength channels for simultaneous two-channel observations.  We discuss strategies, expected outcome, and applications of observations in this mode in the following.

We use the JWST PSF simulator \texttt{webbpsf} and JWST exposure time calculator (ETC) to estimate the high-contrast direct imaging time-series capability of JWST/NIRCam.
The approach that we detail in the following are based on the assumption that the hybrid PSF method achieves the same performance for JWST observations as in HST/WFC3 observations.
We first simulate a scene composed of the primary and companion model PSFs, for which filter choice, separation, and position angle are the input parameters.
The amplitudes of the model PSFs are determined by the brightness and contrast of the simulated system.
We then estimate the SNR of the companion, in which we conservatively assume the noise to be 20\% larger than the photon noise (considering the flux contributions from both the companions and the primary PSFs). For the 2M1207 system, to obtain an SNR of 100 for the companion with JWST F150W filter (central wavelength $1.5\,\micron$), a 15s integration is required, which is about 1/6 of the integration time required for HST/WFC3 to reach the same SNR. This time series can be realized by 15 integrations of two group rapid reads. Figure \ref{fig:NIRCam-Image} shows the expected primary subtracted image and the resulting light curves for one rotation with a planetary wave model light curve \citep{Apai2017} with arbitrary parameters.

The right panel of Figure~\ref{fig:NIRCam-Image} shows the expected SNR of the NIRCam  light curves as a function of angular separation and brightness contrast, assuming a cadence of 100 s and H band brightness of 18.1 magnitude (same as \targetiii). As long as the separation is above 0.5\arcsec, we expect the SNR better than 200. We also calculated the expected SNR for a few representative planetary-mass companions and over plot them on Figure~\ref{fig:NIRCam-Image}. Most of them are brighter than \targetiii, therefore have higher SNRs of greater than 1000. Such precision is comparable to the brown dwarf light curves that resulted in spatially resolved maps \citep[e.g.,][]{Apai2013,Karalidi2015}, we thus expect to retrieve atmosphere maps for planetary-mass companions with JWST.

JWST/NIRSpec's spectroscopic time series mode will also be important for studies of the clouds of directly imaged exoplanets. In this mode, the image of a close companion can fit in the $1\arcsec.6\times1\arcsec.6$ slit together with the hosts. To model and subtract the spectroscopic PSF is crucial for the success in such observations. We note that the NIRCam observations do not perform well with small separation ($<5\arcsec$) or bright host star (due to saturation). Time series with coronagraph or ground-based extreme AO system \citep[e.g.,][]{Apai2016} have the potential to significantly improve the performance in this parameter space.

\section{Summary}
\begin{enumerate}
\item We obtained high-contrast time-resolved observations for three planetary-mass companions, \targeti, \targetii, and \targetiii using HST/WFC3 IR direct-imaging mode. For each target, the observations resulted in light curves in two bands, with temporal baselines of 8.5 hr for \targeti and \targetii, and 28.9 hr for \targetiii.  Using two roll differential imaging and hybrid PSF Photometry technique \citep{Zhou2016}, we achieved sub-percent level precision photometry for the \targeti and \targetii. The standard deviations of the light curves are within 20\% of the photon noise limit. For \targetiii, the precision of the final light curves is limited by instrumental systematics that is primarily related to the pointing drift of the telescope.

\item We marginally detected periodic modulations in the light curves of \targeti in both F125W and F160W filters with consistent period and phase in the two bands. We also marginally detected modulation signals in the F127M light curve of \targetii. The detected signals in \targeti's light curves have a period of 2.1 hr and amplitudes of 0.18\% and 0.14\% in F125W and F160W, respectively. The detections in \targetii's F127M light curve corresponds a period of 6.0 hr and an amplitude of 0.52\%. However, hybrid PSF photometry on simulated data that have injected simulated companions suggests that such low significance periodic signals in the Lomb-Scargle periodograms may be false positives. Therefore, with the current data we identify the detection of the modulations in 2M0122B as tentative. 

\item We compare the condensate cloud properties of these three planetary-mass companions and those of the field brown dwarfs on the basis of  the wavelength dependence of the modulation amplitudes. We do not rule out  that the spectral type matched planetary-mass objects and brown dwarfs have the same vertical cloud structures.

\item We found no additional planet candidates in any of the three systems using two-roll angular differential imaging results. We ruled out any companions that are more massive than 5\mjup and 50 au away from the primary star in all three systems.

\item The rotational velocities for gas giant or more massive substellar objects do not show significant correlations with their mass. The lack of correlation argues against a claim of \citet{Scholz2018} stating that terrestrial planets, ice giant planets, gas giant planets, and brown dwarfs share a universal spin-mass relation..

 \item The hybrid PSF photometry technique will achieve more than one order of magnitude better precision with JWST/NIRCam observations for known planetary-mass companions. Future observations will allow detailed directly-imaged exoplanet cloud maps that tightly constrain the exoplanet cloud and atmospheric circulation models.
 \end{enumerate}

\software{Numpy\&Scipy \citep{VanderWalt2011}, Matplotlib
  \citep{Hunter2007}, IPython \citep{Perez2007}, Astropy
  \citep{Robitaille2013}, Seaborn \citep{Waskom2017}}

 \acknowledgments We thank the referee for a constructive report. We thank Dr. Kaitlin Kratter for pointing out an inadequacy in an earlier version of the manuscript and Dr. Aleks Scholz for useful comments. Y.Z. acknowledges support in part by the NASA Earth and Space Science Fellowship Program - Grant “NNX16AP54H”. D.A. acknowledges support by NASA under agreement No. NNX15AD94G for the program Earths in Other Solar Systems. Support for Program number 14241 was provided by NASA through a grant from the Space Telescope Science Institute, which is operated by the Association of Universities for Research in Astronomy, Incorporated, under NASA contract NAS5-26555. Based on observations made with the NASA/ESA Hubble Space Telescope, obtained in GO program 14241 at the Space Telescope Science Institute.

 \bibliographystyle{yahapj}

\begin{thebibliography}{}
\providecommand\natexlab[1]{#1}
\providecommand\JournalTitle[1]{#1}

\bibitem[{Ackerman \& Marley(2001)}]{Ackerman2001}
Ackerman, A.~S., \& Marley, M.~S. 2001,
  \href{http://dx.doi.org/10.1086/321540}{\JournalTitle{ApJ}, 556, 872}

\bibitem[{Apai {et~al.}(2013)Apai, Radigan, Buenzli, Burrows, Reid, \&
  Jayawardhana}]{Apai2013}
Apai, D., Radigan, J., Buenzli, E., {et~al.} 2013,
  \href{http://dx.doi.org/10.1088/0004-637X/768/2/121}{\JournalTitle{ApJ}, 768,
  121}

\bibitem[{Apai {et~al.}(2016)Apai, Kasper, Skemer, Hanson, Lagrange, Biller,
  Bonnefoy, Buenzli, \& Vigan}]{Apai2016}
Apai, D., Kasper, M., Skemer, A., {et~al.} 2016,
  \href{http://dx.doi.org/10.3847/0004-637X/820/1/40}{\JournalTitle{ApJ}, 820,
  40}

\bibitem[{Apai {et~al.}(2017)Apai, Karalidi, Marley, Yang, Flateau, Metchev,
  Cowan, Buenzli, Burgasser, Radigan, Artigau, \& Lowrance}]{Apai2017}
Apai, D., Karalidi, T., Marley, M.~S., {et~al.} 2017,
  \href{http://dx.doi.org/10.1126/science.aam9848}{\JournalTitle{Science}, 357,
  683}

\bibitem[{Artigau {et~al.}(2009)Artigau, Bouchard, Doyon, \&
  Lafreni{\`{e}}re}]{Artigau2009}
Artigau, {\'{E}}., Bouchard, S., Doyon, R., \& Lafreni{\`{e}}re, D. 2009,
  \href{http://dx.doi.org/10.1088/0004-637X/701/2/1534}{\JournalTitle{ApJ},
  701, 1534}

\bibitem[{Barman {et~al.}(2015)Barman, Konopacky, Macintosh, \&
  Marois}]{Barman2015}
Barman, T.~S., Konopacky, Q.~M., Macintosh, B., \& Marois, C. 2015,
  \JournalTitle{The Astrophysical Journal}, 804, 61

\bibitem[{Barman {et~al.}(2011)Barman, Macintosh, Konopacky, \&
  Marois}]{Barman2011}
Barman, T.~S., Macintosh, B., Konopacky, Q.~M., \& Marois, C. 2011,
  \href{http://dx.doi.org/10.1088/2041-8205/735/2/L39}{\JournalTitle{ApJ}, 735,
  L39}

\bibitem[{Berta-Thompson {et~al.}(2012)Berta-Thompson, Charbonneau,
  D{\'{e}}sert, {Miller-Ricci Kempton}, McCullough, Burke, Fortney, Irwin,
  Nutzman, \& Homeier}]{Berta2012}
Berta-Thompson, Z., Charbonneau, D., D{\'{e}}sert, J.-M., {et~al.} 2012,
  \href{http://dx.doi.org/10.1088/0004-637X/747/1/35}{\JournalTitle{ApJ}, 747,
  35}

\bibitem[{Beuzit {et~al.}(2008)Beuzit, Feldt, Dohlen, Mouillet, Puget, Wildi,
  Abe, Antichi, Baruffolo, Baudoz, {et~al.}}]{Beuzit2008}
Beuzit, J.-L., Feldt, M., Dohlen, K., {et~al.} 2008, in Ground-based and
  airborne instrumentation for astronomy II, Vol. 7014, International Society
  for Optics and Photonics, 701418

\bibitem[{Biller {et~al.}(2015)Biller, Vos, Bonavita, Buenzli, Baxter,
  Crossfield, Allers, Liu, Bonnefoy, Deacon, Brandner, Schlieder, Dupuy,
  Kopytova, Manjavacas, Allard, Homeier, \& Henning}]{Biller2015}
Biller, B.~A., Vos, J.~M., Bonavita, M., {et~al.} 2015,
  \href{http://dx.doi.org/10.1088/2041-8205/813/2/L23}{\JournalTitle{ApJ}, 813,
  L23}

\bibitem[{Biller {et~al.}(2018)Biller, Vos, Buenzli, Allers, Bonnefoy, Charnay,
  B{\'{e}}zard, Allard, Homeier, Bonavita, Brandner, Crossfield, Dupuy,
  Henning, Kopytova, Liu, Manjavacas, \& Schlieder}]{Biller2017}
Biller, B.~A., Vos, J., Buenzli, E., {et~al.} 2018,
  \href{http://dx.doi.org/10.3847/1538-3881/aaa5a6}{\JournalTitle{AJ}, 155, 95}

\bibitem[{Bonnefoy {et~al.}(2010)Bonnefoy, Chauvin, Rojo, Allard, Lagrange,
  Homeier, Dumas, \& Beuzit}]{Bonnefoy2010}
Bonnefoy, M., Chauvin, G., Rojo, P., {et~al.} 2010,
  \href{http://dx.doi.org/10.1051/0004-6361/200912688}{\JournalTitle{A{\&}A},
  512}
\bibitem[Bouvier et al.(2014)]{Bouvier2014} Bouvier, J., Matt, S.~P., Mohanty, S., et al.\ 2014, Protostars and Planets VI, 433

  
\bibitem[{Bowler(2016)}]{Bowler2016}
Bowler, B.~P. 2016,
  \href{http://dx.doi.org/10.1088/1538-3873/128/968/102001}{\JournalTitle{PASP},
  128}

\bibitem[{Bowler {et~al.}(2013)Bowler, Liu, Shkolnik, \& Dupuy}]{Bowler2013}
Bowler, B.~P., Liu, M.~C., Shkolnik, E.~L., \& Dupuy, T.~J. 2013,
  \href{http://dx.doi.org/10.1088/0004-637X/774/1/55}{\JournalTitle{ApJ}, 774,
  55}

\bibitem[{Bryan {et~al.}(2018)Bryan, Benneke, Knutson, Batygin, \&
  Bowler}]{Bryan2018}
Bryan, M.~L., Benneke, B., Knutson, H.~A., Batygin, K., \& Bowler, B.~P. 2018,
  \href{http://dx.doi.org/10.1038/s41550-017-0325-8}{\JournalTitle{Nat.
  Astron.}, 2, 138}

\bibitem[{Buenzli {et~al.}(2014)Buenzli, Apai, Radigan, Reid, \&
  Flateau}]{Buenzli2014}
Buenzli, E., Apai, D., Radigan, J., Reid, I.~N., \& Flateau, D. 2014,
  \href{http://dx.doi.org/10.1088/0004-637X/782/2/77}{\JournalTitle{ApJ}, 782,
  77}

\bibitem[{Buenzli {et~al.}(2012)Buenzli, Apai, Morley, Flateau, Showman,
  Burrows, Marley, Lewis, \& Reid}]{Buenzli2012}
Buenzli, E., Apai, D., Morley, C.~V., {et~al.} 2012,
  \href{http://dx.doi.org/10.1088/2041-8205/760/2/L31}{\JournalTitle{ApJ}, 760,
  L31}

\bibitem[{Burrows {et~al.}(2006)Burrows, Sudarsky, \& Hubeny}]{Burrows2006a}
Burrows, A.~S., Sudarsky, D., \& Hubeny, I. 2006,
  \href{http://dx.doi.org/10.1086/500293}{\JournalTitle{ApJ}, 640, 1063}

\bibitem[{{Chandrasekhar}(1933)}]{Chandrasekhar1933}
{Chandrasekhar}, S. 1933,
  \href{http://dx.doi.org/10.1093/mnras/93.8.539}{\JournalTitle{\mnras}, 93,
  539}


\bibitem[Charnay et al.(2018)]{Charnay2018} Charnay, B., B{\'e}zard, B., Baudino, J.-L., et al.\ 2018, \apj, 854, 172

\bibitem[{Chauvin {et~al.}(2004)Chauvin, Lagrange, Dumas, Zuckerman, Mouillet,
  Song, Beuzit, \& Lowrance}]{Chauvin2004}
Chauvin, G., Lagrange, A.-M., Dumas, C., {et~al.} 2004,
  \href{http://dx.doi.org/10.1051/0004-6361:200400056}{\JournalTitle{A{\&}A},
  425, 29}

\bibitem[{Chauvin {et~al.}(2005)Chauvin, Lagrange, Zuckerman, Dumas, Mouillet,
  Song, Beuzit, Lowrance, \& Bessell}]{Chauvin2005}
Chauvin, G., Lagrange, A.-M., Zuckerman, B., {et~al.} 2005,
  \href{http://dx.doi.org/10.1051/0004-6361:200500111}{\JournalTitle{A{\&}A},
  438, L29}

\bibitem[{Cowan \& Agol(2008)}]{Cowan2008}
Cowan, N.~B., \& Agol, E. 2008,
  \href{http://iopscience.iop.org/article/10.1086/588553/pdf}{\JournalTitle{ApJ},
  678, 129}

\bibitem[{Deustua {et~al.}(2016)Deustua, Baggett, Brammer, Bourque, Bowers,
  Dahlen, Dulude, Deustua, Dressel, Gosmeyer, Gunning, Hammer, Hilbert,
  Khozurina-Platais, Lee, Long, Mack, MacKenty, McCullough, Noeske, Pirzkal,
  Sahu, Sosey, Riess, Sabbi, Taylor, Viana, Deustia, Rajan, {Kim Quijano}, \&
  Bushouse}]{Deustua2016}
Deustua, S., Baggett, S.~M., Brammer, G., {et~al.} 2016,
  \href{http://www.stsci.edu/hst/wfc3}{\JournalTitle{(Baltimore:STScI)}}

\bibitem[{Dressel(2018)}]{Dressel2018}
Dressel, L. 2018, {Wide field camera 3 instrument handbook v 10.0} ((Baltimore:
  STScI))

\bibitem[{Foreman-Mackey {et~al.}(2013)Foreman-Mackey, Hogg, Lang, \&
  Goodman}]{Foreman-Mackey2012}
Foreman-Mackey, D., Hogg, D.~W., Lang, D., \& Goodman, J. 2013,
  \href{http://dx.doi.org/10.1086/670067}{\JournalTitle{PASP}, 125, 306}

\bibitem[{Gagn{\'{e}} {et~al.}(2018)Gagn{\'{e}}, Mamajek, Malo, Riedel,
  Rodriguez, Lafreni{\`{e}}re, Faherty, Roy-Loubier, Pueyo, Robin, \&
  Doyon}]{Gagne2018}
Gagn{\'{e}}, J., Mamajek, E.~E., Malo, L., {et~al.} 2018,
  \href{http://dx.doi.org/10.3847/1538-4357/aaae09}{\JournalTitle{ApJ}, 856,
  23}

\bibitem[{{Gaia Collaboration} {et~al.}(2016)}]{Gaia2016}
{Gaia Collaboration}, {et~al.} 2016,
  \href{http://dx.doi.org/10.1051/0004-6361/201629512}{\JournalTitle{A{\&}A},
  595, A2}

\bibitem[{{Gaia Collaboration} {et~al.}(2018)}]{Gaia2018}
---. 2018,
  \href{http://dx.doi.org/10.1051/0004-6361/201833051}{\JournalTitle{A{\&}A},
  616, A1}

\bibitem[{Ginsburg {et~al.}(2014)Ginsburg, andrew giessel, \&
  Chef}]{Ginsburg2014}
Ginsburg, A., andrew giessel, \& Chef, B. 2014, image\_registration v0.2.1

\bibitem[{Helling {et~al.}(2008)Helling, Ackerman, Allard, Dehn, Hauschildt,
  Homeier, Lodders, Marley, Rietmeijer, Tsuji, \& Woitke}]{Helling2008}
Helling, C., Ackerman, A.~S., Allard, F., {et~al.} 2008,
  \href{http://dx.doi.org/10.1111/j.1365-2966.2008.13991.x}{\JournalTitle{MNRAS},
  391, 1854}

\bibitem[{Hoeijmakers {et~al.}(2018)Hoeijmakers, Schwarz, Snellen, de~Kok,
  Bonnefoy, Chauvin, Lagrange, \& Girard}]{Hoeijmakers2018}
Hoeijmakers, H., Schwarz, H., Snellen, I., {et~al.} 2018, \JournalTitle{arXiv
  preprint arXiv:1802.09721}

\bibitem[{Hunter(2007)}]{Hunter2007}
Hunter, J.~D. 2007,
  \href{http://dx.doi.org/10.1109/MCSE.2007.55}{\JournalTitle{Comput. Sci.
  Eng.}, 9, 90}

\bibitem[{Ingraham {et~al.}(2014)Ingraham, Marley, Saumon, Marois, Macintosh,
  Barman, Bauman, Burrows, Chilcote, {De Rosa}, \& Others}]{Ingraham2014}
Ingraham, P., Marley, M.~S., Saumon, D., {et~al.} 2014, \JournalTitle{ApJL},
  794, L15

\bibitem[{Karalidi {et~al.}(2015)Karalidi, Apai, Schneider, Hanson, \&
  Pasachoff}]{Karalidi2015}
Karalidi, T., Apai, D., Schneider, G., Hanson, J.~R., \& Pasachoff, J.~M. 2015,
  \href{http://dx.doi.org/10.1088/0004-637X/814/1/65}{\JournalTitle{ApJ}, 814,
  65}

\bibitem[{Konopacky {et~al.}(2013)Konopacky, Barman, Macintosh, \&
  Marois}]{Konopacky2013}
Konopacky, Q.~M., Barman, T.~S., Macintosh, B.~A., \& Marois, C. 2013,
  \href{http://dx.doi.org/10.1126/science.1232003}{\JournalTitle{Science}, 339,
  1398}

\bibitem[{Kostov \& Apai(2013)}]{Kostov2013}
Kostov, V., \& Apai, D. 2013,
  \href{http://dx.doi.org/10.1088/0004-637X/762/1/47}{\JournalTitle{ApJ}, 762,
  47}

\bibitem[{Krist {et~al.}(2011)Krist, Hook, \& Stoehr}]{Krist2011}
Krist, J.~E., Hook, R.~N., \& Stoehr, F. 2011, in Optical Modeling and
  Performance Predictions V, Vol. 8127, International Society for Optics and
  Photonics, 81270J

\bibitem[{{Lafreni{\`e}re} {et~al.}(2007){Lafreni{\`e}re}, {Marois}, {Doyon},
  {Nadeau}, \& {Artigau}}]{Lafreniere2007}
{Lafreni{\`e}re}, D., {Marois}, C., {Doyon}, R., {Nadeau}, D., \& {Artigau},
  {\'E}. 2007, \href{http://dx.doi.org/10.1086/513180}{\JournalTitle{\apj},
  660, 770}


\bibitem[Lagrange et al.(2010)]{Lagrange2010} Lagrange, A.-M., Bonnefoy, M., Chauvin, G., et al.\ 2010, Science, 329, 57 

\bibitem[{Lew {et~al.}(2016)Lew, Apai, Zhou, Schneider, Burgasser, Karalidi,
  Yang, Marley, Cowan, Bedin, Metchev, Radigan, \& Lowrance}]{Lew2016}
Lew, B. W.~P., Apai, D., Zhou, Y., {et~al.} 2016,
  \href{http://dx.doi.org/10.3847/2041-8205/829/2/L32}{\JournalTitle{ApJ}, 829,
  L32}

\bibitem[{Lomb(1976)}]{Lomb1976}
Lomb, N.~R. 1976,
  \href{http://dx.doi.org/10.1007/BF00648343}{\JournalTitle{Astrophys. Space
  Sci.}, 39, 447}

\bibitem[{Macintosh {et~al.}(2015)Macintosh, Graham, Barman, {De Rosa},
  Konopacky, Marley, Marois, Nielsen, Pueyo, Rajan, Rameau, Saumon, Wang,
  Patience, Ammons, Arriaga, Artigau, Beckwith, Brewster, Bruzzone, Bulger,
  Burningham, Burrows, Chen, Chiang, Chilcote, Dawson, Dong, Doyon, Draper,
  Duchene, Esposito, Fabrycky, Fitzgerald, Follette, Fortney, Gerard, Goodsell,
  Greenbaum, Hibon, Hinkley, Cotten, Hung, Ingraham, Johnson-Groh, Kalas,
  Lafreniere, Larkin, Lee, Line, Long, Maire, Marchis, Matthews, Max, Metchev,
  Millar-Blanchaer, Mittal, Morley, Morzinski, Murray-Clay, Oppenheimer,
  Palmer, Patel, Perrin, Poyneer, Rafikov, Rantakyro, Rice, Rojo, Rudy, Ruffio,
  Ruiz, Sadakuni, Saddlemyer, Salama, Savransky, Schneider, Sivaramakrishnan,
  Song, Soummer, Thomas, Vasisht, Wallace, Ward-Duong, Wiktorowicz, Wolff, \&
  Zuckerman}]{Macintosh2015a}
Macintosh, B., Graham, J.~R., Barman, T., {et~al.} 2015,
  \href{http://dx.doi.org/10.1126/science.aac5891}{\JournalTitle{Science}, 350,
  64}

\bibitem[{Macintosh {et~al.}(2008)Macintosh, Graham, Palmer, Doyon, Dunn,
  Gavel, Larkin, Oppenheimer, Saddlemyer, Sivaramakrishnan,
  {et~al.}}]{Macintosh2008}
Macintosh, B.~A., Graham, J.~R., Palmer, D.~W., {et~al.} 2008, in Adaptive
  Optics Systems, Vol. 7015, International Society for Optics and Photonics,
  701518

\bibitem[{Mandell {et~al.}(2013)Mandell, Haynes, Sinukoff, Madhusudhan,
  Burrows, \& Deming}]{Mandell2013}
Mandell, A.~M., Haynes, K., Sinukoff, E., {et~al.} 2013,
  \href{http://dx.doi.org/10.1088/0004-637X/779/2/128}{\JournalTitle{ApJ}, 779,
  128}

\bibitem[{Manjavacas {et~al.}(2017)Manjavacas, Apai, Zhou, Karalidi, Lew,
  Schneider, Cowan, Metchev, Miles-P{\'{a}}ez, Burgasser, Radigan, Bedin,
  Lowrance, \& Marley}]{Manjavacas2017}
Manjavacas, E., Apai, D., Zhou, Y., {et~al.} 2017,
\href{http://dx.doi.org/10.3847/1538-3881/aa984f}{\JournalTitle{AJ}, 155, 11}

\bibitem[Manjavacas et al.(2019)]{Manjavacas2019} Manjavacas, E., Apai, D., Zhou, Y., et al.\ 2019, \aj, 157, 101 

\bibitem[{Marley {et~al.}(2012)Marley, Saumon, Cushing, Ackerman, Fortney, \&
  Freedman}]{Marley2012}
Marley, M.~S., Saumon, D., Cushing, M., {et~al.} 2012, \JournalTitle{The
  Astrophysical Journal}, 754, 135

\bibitem[{Marley {et~al.}(2002)Marley, Seager, Saumon, Lodders, Ackerman,
  Freedman, \& Fan}]{Marley2002}
Marley, M.~S., Seager, S., Saumon, D., {et~al.} 2002, \JournalTitle{ApJ}, 568,
  335

\bibitem[{Marley \& Sengupta(2011)}]{Marley2011}
Marley, M.~S., \& Sengupta, S. 2011,
  \href{http://dx.doi.org/10.1111/j.1365-2966.2011.19448.x}{\JournalTitle{MNRAS},
  417, 2874}

\bibitem[{Marois {et~al.}(2008)Marois, Macintosh, Barman, Zuckerman, Song,
  Patience, Lafreni{\`e}re, \& Doyon}]{Marois2008}
Marois, C., Macintosh, B., Barman, T., {et~al.} 2008, \JournalTitle{Science},
  322, 1348

\bibitem[{Metchev {et~al.}(2015)Metchev, Heinze, Apai, Flateau, Radigan,
  Burgasser, Marley, Artigau, Plavchan, \& Goldman}]{Metchev2015}
Metchev, S.~A., Heinze, A., Apai, D., {et~al.} 2015,
  \href{http://dx.doi.org/10.1088/0004-637X/799/2/154}{\JournalTitle{ApJ}, 799,
  154}

\bibitem[Moore et al.(2019)]{Moore2019} Moore, K., Scholz, A., \& Jayawardhana, R.\ 2019, arXiv:1901.05523

  
\bibitem[{Morley {et~al.}(2012)Morley, Fortney, Marley, Visscher, Saumon, \&
  Leggett}]{Morley2012}
Morley, C.~V., Fortney, J.~J., Marley, M.~S., {et~al.} 2012,
  \href{http://dx.doi.org/10.1088/0004-637X/756/2/172}{\JournalTitle{ApJ}, 756,
    172}
  

\bibitem[{Patience {et~al.}(2010)Patience, King, {De Rosa}, \&
  Marois}]{Patience2010}
Patience, J., King, R.~R., {De Rosa}, R.~J., \& Marois, C. 2010,
  \href{http://dx.doi.org/10.1051/0004-6361/201014173}{\JournalTitle{A{\&}A},
  517, A76}

\bibitem[{Patience {et~al.}(2012)Patience, King, Rosa, Vigan, Witte, \&
  Rice}]{Patience2012}
Patience, J., King, R.~R., Rosa, R. J.~D., {et~al.} 2012,
  \JournalTitle{A{\&}A}, 85, 1

\bibitem[{Perez \& Granger(2007)}]{Perez2007}
Perez, F., \& Granger, B.~E. 2007,
  \href{http://dx.doi.org/10.1109/MCSE.2007.53}{\JournalTitle{Comput. Sci.
  Eng.}, 9, 21}

\bibitem[{Rackham {et~al.}(2018)Rackham, Apai, \& Giampapa}]{Rackham2018}
Rackham, B.~V., Apai, D., \& Giampapa, M.~S. 2018, \JournalTitle{The
  Astrophysical Journal}, 853, 122

\bibitem[{Radigan {et~al.}(2012)Radigan, Jayawardhana, Lafreni{\`{e}}re,
  Artigau, Marley, \& Saumon}]{Radigan2012}
Radigan, J., Jayawardhana, R., Lafreni{\`{e}}re, D., {et~al.} 2012,
  \href{http://dx.doi.org/10.1088/0004-637X/750/2/105}{\JournalTitle{ApJ}, 750,
  105}

\bibitem[{{Rajan} {et~al.}(2015){Rajan}, {Barman}, {Soummer}, {Hagan},
  {Patience}, {Pueyo}, {Choquet}, {Konopacky}, {Macintosh}, \&
  {Marois}}]{Rajan2015}
{Rajan}, A., {Barman}, T., {Soummer}, R., {et~al.} 2015,
  \href{http://dx.doi.org/10.1088/2041-8205/809/2/L33}{\JournalTitle{\apjl},
  809, L33}

\bibitem[{Rajan {et~al.}(2017)Rajan, Rameau, Rosa, Marley, Graham, Macintosh,
  Marois, Morley, Patience, Pueyo, Saumon, Ward-Duong, Ammons, Arriaga, Bailey,
  Barman, Bulger, Burrows, Chilcote, Cotten, Czekala, Doyon, Duch{\^{e}}ne,
  Esposito, Fitzgerald, Follette, Fortney, Goodsell, Greenbaum, Hibon, Hung,
  Ingraham, Johnson-Groh, Kalas, Konopacky, Lafreni{\`{e}}re, Larkin, Maire,
  Marchis, Metchev, Millar-Blanchaer, Morzinski, Nielsen, Oppenheimer, Palmer,
  Patel, Perrin, Poyneer, Rantakyr{\"{o}}, Ruffio, Savransky, Schneider,
  Sivaramakrishnan, Song, Soummer, Thomas, Vasisht, Wallace, Wang, Wiktorowicz,
  \& Wolff}]{Rajan2017}
Rajan, A., Rameau, J., Rosa, R. J.~D., {et~al.} 2017,
  \href{http://dx.doi.org/10.3847/1538-3881/aa74db}{\JournalTitle{AJ}, 154, 10}

\bibitem[{Robitaille {et~al.}(2013)Robitaille, Tollerud, Greenfield,
  Droettboom, Bray, Aldcroft, Davis, Ginsburg, Price-Whelan, Kerzendorf,
  Conley, Crighton, Barbary, Muna, Ferguson, Grollier, Parikh, Nair,
  G{\"{u}}nther, Deil, Woillez, Conseil, Kramer, Turner, Singer, Fox, Weaver,
  Zabalza, Edwards, {Azalee Bostroem}, Burke, Casey, Crawford, Dencheva, Ely,
  Jenness, Labrie, Lim, Pierfederici, Pontzen, Ptak, Refsdal, Servillat, \&
  Streicher}]{Robitaille2013}
Robitaille, T.~P., Tollerud, E.~J., Greenfield, P., {et~al.} 2013,
  \href{http://dx.doi.org/10.1051/0004-6361/201322068}{\JournalTitle{A{\&}A},
  558, A33}

\bibitem[{Samland {et~al.}(2017)Samland, Molli{\`e}re, Bonnefoy, Maire,
  Cantalloube, Cheetham, Mesa, Gratton, Biller, Wahhaj, {et~al.}}]{Samland2017}
Samland, M., Molli{\`e}re, P., Bonnefoy, M., {et~al.} 2017,
  \JournalTitle{Astronomy \& Astrophysics}, 603, A57

\bibitem[{{Saumon} \& {Marley}(2008)}]{Saumon2008}
{Saumon}, D., \& {Marley}, M.~S. 2008,
  \href{http://dx.doi.org/10.1086/592734}{\JournalTitle{\apj}, 689, 1327}

\bibitem[{Scargle(1982)}]{Scargle1982a}
Scargle, J.~D. 1982,
  \href{http://dx.doi.org/10.1086/160554}{\JournalTitle{ApJ}, 263, 835}

\bibitem[{Scholz {et~al.}(2018)Scholz, Moore, Jayawardhana, Aigrain, Peterson,
  \& Stelzer}]{Scholz2018}
Scholz, A., Moore, K., Jayawardhana, R., {et~al.} 2018,
  \href{http://dx.doi.org/10.3847/1538-4357/aabfbe}{\JournalTitle{ApJ}, 859,
  153}

\bibitem[{Skemer {et~al.}(2011)Skemer, Close, Szűcs, Apai, Pascucci, \&
  Biller}]{Skemer2011}
Skemer, A.~J., Close, L.~M., Szűcs, L., {et~al.} 2011,
  \href{http://dx.doi.org/10.1088/0004-637X/732/2/107}{\JournalTitle{ApJ}, 732,
  107}

\bibitem[{Snellen {et~al.}(2014)Snellen, Brandl, de~Kok, Brogi, Birkby, \&
  Schwarz}]{Snellen2014}
Snellen, I. A.~G., Brandl, B.~R., de~Kok, R.~J., {et~al.} 2014,
  \href{http://dx.doi.org/10.1038/nature13253}{\JournalTitle{Nature}, 509, 63}

\bibitem[{Song {et~al.}(2006)Song, Schneider, Zuckerman, Farihi, Becklin,
  Bessell, Lowrance, \& Macintosh}]{Song2006}
Song, I., Schneider, G.~H., Zuckerman, B., {et~al.} 2006,
  \href{http://dx.doi.org/10.1086/507831}{\JournalTitle{ApJ}, 652, 724}

\bibitem[{Soummer {et~al.}(2012)Soummer, Pueyo, \& Larkin}]{Soummer2012}
Soummer, R., Pueyo, L., \& Larkin, J. 2012, \JournalTitle{The Astrophysical
  Journal Letters}, 755, L28

\bibitem[{Tan \& Showman(2018)}]{Tan2018}
Tan, X., \& Showman, A.~P. \href{http://arxiv.org/abs/1809.06467}{2018},
  \href{http://arxiv.org/abs/1809.06467}{{\sffamily arXiv:1809.06467}}

\bibitem[{van~der Walt {et~al.}(2011)van~der Walt, Colbert, \&
  Varoquaux}]{VanderWalt2011}
van~der Walt, S., Colbert, S.~C., \& Varoquaux, G. 2011,
  \href{http://dx.doi.org/10.1109/MCSE.2011.37}{\JournalTitle{Comput. Sci.
  Eng.}, 13, 22}

\bibitem[{VanderPlas(2018)}]{VanderPlas2018}
VanderPlas, J.~T. 2018,
  \href{http://dx.doi.org/10.3847/1538-4365/aab766}{\JournalTitle{ApJS}, 236,
  16}

\bibitem[{Vos {et~al.}(2018)Vos, Allers, Biller, Liu, Dupuy, Gallimore,
  Adenuga, \& Best}]{Vos2017}
Vos, J.~M., Allers, K.~N., Biller, B.~A., {et~al.} 2018,
  \href{http://dx.doi.org/10.1093/mnras/stx2752}{\JournalTitle{MNRAS}, 474,
    1041}

  \bibitem[Vos et al.(2019)]{Vos2018b} Vos, J.~M., Biller, B.~A., Bonavita, M., et al.\ 2019, \href{http://dx.doi.org/10.1093/mnras/sty3123}{\JournalTitle{\mnras}, 483, 480}

\bibitem[{Waskom {et~al.}(2017)Waskom, Botvinnik, O'Kane, Hobson, Lukauskas,
  Gemperline, Augspurger, Halchenko, Cole, Warmenhoven, de~Ruiter, Pye, Hoyer,
  Vanderplas, Villalba, Kunter, Quintero, Bachant, Martin, Meyer, Miles, Ram,
  Yarkoni, Williams, Evans, Fitzgerald, Brian, Fonnesbeck, Lee, \&
  Qalieh}]{Waskom2017}
Waskom, M., Botvinnik, O., O'Kane, D., {et~al.} 2017, mwaskom/seaborn: v0.8.1
  (September 2017)

\bibitem[{Yang {et~al.}(2014)Yang, Apai, Marley, Saumon, Morley, Buenzli,
  Artigau, Radigan, Metchev, Burgasser, Mohanty, Lowrance, Showman, Karalidi,
  Flateau, \& Heinze}]{Yang2014}
Yang, H., Apai, D., Marley, M.~S., {et~al.} 2014,
  \href{http://dx.doi.org/10.1088/2041-8205/798/1/L13}{\JournalTitle{ApJ}, 798,
  L13}

\bibitem[{Yang {et~al.}(2016)Yang, Apai, Marley, Karalidi, Flateau, Showman,
  Metchev, Buenzli, Radigan, Artigau, Lowrance, \& Burgasser}]{Yang2016}
---. 2016,
  \href{http://dx.doi.org/10.3847/0004-637X/826/1/8}{\JournalTitle{ApJ}, 826,
  8}

\bibitem[{Zhou {et~al.}(2017)Zhou, Apai, Lew, \& Schneider}]{Zhou2017}
Zhou, Y., Apai, D., Lew, B. W.~P., \& Schneider, G.~H. 2017,
  \href{http://dx.doi.org/10.3847/1538-3881/aa6481}{\JournalTitle{AJ}, 153,
  243}

\bibitem[{Zhou {et~al.}(2016)Zhou, Apai, Schneider, Marley, \&
  Showman}]{Zhou2016}
Zhou, Y., Apai, D., Schneider, G.~H., Marley, M.~S., \& Showman, A.~P. 2016,
  \href{http://dx.doi.org/10.3847/0004-637X/818/2/176}{\JournalTitle{ApJ}, 818,
  176}

\end{thebibliography}

\end{document}